\documentclass[brazil,12pt,a4paper,openany]{book}
\usepackage{anysize}
\usepackage[T1]{fontenc}
\usepackage[utf8]{inputenc}
\usepackage{makeidx}
\usepackage{subfig}
\usepackage{amsmath}
\usepackage{amssymb}
\usepackage{amsthm}
\usepackage{mathrsfs}
\usepackage{fancyhdr}
\usepackage{graphicx}
\usepackage{geometry}
\usepackage{simplewick}
\usepackage{latexsym}
\usepackage[all]{xy}
\usepackage{enumerate}
\usepackage{dsfont}
\usepackage{slashed}

\usepackage[style=authoryear]{biblatex}
\usepackage{csquotes}
\usepackage{pdfpages}
\usepackage{enumitem}
\usepackage{bm}

\makeindex

\graphicspath{{figs/}} %Setting the graphicspath

\newcommand{\T}{\mathrm{T}}
\newcommand{\mathbbm}[1]{\text{\usefont{U}{bbm}{m}{n}#1}} % from mathbbm.sty
\newcommand{\teta}{\tau_{\eta}}

\marginsize{25mm}{25mm}{25mm}{25mm}

\begin{document}

\setlength{\abovecaptionskip}{0.0cm}
\setlength{\belowcaptionskip}{0.0cm}
\setlength{\baselineskip}{24pt}

\pagestyle{fancy}
\lhead{}
\chead{}
\rhead{\thepage}
\lfoot{}
\cfoot{}
\rfoot{}

\fancypagestyle{plain}
{
	\fancyhf{}
	\lhead{}
	\chead{}
	\rhead{\thepage}
	\lfoot{}
	\cfoot{}
	\rfoot{}
}

\renewcommand{\headrulewidth}{0pt}

%%%%%%%%%%%%%%%%%%%%%%%%%%%%%%%%% Capa %%%%%%%%%%%%%%%%%%%%%%%%%%%%%%%%%%

\frontmatter

\thispagestyle{empty}

\begin{figure}[h]
	\includegraphics[scale=0.8]{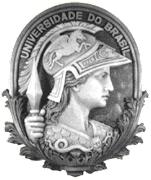}
\end{figure}

\vspace{15pt}

\begin{center}

\textbf{UNIVERSIDADE FEDERAL DO RIO DE JANEIRO}

\textbf{INSTITUTO DE FÍSICA}

\vspace{30pt}

{\Large \bf Large Fluctuations in Stochastic Models of Turbulence}

\vspace{25pt}

{\large \bf Gabriel Brito Apolinário}

\vspace{35pt}

\begin{flushright}
\parbox{10.3cm}{Tese de Doutorado apresentada ao Programa de Pós-Graduação em Física do Instituto de Física da Universidade Federal do Rio de Janeiro - UFRJ, como parte dos requisitos necessários à obtenção do título de Doutor em Ciências (Física)}.

\vspace{18pt}

{\large \bf Orientador: Luca Moriconi}

%\vspace{12pt}

%{\large \bf Coorientador: Nome do Coorientador}}
\end{flushright}

\vspace{90pt}

\textbf{Rio de Janeiro}

\textbf{Abril de 2020}

\end{center}

%%%%%%%%%%%%%%%%%%%%%%%%%%%%%%%%%%%%%%%%%%%%%%%%%%%%%%%%%%%%%%%%%%%%%%%%%

%%%%%%%%%%%%%%%%%%%% Folha de Assinaturas da Banca %%%%%%%%%%%%%%%%%%%%%%

\newpage

\thispagestyle{empty}

\noindent

%\begin{figure}[h]
%  \begin{center}
%	\includegraphics[scale=0.45]{assinaturas.png}
%  \end{center}
%\end{figure}

\clearpage

%%%%%%%%%%%%%%%%%%%%%%%%%%%%%%%%%%%%%%%%%%%%%%%%%%%%%%%%%%%%%%%%%%%%%%%%%

%\includepdf[pages=-]{ficha.pdf}
%\includepdf[pages=-]{aprovacao.pdf}

%%%%%%%%%%%%%%%%%%%%%%%%%%%%%%%%%%%%%%%%%%%%%%%%%%%%%%%%%%%%%%%%%%%%%%%%%

%%%%%%%%%%%%%%%%%%%%%%%%%%%%%%%% Resumo %%%%%%%%%%%%%%%%%%%%%%%%%%%%%%%%%

\newpage

\noindent

\vspace*{20pt}
\begin{center}
{\LARGE\bf Resumo}\\
\vspace{15pt}
{\Large\bf Large Fluctuations in Stochastic Models of Turbulence}\\
\vspace{6pt}
{\bf Gabriel Brito Apolinário}\\
\vspace{12pt}
{\bf Orientador: Luca Moriconi}\\
%{\bf Coorientador: Nome do Coorientador}\\
\vspace{20pt}
\parbox{14cm}{Resumo da Tese de Doutorado apresentada ao Programa de Pós-Graduação em Física do Instituto de Física da Universidade Federal do Rio de Janeiro - UFRJ, como parte dos requisitos necessários à obtenção do título de Doutor em Ciências (Física).}
\end{center}
\vspace*{35pt}

As primeiras observações de flutuações estatísticas em turbulência, descritas pelas equações de Navier-Stokes, foram feitas na década de 1940 e já revelavam a presença de intermitência: Dis\-tri\-bui\-ções de probabilidade de caudas largas em observáveis como gradientes de velocidade, diferenças de velocidade, e dissipação de energia. Hoje em dia, com experimentos cada vez mais precisos e simulações numéricas em números de Reynolds cada vez maiores, reitera-se as observações de grandes flutuações, mas uma descrição teórica que conecta as equações de Navier-Stokes à estatística observada ainda é um problema em aberto. Esta tese discute o fenômeno da intermitência em três modelos de turbulência, se valendo de técnicas analíticas e numéricas para a análise de processos estocásticos e as distribuições estatísticas que estes geram.

Os capítulos iniciais apresentam uma revisão da teoria estat\'{i}stica da turbulência e da teoria estatística de campos aplicada à física fora do equilíbrio. É discutida a história das equações de Navier-Stokes e a necessidade de uma formulação estocástica destas para explicar os fenômenos da turbulência. Em seguida, os sucessos e limitações da fenomenologia de Kolmogorov de 1941 são apresentados. A principal dessas limitações é não incluir os efeitos das flutuações, que são ponto central da teoria de Kolmogorov-Obukhov e da teoria multifractal, abordadas em seguida. Também é feita uma revisão dos métodos funcionais aplicados a sistemas de equações diferenciais estocásticas.

O modelo Recent Fluid Deformation (RFD), que descreve a dinâmica de gradientes de velocidade lagrangeanos, e o modelo de Burgers, que descreve a formação de ondas de choque em fluidos unidimensionais, são investigados através do método funcional Martin-Siggia-Rose-Janssen-de Dominicis (MSRJD). Nesta formulação, os instantons (soluções das equações de Euler-Lagrange obtidas a partir da ação de MSRJD) já tinham sido identificados como a contribuição principal à distribuição de probabilidade de gradientes de velocidade nos dois modelos. Nesta tese é feita uma análise cuidadosa das hipóteses e resultados que levaram a essas observações anteriores sobre os instantons e sobre correções perturbativas correspondentes a flutuações nos dois modelos.

No caso do RFD, é feita uma contagem de potências dos diagramas perturbativos associados às flutuações ao redor dos instantons e uma verificação da validade do instanton aproximado usado em cálculos anteriores. No modelo de Burgers, foram identificados dois diagramas perturbativos até segunda ordem na expansão em cumulantes, que são cruciais para entender um procedimento \textit{ad hoc} de renormalização já observado em discussões anteriores: A distribuição de probabilidade gerada pelos instantons descreve bem as distribuições de probabilidade numéricas, mas é necessário adicionar um fator de correção ao parâmetro que regula a intensidade das flutuações. Foi investigada a contribuição das flutuações para induzir esse fator de correção.

Por último, é apresentado um processo estocástico estacionário e unidimensional como um modelo das flutuações da pseudo-dissipação lagrangeana. Esse processo apresenta ruído aleatório na escala microscópica, construído a partir de um processo multifractal discreto, assim como uma dinâmica regular em escalas ainda menores. Foram verificadas as propriedades estatísticas desse processo aleatório, que apresenta uma distribuição de probabilidade lognormal, e correlações de longo alcance na forma de leis de potência, de acordo com as propriedades estatísticas experimentais da pseudo-dissipação.

\vspace{15pt}

\textbf{Palavras-chave:} Turbulência, modelos estocásticos, eventos extremos, formalismo Martin-Sig\-gia-Ro\-se, renormalização.

%%%%%%%%%%%%%%%%%%%%%%%%%%%%%%% Abstract %%%%%%%%%%%%%%%%%%%%%%%%%%%%%%%%

\newpage

\noindent

\vspace*{20pt}
\begin{center}
{\LARGE\bf Abstract}\\
\vspace{15pt}
{\Large\bf Large Fluctuations in Stochastic Models of Turbulence}\\
\vspace{6pt}
{\bf Gabriel Brito Apolinário}\\
\vspace{12pt}
{\bf Orientador: Luca Moriconi}\\
%{\bf Coorientador: Name of the Coadvisor}\\
\vspace{20pt}
\parbox{14cm}{\emph{Abstract} da Tese de Doutorado apresentada ao Programa de Pós-Graduação em Física do Instituto de Física da Universidade Federal do Rio de Janeiro - UFRJ, como parte dos requisitos necessários à obtenção do título de Doutor em Ciências (Física).}
\end{center}
\vspace*{35pt}

The first observations of statistical fluctuations in turbulence, described by the Navier-Stokes equations, were made in decade of 1940 and already revealed the presence of intermittency: The probability distribution functions of observables such as velocity gradients, velocity differences and kinetic energy dissipation are heavy tailed. Nowadays, with more precise experiments and numerical simulations at larger Reynolds numbers, the observations of large fluctuations are reinforced, but a theoretical description that connects the Navier-Stokes equations to the observed statistics is still an open problem. This dissertation discusses the intermitency phenomenon in three models of turbulence, employing analytical and numerical techniques in the analysis of stochastic processes and the probability distributions which they induce.

The initial chapters present a review of the statistical theory of turbulence and of statistical theory of fields as applied to out of equilibrium physics. The history of the Navier-Stokes equation is discussed and associated to the need for a stochastic formulation of them in order to explain turbulent phenomena. Then, the successes and limitations of the Kolmogorov 1941 phenomenology are presented. The predominant limitation is not to include fluctuation effects, the central point of the Kolmogorov-Obukhov and multifractal theories, which are reviewed. A summary of functional methods applied to systems of stochastic differential equations is also presented.

The RFD model, which describes the dynamics of Lagrangian velocity gradients, and the Burgers model, describing the formation of shock waves in one-dimensional fluids, are investigated through the Martin-Siggia-Rose-Janssen-de Dominicis (MSRJD) functional method. In this formulation, the instantons (solutions of the Euler-Lagrange equations obtained from the MSRJD action) had already been identified as the leading contribution to the velocity gradient probability distribution function in both models. In this dissertation, a careful analysis of the previous hypothesis and results is undertaken, in order to verify the observations of perturbative corrections corresponding to fluctuations around the instanton in both models.

In the Recent Fluid Deformation (RFD) model, a power-counting procedure on the perturbative diagrams associated to fluctuations around the instantons is pursued, along with a validation of the approximate instanton used in prior calculations. In the Burgers model, two perturbative diagrams were identified up to second order in the cumulant expansion. They are revealed to be crucial in understanding an \textit{ad hoc} renormalization procedure reported in the literature: The probability distribution induced by the instantons describes velocity gradient fluctuations well, if a correction factor is added to the parameter associated to the intensity of fluctuations. The contribution of fluctuations to cause this correction factor is investigated.

In the end, a stationary one-dimensional stochastic process is presented as a model of Lagrangian pseudo-dissipation fluctuations. This model displays random noise in the microscopic scale, built from a discrete multifractal process, but at smaller scales its dynamics is regular. The statistical properties of this random process are verified, and it is observed that, in agreement with established properties of the pseudo-dissipation, this process has a lognormal probability distribution and long range power-law correlations.

\vspace{15pt}

\textbf{Keywords:} Turbulence, stochastic models, extreme events, Martin-Siggia-Rose formalism, renormalization.

%%%%%%%%%%%%%%%%%%%%%%%%%%%% Agradecimentos %%%%%%%%%%%%%%%%%%%%%%%%%%%%%

\newpage

\noindent

\vspace*{20pt}

\begin{center}

{\LARGE\bf Agradecimentos}

\end{center}

\vspace*{40pt}

Agradeço aos meus pais pelo apoio e pela confiança ao longo destes anos de estudo. Nada disso seria possível sem esse incentivo e amparo.

Pelo aprendizado sobre a prática da ciência, que inclui as técnicas, a leitura, a escrita, as discussões e o pensamento crítico, e pelos riquíssimos problemas científicos que me propôs, agradeço ao Prof. Luca Moriconi.

Agradeço também aos companheiros semanais no journal club de turbulência e dinâmica de fluidos, Rodrigo Arouca, Elvis Soares, Bruno Magacho, Victor Valadão, Mirlene Oliveira, Marina Moesia e Maiara Neumann. E agradeço ao Rodrigo Pereira pela valiosa colaboração, e pelos ensinamentos em programação.

Agradeço aos novos amigos que fiz nesse período na UFRJ, aos amigos de longa data da UFF, e aos amigos que fiz nas viagens, ao longo do doutorado. Vinicius Henning, Pedro Foster, Reginaldo Junior, Luis Fernando, Rodrigo Bruni, Kainã Diniz, Flavianna Siller, Lucas Hutter, Lucas Torres,  Patrícia Abrantes, Larissa Inácio, Ethe Costa, Evelyn Caso, Daniel Martin, Irene Lamberti, Fabiana Monteiro, Humberto Medeiros, Gleice Germano, Leonardo Pio, Ruslan Guerra: Obrigado.

Aos funcionários Igor Silva, Khrisna Teixeira, César Chagas e Mariana Sampaio, agradeço pelo apoio em questões técnicas e administrativas.
E à Luana Serpa pela ajuda com o catálogo da exposição \enquote{Turbulência e Arte} e as respectivas fotos que abrem alguns dos capítulos.

Por fim, agradeço ao CNPq pelo surporte financeiro, sem o qual este doutoramento não teria sido possível.

%%%%%%%%%%%%%%%%%%%%%%%%%%%%%%%%%%%%%%%%%%%%%%%%%%%%%%%%%%%%%%%%%%%%%%%%%

%%%%%%%%%%%%%%%%%%%%%%%%%%%%%%%%%%%%%%%%%%%%%%%%%%%%%%%%%%%%%%%%%%%%%%%%%

\newpage
%\addcontentsline{toc}{chapter}{Contents}
\tableofcontents

\newpage
\addcontentsline{toc}{chapter}{List of Figures}
\listoffigures

%\newpage
%\addcontentsline{toc}{chapter}{Lista de Tabelas}
%\listoftables

\newpage
\addcontentsline{toc}{chapter}{List of Abbreviations}

\chapter*{List of Abbreviations}

\begin{table}[h]
\begin{tabular}{ll}
DNS & Direct Numerical Simulation \\
LHS & Left-hand side \\
MSRJD & Martin-Siggia-Rose-Janssen-de Dominicis \\
PDF  & Probability Density Function \\
RFD & Recent Fluid Deformation \\
RHS & Right-hand side \\
SDE & Stochastic Differential Equation \\
SPDE & Stochastic Partial Differential Equation \\
\end{tabular}
\end{table}

\mainmatter
\begin{chapter}{Introduction}
\label{cap1}

\hspace{5 mm}

In the beginning of the XXth century, the Isar Company of Munich
had the task of building banks for the Isar River, to prevent it from
flooding. They contacted Arnold Sommerfeld with the question:
At what point would the river flow change
from smooth (laminar) to irregular (turbulent)?
Sommerfeld and Werner Heisenberg, then, performed an
analysis of the equations of flow and predicted a limit of
stability for the smooth solution.
As a follow-up to this story, there is an apocryphal quote attributed
to Heisenberg (cited by \textcite{ball2014scientific}:
\begin{displayquote}
When I meet God, I am going to ask him two questions. Why relativity?
And why turbulence? I really believe he will have an answer for the first.
\end{displayquote}
The same citation has been attributed as well to Richard Feynman and to Horace Lamb,
and relativity is sometimes replaced with quantum electrodynamics.
For more details on these stories, the
reader is referred to \textcite{ball2014scientific,eckert2017}.

This tale illustrates the lasting interest that turbulence
has had as a scientific problem, both in pure and applied research.
Turbulent flows are ubiquitous in daily and industrial flows,
such as the atmosphere, the ocean, combustion engines and wind tunnels.
It is easy to imagine that mankind has tried to understand
and control it since the beginning of science.
The first theoretical contributions to this problem came from
the founders of classical mechanics:
Newton, Euler, Bernoulli, Lagrange, Navier and Stokes, among others, who
determined the equations which dictate the evolution
of velocity and pressure in viscous and inviscid flows,
respectively the Navier-Stokes equation and the Euler equation.

The modern understanding of the problem
% of turbulence
only came with
the experiments and theories developed in the XX century, by scientists such as
Osborne Reynolds (1842-1912), Geoffrey I. Taylor (1886-1975),
Lewis Fry Richardson (1881-1953), Ludwig Prandtl (1875-1953)
and Theodore von Kármán (1881-1963),
who described the phenomena of the laminar-turbulent transition,
the formation of vortical structures, the energy cascade
and the boundary layer.
Their contributions served as inspirations to theorists
such as Andrey Kolmogorov (1903-1987), Lev Landau (1908-1968)
and Lars Onsager (1903-1976), who built the first models
to describe these phenomena and still inspire much of
the current developments in this research area.
% other names: Mandelbrot, Kraichnan, Hopf}

The Navier-Stokes and Euler equations have attracted the interest of
mathematicians, as well, for they provide seemingly simple equations,
but with complex nonlinearities, still not fully understood.
Even the basic problem of existence and smoothness of solutions of these
equations remains open. This is a famous open problem
in mathematics, one of the Millenium Prize Problems,
with a US\$ 1 million prize offered by the Clay Mathematics Institute
for its solution. The formal statement asks for a proof of existence (or non-existence) of global regularity in the Navier-Stokes equations. In other words, if starting with smooth initial conditions, do the velocity and pressure fields remain regular and analytical for any finite time or do they develop singularities?

The global regularity problem is also open for the Euler equation in three dimensions, but partial results in different settings have been obtained. For Navier-Stokes in two dimensions, regularity was proven in \textcite{ladyzhenskaya1969mathematical}.
In three dimesions, the existence of \textit{weak solutions} to the Navier-Stokes equations was proved in \textcite{leray1934} (weak solutions are briefly discussed in Sec.~\ref{sec:onsager}).
%And in \textcite{tao2016finite}, a finite time blowup solution was found for an averaged version of Navier-Stokes, which hints at the absence of global regularity.
A rigorous description of this problem can be found
in \textcite{fefferman2006existence}, and a mathematical discussion
of the theory of turbulence is available in \textcite{temam2001}.

\begin{figure}[t]
    \centering
    \includegraphics[width=.69\textwidth]{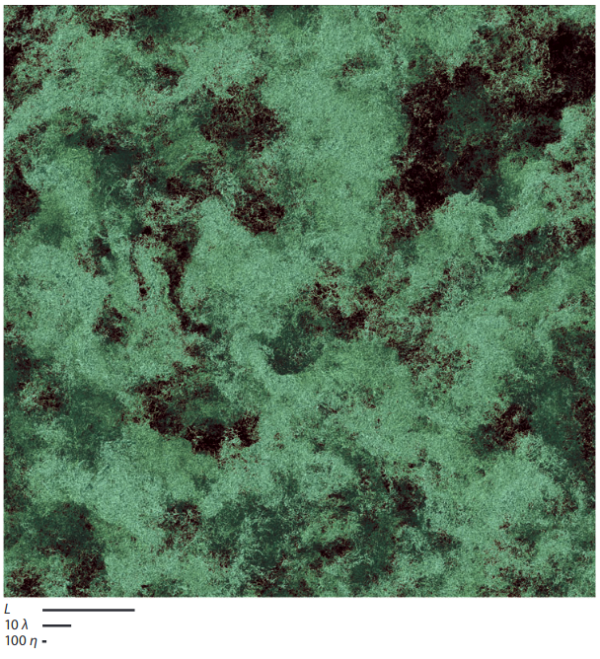}
    \caption[Intense vorticity isosurfaces in a large-scale DNS]
    {Large scale numerical simulation of a turbulent flow from \textcite{ishihara2009}.
    Intense vorticity isosurfaces are depicted, the vorticity is
    displayed if it deviates more than four standard deviations from the mean. Clouds of small eddies interleaved with void regions can be seen, illustrating the non-homogeneity of turbulent fiels.}
    \label{fig:cross-section}
\end{figure}

Another area of study with rich interactions with turbulence
is that of numerical simulations of partial differential equations.
Various numerical schemes for their solution have been developed, most notably the spectral methods,
which rely on fast algorithms for the numerical Fourier transform
to obtain higher precision than would be possible with
numerical simulations of the same size in real space.
Spectral methods were first applied in turbulence in \textcite{orszag1972},
on a grid of $32^3$ points, and have evolved
to simulations with $16\ 384^3$ points \parencite{iyer2019},
where very detailed structures can be identified. This is illustrated in Fig.~\ref{fig:cross-section}, extracted from \textcite{ishihara2009}. In this work, snapshots of a simulation with $4096^3$ grid points are shown, and structures such as vortex tubes are clearly seen.
%An initiative created for the analysis of The Johns Hopkins Turbulence Database is
%described in \cite{li2008} and available in \cite{jhtdb}.

Notable as well is the importance that the study of turbulence has
on technological applications. In some industrial areas, it is desired to suppress turbulence, such as in flows through pipes, where the external pressure difference applied to the ends of the pipe is greatly reduced, for a fixed flow rate,
if the fluid is calm and laminar, rather than irregular and turbulent.
In other situations turbulence is beneficial, for instance in the efficient
mix of chemical reactants, where turbulent flows vastly accelerate
the dispersal of chemicals and the occurrence of reactions.
For a review of engineering problems and turbulence, the reader
is referred to \textcite{dewan2010,ting2016}.

To summarize, the prominent features of turbulent flows, which permit us to understand the challenges in their study and the potential for applications are:
\begin{enumerate}
\item enhanced energy dissipation, even in the limit of vanishing viscosity;
\item strong mixing of energy, momentum, mass and heat;
\item unpredictability of flow configurations.
\end{enumerate}
A discussion of these properties is pursued in Chap.~\ref{chap:turb}, where the statistical theory of turbulence is discussed, and the topics investigated in this dissertation are different manifestations of these phenomena. The inclusion of this general discussion has the objective of making this text self-contained and accessible to researchers and students unfamiliar with the topic, since the theory of turbulence is seldom included in graduation curricula. The content of the review chapter is based on the discussions of
\textcite{frisch1995,foias2001navier,moriconi2008introducao,eyink2008turbulence} with references to further sources.

Chap.~\ref{chap:stoc} is, likewise, a review on the theory of stochastic calculus, functional methods and instantons, where its history and some techniques are discussed. These techniques are employed in the next chapters, which cover the original contributions discussed in this dissertation. In this chapter, the discussion is mostly based on \textcite{gardiner2009,grafke2015instanton,canet2019leshouches}.

In Chap.~\ref{chap:rfd}, Lagrangian turbulence and statistical closures are introduced, along with a discussion of the results of \textcite{apolinario2019instantons}. To predict transport properties, such as mixing and spread of dispersed particles, it is necessary to understand Lagrangian velocity gradients, but the equation for their dynamics, derived from the Navier-Stokes equations, is unclosed. Consequently, many closure approximations have been developed, particularly the Recent Fluid Deformation (RFD), in \textcite{ChevPRL}, which was investigated with functional methods in \textcite{moriconi2014}.
An extension of these analytical results is pursued in two ways: A hierarchical classification of perturbative contributions is performed and the most relevant diagrams are integrated into the renormalized effective action. The approximate instanton hypothesis is also verified to be true.
%It is observed that renormalization plays a significant role in the description of non-Gaussian cores of velocity gradient PDFs.

The next chapter,~\ref{chap:burgers}, discusses the onset
of intermittency in Burgers turbulence along the lines of the functional formalism. The Burgers model is a one-dimensional version of the Navier-Stokes equation, which shares many qualitative features with the original model, hence it is sometimes used as test case for analytical and numerical techniques in turbulence. For this equation, the relevance of the instanton approach in the description of large fluctuations of the velocity gradient has been verified in previous works \parencite{grafke2015relevance}. Nevertheless, it was also revealed the need of an empirical noise renormalization. In \textcite{apolinario2019onset}, on which this chapter is based, a theoretical explanation for the mechanism of noise renormalization is presented.
% as arising from fluctuations around the instanton description.

In Chap.~\ref{chap:shotnoise}, a different route is taken in the investigation of Lagrangian pseudo-dissipation. A stochastic differential equation with a stationary solution is used to model the fluctuations of this observable. This stochastic process is driven by shot noise, a periodic and discrete source of randomness, inspired by the discrete and exactly multifractal causal process described in \textcite{perpete2011}. It is verified that this discrete dynamics leads to multifractal statistics and long-range correlations compatible with known models of pseudo-dissipation. This chapter is based on 
\textcite{apolinario2020shotnoise}.

\end{chapter}

\begin{chapter}{Statistical Theory of Turbulence}
\label{chap:turb}

\hspace{5 mm}

The pursuit of a complete statistical theory of turbulence
has been singular for some reasons. Its phenomena have been
recognized since ancient times, yet a descriptive
approach which stems from the Navier-Stokes equations,
and derives
the results known nowadays still seems far from being realized.
%In Eq.~\eqref{eq:ns}, $u_i$ is the velocity of the fluid,
%$\nu$ the kinematic viscosity, $p$ is the pressure, and $f_i$
%is an external force.
%The second equation describes the incompressibility condition.
The most successful attempts have been to build phenomenological
theories, and their use was aptly described by Feynman, talking about
superfluidity:
\begin{displayquote}
Rather than look at the Hamiltonian we shall \enquote{wave our hands},
use analogies
with simpler systems, draw pictures, and make plausible guesses based on physical
intuition to obtain a qualitative picture of the solutions (wave functions). This
qualitative approach will prove singularly successful.
\parencite[p.321]{feynman1972}
\end{displayquote}
Just as described, the theory of turbulence has relied strongly
on physical pictures and analogies.
One of the main ideas that have inspired the construction of theories,
and forms the basic picture of turbulence, is the
energy cascade. First stated in \textcite{richardson1922weather},
it described turbulent flows as composed of eddies of different sizes, ranging
from the largest scales in the flow to the smallest.
Kinetic energy is injected into the flow at the large scales,
through the external force, and generates
large vortices, which are unstable and break up.
Energy is then transferred progressively and without loss to smaller eddies,
that go through the same breakup process.
After reaching the smallest scales, energy is finally dissipated by viscosity. This process is illustrated in Fig.~\ref{fig:cascade}

\begin{figure}[ht]
    \centering
    \includegraphics[width=1.\textwidth]{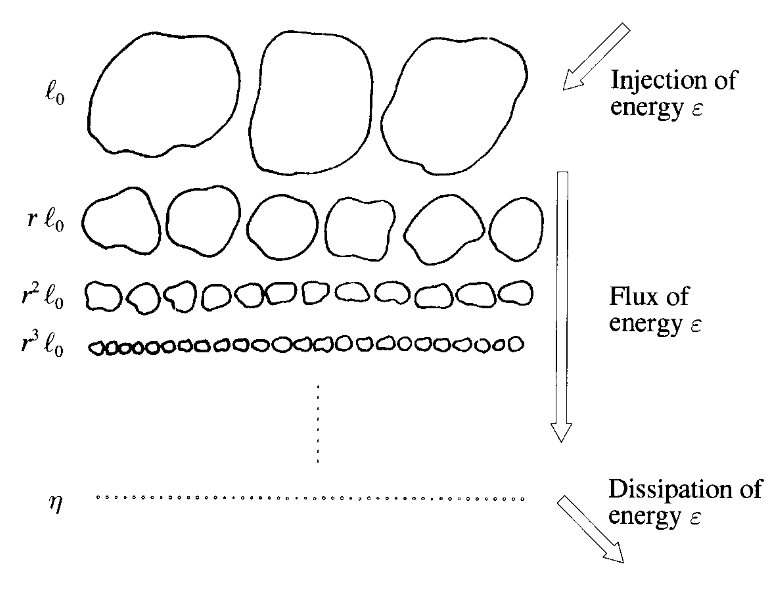}
    \caption[An illustration of the Richardson cascade]
    {The Richardson cascade is illustrated in this figure from \textcite{frisch1995}. Vortices of different scales are shown, from the integral scale down to the dissipative scale. The transference of kinetic energy across scales is local and inviscid, that is, the same amount of energy being is transferred between neighboring scales at every step.}
    \label{fig:cascade}
\end{figure}

The cascade picture was the basic inspiration of Kolmogorov in building
his theory, which proposed a self-similar velocity
field as approximate solution to the Navier-Stokes equations.
The approach of Kolmogorov was to incorporate experimental
and theoretical knowledge into a theory in order to build
a consistent framework from which further predictions could
be made. Some of these predictions are the scaling behavior
of statistical observables, such as velocity differences,
energy dissipation and enstrophy. All of these observables will
be defined in later sections.

From the energy cascade picture, a large separation of scales can be inferred.
The energy containing range is placed in one end of the spectrum, the large scale,
and the dissipation range in the other end, the small scale.
This separation of scales also induces the idea of scale invariance
of some observables, which, together with its anomalous breaking, is one of the main ingredients of the current understanding of turbulent flows.

\section{The Navier-Stokes Equations}

Fluids are treated as a continuous medium, despite
the knowledge, confirmed in the beginning of the XX century that matter
is made of atoms and is discontinuous.
A successful continuum description relies on the large separation
between the molecular scale (set by the mean free path of the molecules of the fluid, $\lambda$) and the typical scales of the flow
(the integral scale $L$, defined by the external force on the fluid and by the geometry).
In quantitative terms, the Knudsen number, $\mathrm{Kn} = \lambda/L$, must satisfy $\mathrm{Kn} \ll 1$ for a valid continuum formulation.
In this regime, the fluid can be described by quantities which
vary with continuous position and time indexes: A single instant of a flow is characterized by its density, velocity and pressure at each point.
This is called the Eulerian point of view in fluid dynamics, in which the
coordinates refer to a fixed reference frame.
There is also the Lagrangian frame, in which individual parcels
of fluid are tracked, and the position index refers to the position
of this \enquote{fluid particle} in the initial time.
By \enquote{fluid particle} it is not meant a molecule, but
an amount of fluid which can be treated as an individual particle,
it is much smaller than all relevant scales in the flow, yet much larger
than the mean free path of its molecules. This is only possible
because of the large separation of scales.

Fluid dynamics is then the result of applying conservation of mass,
energy and momentum to a continuous medium.
Mass conservation in this context is also called the equation of continuity:
\begin{equation} \label{eq:mass}
    \partial_t \rho + \partial_i (\rho u_i) = 0 \ .
\end{equation}
Notice that, in this equation and in the rest of this dissertation, latin
indices stand for spatial coordinates, ranging from one to three.
Derivatives with respect to time are represented by $\partial_t$
and derivatives with respect to space coordinate $x_i$ by $\partial_i$.
And the Einstein convention for
summation over repeated indices is adopted, unless otherwise specified.

Momentum conservation is written through Newton's second law,
in terms of the acceleration $\gamma_i$, at a certain point,
and the Cauchy stress tensor $\sigma_{ij}$, which describes
the interaction of different fluid parcels with each other:
\begin{equation}
    \rho \gamma_i = \partial_j \sigma_{ij} + f_i \ ,
\end{equation}
where $f_i$ is an external, divergenceless, body force, acting on the whole
fluid, with a possible dependence on position and time.
The acceleration is the material derivative of velocity:
\begin{equation}
    \gamma_i = D u_i / Dt = \partial_t u_i + u_j \partial_j u_i \ .
\end{equation}
The material derivative of the velocity is the acceleration experienced by a particle moving with the flow, which is different from the acceleration of the velocity at this point ($\partial_t u_i$).

The material derivative computes the time rate of change of any quantity such as temperature or velocity (which gives acceleration) for a portion of a material moving with a velocity, v. If the material is a fluid, then the movement is simply the flow field.

To proceed further, some knowledge of the properties of the fluid,
modelled through the $\sigma_{ij}$ tensor,
is needed. The simplest model for the stress tensor
is a linear dependence on velocity, which describes a Newtonian fluid:
\begin{equation} \label{eq:gen-stress}
    \sigma_{ij} = \mu (\partial_j u_i + \partial_i u_j )
    + (\lambda \partial_m u_m - p) \delta_{ij} \ .
\end{equation}
The variable $p$ is the pressure and it varies with position and time,
and the constants
$\mu$ and $\lambda$ are specific to each fluid, $\mu$ is
the molecular (or dynamic) viscosity
coefficient, and $3\lambda+2\mu$ is called the viscous dilatation coefficient.
The linear form was proposed by Newton, who also
performed the first experiments to measure the viscosity coefficient.
Eq.~\ref{eq:gen-stress} is the most straightforward model of a viscous fluid,
capturing the behavior of actual substances
such as air and water remarkably well in most quotidian or
industrial situations \parencite{kremer2010}.

In these common settings, a useful approach is to consider
the typical speeds in the flow much smaller than the speed
of sound in the medium. In this limit,
spatial differences in density adjust quickly compared to the motion of the
fluid, which means the density can be considered constant in the whole
flow. Such a flow is called incompressible.
In this special situation, Eq. \eqref{eq:mass} is altered to:
\begin{equation} \label{eq:incompressibility}
    \partial_i u_i = 0 \ ,
\end{equation}
and this is called the \textit{incompressibility condition}.
In turn, this also simplifies Eq. \eqref{eq:gen-stress}:
\begin{equation}
    \sigma_{ij} = \mu (\partial_j u_i + \partial_i u_j ) - p \delta_{ij}\ .
\end{equation}

With the above expressions the Navier-Stokes equations
for incompressible flows are obtained:
\begin{equation}
    \partial_t u_i + u_j \partial_j u_i
    = - \frac{1}{\rho} \partial_i p + \frac{\mu}{\rho} \partial^2 u_i \ ,
\end{equation}
where $\partial^2$ represents the Laplacian operator, $\partial_i \partial_i$.
Since the density is constant, it is customary to rewrite
\begin{equation} \label{eq:change-vars}
p/\rho \to p \qquad \mbox{and} \qquad \mu/\rho \to \nu \ ,
\end{equation}
where $\nu$ is called the kinematic viscosity.
Its value is known for all common fluids and is more directly relevant
to applications than the dynamic viscosity.
As an example, the kinematic viscosity of water at 20ºC is
$1.00 \times 10^{-6}$ m$^2$/s and $1.51 \times 10^{-5}$ m$^2$/s for air at 20ºC
(see \textcite{visc-water} and \textcite{visc-air}).
These values are not so distant from each other, despite the difference in the respective dynamic viscosity of nearly three decades.
% explanation: mu (Water) = 8.9 x 10^-3 \sim 10^-2
%              mu (Air)   = 1.8 x 10^-5 \sim 10^-5

Under the change of variables in Eq.~\eqref{eq:change-vars},
the incompressible Navier-Stokes equations become:
\begin{equation} \label{eq:ns}
\begin{split}
    \partial_t u_i + u_j \partial_j u_i
    &= - \partial_i p + \nu \partial^2 u_i + f_i \ , \\
    \partial_i u_i &= 0 \ .
\end{split}
\end{equation}

A complete description of a fluid dynamical requires these
equations,
%together with the incompressibility condition, Eq. \eqref{eq:incompressibility},
together with
an initial condition for the velocity field, and the no-slip boundary condition:
the velocity of the fluid at the boundaries is
always null.

The no-slip boundary condition was proposed by Stokes, who performed experiments in fluids and found no slip velocity. Later, a kinetic theory argument was used by Maxwell to demonstrate that the slip velocity is of the order of magnitude of the mean free path of the molecules of the fluid, hence for macroscopic fluids, the no-slip boundary condition is well-suited. More details on this topic can be found in \textcite{denniston2006,shen2007} and in \textcite[Chap.6]{eyink2008turbulence}.

Observe that the incompressibility condition renders the pressure field
non-local. This effect can be seen by deriving
the Navier-Stokes equations with respect to coordinate $i$ and obtaining
\begin{equation}
    \partial^2 p = - (\partial_i u_j) (\partial_j u_i) \ ,
\end{equation}
which is the Poisson equation. It can be solved with the use of
Green's functions and its solution for $p$ at each point is an integral
over the whole domain, hence is it nonlocal.
This already illustrates
some of the difficulties in determining mathematical
solutions to the Navier-Stokes equations.

\section{Reynolds Similarity} \label{sec:similarity}

Important contributions to the phenomenology of turbulence
came from the British scientist Osborne Reynolds (1842 - 1912), who
performed paramount experiments in this subject.
Many of the methods developed by him at this time have a direct
connection with modern experimental techniques for visualization
of flows. Reynolds investigated flows on long pipes,
which be seen in Fig.~\ref{fig:reynolds-draw}, in drawings
from his notebooks.
Using small jets of dyed water introduced into
the center of the flow and electric sparks, he could visualize
structures in the motion of water in the pipe.

\begin{figure}[t]
    \centering
    \begin{minipage}{0.45\textwidth}
        \centering
        \includegraphics[width=.9\textwidth]{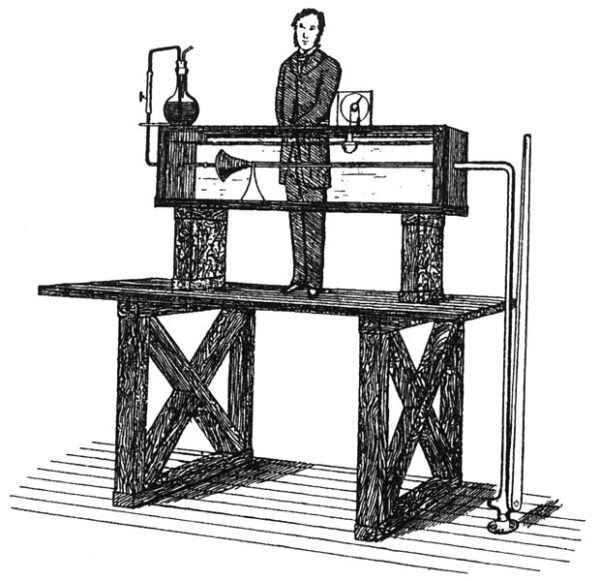} % first figure itself
    \end{minipage}\hfill
    \begin{minipage}{0.45\textwidth}
        \centering
        \includegraphics[width=.9\textwidth]{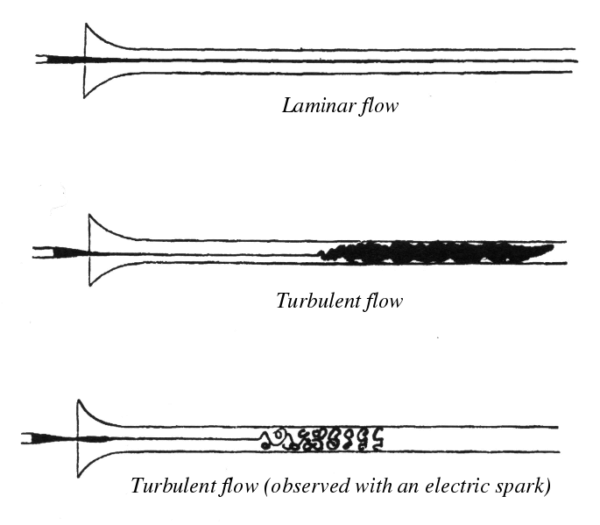} % second figure itself
    \end{minipage}
    \caption
    [Original drawings of Osborne Reynolds depicting the la\-mi\-nar-turbulent transition]
    {The original drawings of \textcite{reynolds1883} describe his experimental apparatus: A long pipe, in the middle of which a jet of dyed water was inserted. A valve allowed him to control the input velocity of the fluid, and thus demonstrate the transition from a laminar state to a turbulent flow. With an electric spark, vortical structures in the pipe could also be visualized.}
    \label{fig:reynolds-draw}
\end{figure}

In an experiment from 1883, he
demonstrated the transition of a flow from an ordered state to another
which is irregular and unpredictable.
The first is called a laminar flow, because different streams in
the fluid seem to form
independent layers, barely interacting with each other.
And the latter is a turbulent flow. It displays complex
swirly patterns, called vortices, which mix
all layers of the stream.
A flow control
valve was used to regulate the inlet velocity of water in the tube and Reynolds noticed
that, as the velocity increases, the behavior of the flow changes from
the laminar to the turbulent state.

The scientist also noticed that a single dimensionless number is responsible
for this transition. In his memory this parameter of the flow is
called the Reynolds number, defined as
\begin{equation} \label{eq:reynolds}
 \mathrm{Re} = U L /\nu \ .
\end{equation}
In this definition, $U$ and $L$ are respectively a typical
velocity and length in the flow, and $\nu$ is the kinematic viscosity,
which can be measured for each different fluid.
The typical velocity may be defined by the inlet velocity of the fluid,
or its mean velocity in the flow. And the typical length is usually
defined by the geometry, such as the length of the boundaries
containing the flow, or by the characteristic length of the external force.

The concept of the Reynolds number only gained popularity after these experiments, but it was introduced theoretically in 1851 by George Stokes (for a discussion on the history of the Reynolds
number, \textcite{rott1990note} should be consulted).
Stokes noticed that, in the Navier-Stokes equations, there is only one relevant dimensionless parameter.
The relevance of the Reynolds number may be noticed by rewriting the
Navier-Stokes equations in dimensionless form, with
the following changes of variables:
\begin{equation}
    \partial_t = \frac{U}{L} \partial_{t}' \ , \quad
    \partial_i = \frac1L \partial_{i}' \ , \quad
    u = U u' \ , \quad
    p = U^2 p' \quad \mbox{and} \quad
    f = \frac{U^2}{L} f' \ ,
\end{equation}
with the same typical scales $U$ and $L$
as in Eq.~\eqref{eq:reynolds}.

After these transformations, the Navier-Stokes equations become:
\begin{equation}
    \partial_t' u_i' + u_j' \partial_j' u_i' =
    - \partial_i' p' + \frac{1}{\mathrm{Re}} \partial'^2 u_i' + f_i' \ ,
\end{equation}
with the Reynolds number as defined in Eq.~\eqref{eq:reynolds}.
It can be seen that, in the regime of high Reynolds number, the dissipation term $\partial'^2 u_i' / \mathrm{Re}$ becomes less relevant in comparison with the nonlinear term.
On the contrary, when the Reynolds number is very small, dissipation gains relevance relative to the inertial contribution.

A simple dimensional argument exists to reinforce this conclusion: Looking at the flow only at a characteristic scale $\ell$, the nonlinear term has a typical intensity $u^2 /\ell$, and the dissipative term, in turn, has a typical dimension $u / (\ell^2 \mbox{Re})$. At large scales, the nonlinear term prevails, and creates complex vortical structures in the flow, even if the initial state is smooth. And at small scales, the dissipation term dominates, and acts in order to smooth out these turbulent structures.

Nevertheless, one could naively imagine, from this argument, that viscous forces can be ignored at large Reynolds numbers.
But ignoring the viscous contribution changes the Navier-Stokes equations from second order in the spatial derivatives to only first order, and this change renders impossible the realization of the no-slip boundary condition.
Therefore, there is an abrupt difference in behavior between the case of vanishing viscosity (very large Reynolds) and the case of exactly zero viscosity (corresponding to the Euler equations). In the first situation, fluid motion is highly turbulent, whereas in the second there can be laminar solutions irrespective of any scale of the flow.
This abrupt change is a singular limit.

Consequently, one should tread with care in using a scaling argument, for their validity is restricted to intermediate length scales, far from the injection or dissipation scales, illustrated in the cascade picture. This regime of intermediate length scales is called the inertial range, and a more precise description of it is given in Sec.~\ref{sec:k41}.
A scale-based analysis, though, can be made rigorous and is a useful tool in studies of turbulence. Several mathematical and numerical techniques decompose turbulent flows according to their scales of motion, such as Fourier analysis, wavelet methods and coarse-graining in real space. For more details, the reader is referred to \textcite{farge1992wavelet,pereira2012}.

The singular limit is also responsible for the formation of the \textit{boundary layer} in wall flows, such as flows around airplanes and ships, or over a landscape. There is a strong mean component in these flows, which are often treated as idealized and frictionless (solutions of the Euler equation), but near the walls rapid variations of velocity and the appearance of complex vortical structures occur, which cannot be understood without the viscous contribution. This phenomenon, the boundary layer, is of great relevance to engineering applications, as can be imagined. On this topic, the reader is referred to \textcite{schlichting2016}.

\section{The Random External Force} \label{sec:random}

In the experiments on turbulent flows, such as those performed by Reynolds,
%and even before him, with Da Vinci,
one of the first features that was noticed was the seemingly random nature of the solutions. The same phenomenon happens in numerical simulations of the Navier-Stokes equations:
Minute differences in the initial or boundary conditions, or small instabilities due to truncation errors in the numerical routines, generate solutions which, after some time, look nothing like each other. This is also one of the main features of chaotic systems, discovered in \textcite{lorenz1963}, a parallel which has led to numerous investigations on the connection between turbulence and chaos \parencite{eyink2011,boffetta2017,berera2018}.

Quantitative evidence on the effect of small perturbations in turbulence have been investigated, for instance, through the dispersion of particle pairs in turbulent flows since \textcite{richardson1926atmospheric}.
This work describes several experiments with particles in atmospheric flows, and its main result, called the Richardson dispersion law, describes the growth of the mean square distance of two particles in a turbulent flow:
\begin{equation} \label{eq:richardson}
    \langle r^2 \rangle \propto t^3 \ .
\end{equation}
This result shows that the dispersion of trajectories in turbulent flows is superdiffusive: Much faster than in Brownian motion,
in which $\langle r^2 \rangle \propto t$.
In the case of Brownian motion, though, the path taken by the particles is completely uncorrelated and memoryless, and to explain
Eq.~\eqref{eq:richardson}, the complex correlations in time and space
of turbulent fields have to be taken into account.
Modern verifications of this law can be found in \textcite{boffetta2002,bourgoin2006,salazar2009}.

Nevertheless, chaotic behavior is characterized by an exponential separation of trajectories, rather than the algebraic growth of Eq.~\eqref{eq:richardson}. It has been argued that at small scales (the dissipative range), particles separate exponentially fast \parencite{furstenberg1963,zeldovich1984}, while in the inertial range, separation is algebraic, given by Richardson's dispersion. The algebraic separation is an altogether different phenomenon which produces randomness in turbulent flows. Velocity fields in turbulent flows are irregular and not continuously differentiable, which leads to the phenomenon of spontaneous stochasticity of Lagrangian trajectories: Only a statistical description of these trajectories is possible, since the equations describing their evolution display multiple solutions. This phenomenon and Richardson's dispersion law are deeply linked, as observed in \textcite{bernard1998scaling,falkovich2001particles}.
The first description of spontaneous stochasticity were presented in \textcite{bernard1998scaling,gawedzki2000,gawedzki2002}, and a rigorous proof in the context of advection by a random field (the Kraichnan model) was presented in \textcite{lejan2002}. More details on these topics are available in \textcite{falkovich2001particles,eyink2008turbulence}. The roughness of the turbulent velocity field is also further discussed in Sec.~\ref{sec:onsager}.
% from wikipedia:
% Continuously differentiable ⊂ Lipschitz continuous ⊂ α-Hölder continuous ⊂ uniformly continuous = continuous, 0 < α < 1

% "infinitesimally close trajectories still separate in a finite time. This makes a marked difference in comparison to the smooth chaotic regime"

Due to this inevitable randomness, both from the recent theories, and from the early observations of unpredictable behavior, it is common in theoretical and numerical investigations to define the external force $f_i$ in the Navier-Stokes equations, Eq.~\eqref{eq:ns}, as a random source of Gaussian nature, with zero mean, correlated on the large scales and delta-correlated in time. This is formally described as
\begin{equation}
\begin{split}
    \langle f_i(\mathbf{r},t) \rangle &= 0 \ , \\
    \langle f_i(\mathbf{r},t) f_j(\mathbf{r'},t') \rangle
    &= \chi(|\mathbf{r}-\mathbf{r'})|) \delta(t-t') \delta_{ij} \ .
\end{split}
\end{equation}
In this equation, $\delta(t)$ is the Dirac delta function
and $\delta_{ij}$ is the Kronecker delta symbol.
$\chi(r)$ is the correlation function of the external force, its characteristic
length is $L$ and this function decays fast at distances larger than $L$.

In this manner, the observed randomness is produced artificially.
But the statistical properties of the resulting flow
are expected to be the same for two main reasons.
The first, of a physical nature, is that flows at very high
Reynolds numbers display universal behavior, particularly the exponents of statistical moments of Galilean invariant quantities.
The second reason is in connection with the theory of dynamical
systems: Birkhoff's theorem shows that, for ergodic systems, there is
an underlying probability distribution which reproduces its
statistical features, even if the original dynamical system is deterministic \parencite{frisch1995,cornfeld2012ergodic}.

Resorting to Birkhoff's theorem requires the assumption that turbulence
is ergodic.
In experiments, measurements are often made
of time-averaged quantities. Theoretical considerations, on the other side,
rely on ensemble averages, which are equivalent to averages
over some probability distribution function. To connect the experimental
and theoretical results, ergodicity has also been proposed without proof.
It is important to mention that the same assumption has to be made in many other
systems, in equilibrium or not, since there are few
problems where a proof of ergodicity exists.
One of these is the hard sphere billiard, its ergodicity
was proven in \textcite{sinai1963}.
%As in equilibrium statistical mechanics, to connect the
%experimental and theoretical results, the ergodicity of turbulent
%flows has been proposed without proof.

Studies of the Navier-Stokes equations as a dynamical system
have been pursued since
\textcite{landau1944problem,hopf1948mathematical},
and a mathematical foundation has been established
for the rigorous meaning of the ensemble average
and the statistics of a turbulent stationary state.
For a discussion of these results, the reader is
referred to
\textcite{fursikov1988,foias2001navier}.
% neither ruelle1971 or ruelle1982 are mentioned here
With the availability of large scale numerical
simulations, where both time and ensemble averages can
be calculated, numerical verifications of ergodicity
have also been done, in \textcite{galanti2004turbulence},
supporting the hypothesis.

All of these considerations are drawn as basis for the use of a stochastic version of the Navier-Stokes equations and its variants, although an explanation of the clear connection between noise and Navier-Stokes dynamics is still missing \parencite{eyink2008turbulence}.

\section{Symmetries of Fluid Dynamical Equations} \label{sec:symmetries}

\begin{figure}[ht]
    \centering
    \includegraphics[width=1.\textwidth]{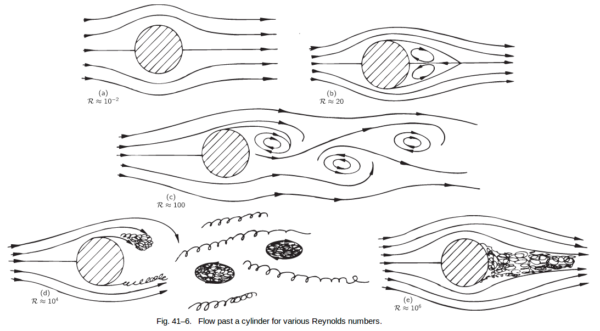}
    \caption[The laminar-turbulent transition in
    a drawing from Feynman]
    {This image, extracted from \textcite{feynman2011feynmanvol2},
    depicts the flow past a cylinder. It starts regular at low
    Reynolds numbers, this is the laminar flow, shown in the top left.
    At higher values of Re, instabilities set in
    and periodic solutions, with rotating vortices arise,
    this is the von Karman vortex street, shown in the middle.
    With even higher Re, the flow becomes turbulent, a situation
    in which all layers of the flow are mixed, seen in the bottom right.
    The Reynolds number is the only parameter that governs this transition in typical flows.}
    \label{fig:reynolds-transition}
\end{figure}

Since few exact results can be drawn directly from the Navier-Stokes
equations, an analysis of its symmetries becomes all
the more relevant to the study of turbulence.
In Fig.~\ref{fig:reynolds-transition}, a drawing from
\textcite{feynman2011feynmanvol2} depicts the flow around a cylinder
at increasing Reynolds numbers, a setting where
symmetries and symmetry breaking can be recognized.

\begin{figure}[t]
    \centering
    \includegraphics[width=1.\textwidth]{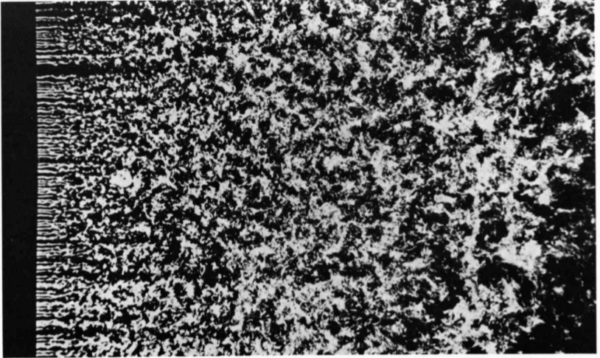}
    \caption
    [A flow coming from a rectangular grid becomes approximately
    homogeneous and isotropic]
    {In this image from \textcite{van1982album},
    homogeneous turbulence is generated by a flow incoming from
    the left and impinging on a rectangular grid.
    This setting can be used to study the properties
    of homogeneous isotropic turbulence, since, far from
    the grid, the large scale features of the flow
    are weaker.}
    \label{fig:grid-turbulence}
\end{figure}

At small Reynolds numbers, this flow is stationary, which means it
is invariant under time translations.
As the Reynolds increases,
instabilities develop and rotating vortices appear behind
the cylinder, breaking time translation to a discrete symmetry.
These vortices alternate between swirling with the flow and against it,
in a periodic manner, but in this situation, the behavior of the flow
is still orderly and predictable. At even higher Reynolds
numbers (Fig.~\ref{fig:reynolds-transition}d), this periodicity
is lost and random vortices begin to populate the flow.
The last state, (Fig.~\ref{fig:reynolds-transition}e), at the highest Reynolds number,
is called fully developed turbulence.
Invariance under time translations is hopeless in this situation,
as are attempts at predicting the future state of the flow, due
to its irregular behavior. But in a statistical description,
since it is observed that stationarity is recovered when looking at
ensemble averages. This is what is meant by a stationary flow:
The statistical symmetry of translation in time of an ensemble of similarly prepared flows.

A similar phenomenon happens regarding spatial symmetries.
The presence of the cylinder makes translational invariance seem impossible,
since it does not exist even at small Reynolds numbers.
Nevertheless, there is such a symmetry if one looks at the small scales of the
turbulent flow in Fig.~\ref{fig:reynolds-transition}e.
Away from the boundaries and at such small length scales, the
overall properties of the flow lose relevance.
Then, spatial symmetries are recovered in this limit:
Both rotational and translational invariance, in a statistical sense, are properties of the fully turbulent flow

Another setup, which is commonly used in experiments,
is turbulence generated by a spatial grid, illustrated
in Fig.~\ref{fig:grid-turbulence}.
In the figure, the fluid is flowing from left to right
and at the left side there is a regular grid. The unpredictable
properties of the flow quickly manifest behind the grid, where
individual streaks of flow, coming out of the empty spacings, start
to mix. The pattern generated away from the grid does not inherit
any of the symmetries of the setup, and instead displays statistical invariance
under translations and rotation.
The name given to a statistical invariance under translations is
homogeneity, and to statistical invariance under rotations, isotropy.

These symmetries are manifest in all flows at sufficiently high
Reynolds numbers, and at small scales,
irrespective of the symmetries of the external force
or the geometry of the flow.
For this reason, the basic paradigm in the study of turbulence
are the properties of stationary, homogeneous and isotropic flows.
This is the simplest configuration of a turbulent flow, and the
view of restored symmetries provides a theoretical vindication
to studying such flows.

\begin{figure}[t]
    \centering
    \includegraphics[width=1.\textwidth]{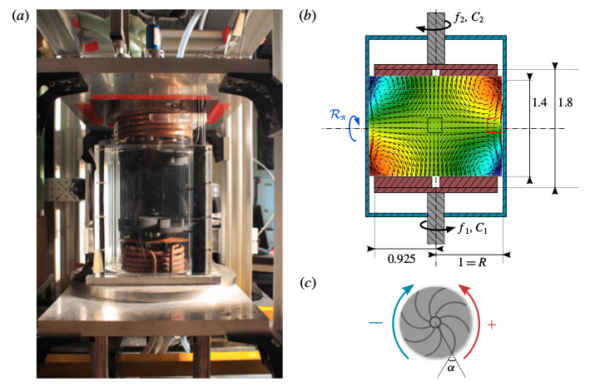}
    \caption[Experimental setup for a Von Karman flow]
    {This figure, extracted from \textcite{dubrulle2019beyond},
    displays an experimental setup for the generation
    of von Karman flows (a). In (b) and (c), details of the
    working of the rotating paddles are shown.
    The black and red boxes in the (b) are the two regions
    observed in the experiments, one in the middle
    of the flow, where mixing from several layers of flow
    generates turbulence which is very accurately homogeneous
    and isotropic, and one very close to the boundary layer,
    yet the statistical observations are equivalent in this
    region.}
    \label{fig:von-karman-flow}
\end{figure}

An experimental setup which is very commonly used
to generate homogeneous isotropic turbulence is shown in
Fig.~\ref{fig:von-karman-flow}: two counter-rotating propellers
at the bottom and the top impel the motion of fluid.
In the middle plane of the setup, there is a mixing layer.
Measurements in the mixing layer display all properties of
homogeneous and isotropic flows with good accuracy.
Recent experiments with this
arrangement have reached Reynolds numbers of the order of $10^5$
\parencite{debue2018experimental}.

The statistical symmetries are inherited from the global
symmetries of the Navier-Stokes equations, which are more
than simply time translations, spatial translations and rotations.
And they form the basis of the modern view of Kolmogorov's theory.
Some of these symmetries are only observed in the regime
of infinite Reynolds (equivalent to vanishing viscosity).
This is called the inviscid regime, and it corresponds formally to
the Euler equation:
\begin{equation} \label{eq:euler}
\begin{split}
    \partial_t u_i + u_j \partial_j u_i
    &= - \partial_i p \ , \\
    \partial_i u_i = 0 \ .
\end{split}
\end{equation}

As discussed, this is a singular limit, and
strong qualitative differences exist
in the solutions of the Euler equation, often smooth
and symmetric, and the Navier-Stokes equation at vanishing
viscosity, corresponding to fully developed turbulence.
But at intermediate length scales, where the dissipative (viscous) effects
can be ignored, the symmetries of the Euler equation are relevant
in the study of fully developed viscous flows.

Then, the global symmetries of the Euler equations are the following:
\begin{enumerate}
\item Space translations: $t, r_i, v_i(\bm{r},t) \to t, r_i+\rho_i, v_i(\bm{r},t)$
\item Time translations: $t, r_i, v_i(\bm{r},t) \to t+\tau, r_i, v_i(\bm{r},t) $
\item Galilean transformations: $t,r_i,v_i(\bm{r},t) \to t, r_i + U_i \ t, v_i(\bm{r},t) + U_i$
\item Parity: $t,r_i,v_i(\bm{r},t) \to t,-r_i,-v_i(\bm{r},t)$
\item Rotations: $t,r_i,v_i(\bm{r},t) \to t, \Lambda_{ij} r_j, \Lambda_{ij} v_j(\bm{r},t)$,
 where $\Lambda \in SO(3)$.
\item Time reversal: $t,r_i,v_i(\bm{r},t) \to -t,r_i,-v_i(\bm{r},t)$ \label{it:time-rev}
\item Scaling: $t,r_i,v_i(\bm{r},t) \to \lambda^{1-h} t, \lambda r_i, \lambda^h v_i(\bm{r},t)$, $\lambda > 0$ and $h$ is a real number. \label{it:scaling}
\end{enumerate}
Proofs can be found in \textcite{moriconi2008introducao}.
%,eyink2008turbulence,frisch1995}.
The last two are symmetries exclusively of the Euler equation,
while the others are symmetries of the Navier-Stokes equation as well.

Applying the time reversal transformations (item \ref{it:time-rev} above) to the Navier-Stokes equations,
it can be noticed that the term responsible for breaking this symmetry is the
viscous contribution, $\nu \partial^2 u_i$. This is one more evidence that the viscous
term is responsible for the system being dissipative, and for this reason
there is no symmetry under time reversal.

Regarding the last symmetry, the Euler equations are invariant under any $h$ rescaling, whereas the Navier-Stokes equation are only invariant under rescaling if $h=-1$.
Nonetheless, the symmetries in the Navier-Stokes equations are not necessarily manifest in the solutions, as can be observed in the flows of Fig.~\ref{fig:von-karman-flow}.

These symmetries inspired
Kolmogorov to build a theory based on three basic hypothesis.
In the next sections, these hypothesis and their limitations are going to be discussed.

\section{The Theory of 1941} \label{sec:k41}

Andrey Kolmogorov was one the most important mathematicians of the twentieth century.
He was responsible for the modern development of the theory of probability
and of turbulence, along with paramount contributions
in areas such as topology, logics, classical mechanics and computational complexity.
In 1941, he published three articles which laid the foundations
for the statistical theory of turbulence:
\textcite{kolmogorov1941dissipation,kolmogorov1941degeneration,kolmogorov1941local}.
For this reason they form what is called the K41 theory.

The picture of the energy cascade inspired Kolmogorov to
build three hypotheses as the basis of a phenomenological theory of fully developed turbulence.
At the time, these hypotheses were formulated in terms of universality
of statistical observables at sufficiently high Reynolds numbers.
In this text, instead, the point of view of
restored symmetries is adopted. This is a contemporary interpretation of the
Kolmogorov hypotheses, described in
\textcite{frisch1995}. Both interpretations are equivalent, portaying the same theory, but the symmetry perspective enables to investigate more deeply their range of validity and limitations.
%One of these advantages is that the language of symmetry remains
%valid if we take fluctuations into account, which the K41 theory
%does not include.
These hypotheses are still some of the foundations of the statistical theory of turbulence, even with all the posterior developments.
They operate as good approximations to reality in the limit of infinite Reynolds number, at small scales and away from boundaries. Their statement follows.

The \textit{first hypothesis} of K41 is that, under the given assumptions, all possible symmetries of the Navier-Stokes equations are statistically restored.
As was discussed in Sec.~\ref{sec:symmetries}, these symmetries are usually broken by the mechanisms producing the turbulent flow. The key point in this hypothesis is that the details of the large scale forcing and flow boundaries lose relevance in the infinite Reynolds limit.

Under the same assumptions, the \textit{second hypothesis} states that
a turbulent flow is self-similar at small
scales. This means the flow possesses a unique scaling exponent $h$,
such that the velocity is a statistically self-similar field:
\begin{equation} \label{eq:v-scale}
	\delta \mathbf{v} ( \mathbf{r}, \lambda \ell )
	\overset{d}{=}
	\lambda^h \delta \mathbf{v} ( \mathbf{r}, \ell ) \ .
\end{equation}
In this equation, $\delta \mathbf{v} ( \mathbf{r}, \ell ) = \mathbf{v} ( \mathbf{r} +  \ell \mathbf{x} ) - \mathbf{v} ( \mathbf{r})$
is the velocity difference along the direction of $\mathbf{x}$. The equality in distribution in Eq.~\eqref{eq:v-scale}, indicated by $\overset{d}{=}$, means that both sides follow the same probability distribution. Hence, this is not a strict
equality, but only a statistical correspondence, valid when an ensemble of flows is considered.

Finally, the \textit{third hypothesis}, valid under the same assumptions, is that
turbulent flows have a finite
nonvanishing mean rate of kinetic energy dissipation per unit mass.
This quantity is denoted by the symbol $\varepsilon$.
This means that, if the integral scale $L$ and the large scale velocity $U$ are
kept constant, and the limit $\nu \to 0$ is taken, $\varepsilon$ reaches a constant, non-zero, value.

Overall, these hypotheses define the general statistical features of the
velocity field in a turbulent flow. For instance, this explains why all turbulent
flows display the same statistical features at small scales, regardless of the
specific geometry of the flow, or the forcing conditions.

An equivalent dimensional argument, which Kolmogorov employed,
is that the only dimensional quantities which influence
the flow at small scales are the mean energy dissipation $\varepsilon$ and the kinematic
viscosity $\nu$. From these quantities, a single characteristic
length scale can be built. It is a microscopic length called the Kolmogorov
scale,
\begin{equation}
	\eta_K = (\nu^3/\varepsilon)^{1/4} \ .
\end{equation}
At this scale, dissipation gains relevance relative to the nonlinear advection, and the vortices are
smoothed out of the flow. For this reason, the scales
below $\eta_K$ are called the dissipative range.
If the Reynolds number is large enough, there is a large difference between the
small scale $\eta_K$ and the large scale $L$.
In the intermediate scale, the relevance of both the large scale effects (anisotropy)
and the small scale effects (dissipation) can be discarded, and
universal behavior is observed. With a precise definition of the dissipation scale, the inertial range also receives a more rigorous characterization: It contains the length scales $\ell$ such that $\eta_K \ll \ell \ll L$. The statistical hypotheses of Kolmogorov were designed to describe turbulent fields in this interval, where neither dissipation nor large scale geometry are relevant, thus the flow properties may only depend on a single property: the kinetic energy dissipation, $\varepsilon$.

\begin{figure}[ht]
	\centering
	\includegraphics[width=0.8\textwidth]{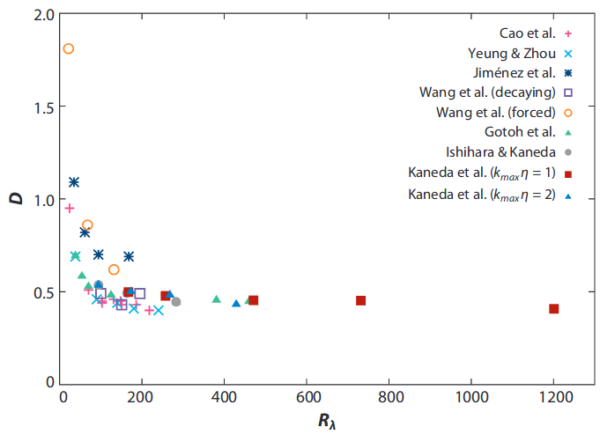}
	\caption
	[Measurements from several different numerical simulations support enhanced dissipation of kinetic energy at large Rey\-nolds numbers]
	{This figure, extracted from
	\textcite{ishihara2009}, displays the nondimensionalized energy dissipation
	for several different numerical simulations.
	The energy dissipation, made dimensionless by
	the respective input parameters of each simulation, evolve
	to the same constant value as the Reynolds number increases.
	}
	\label{fig:zero-law}
\end{figure}

The rate of dissipation $\varepsilon$ is constant across scales,
since energy loss only occurs in the dissipative range.
Therefore, the value of $\varepsilon$ can be determined from
the large scale features, where vortices have a typical kinetic energy of $U^2$
and a turnover time scale of $T=L/U$. Then, the energy transfer rate per unit mass is dimensionally equivalent to $K/T = U^3 / L$.
The constant of proportionality between the actual value of the energy dissipation and $U^3/L$ is expected to be a universal quantity, independent on the flow, in the infinite Reynolds limit. This property, referred to as the zeroth law of turbulence, or dissipative anomaly, was first discussed in \textcite{dryden1943review}, and in the reformulation of the K41 theory in terms of restored symmetries, pursued in \textcite{frisch1991global}, it enters as the third hypothesis.
Several measurements, both in experiments and numerical
simulations, have been made and they support this claim,
as can be seen in Fig.~\ref{fig:zero-law}.
For discussions on this topic, the reader is referred to
\textcite{cadot1997energy,sreenivasan1984scaling,sreenivasan1998update,pearson2002measurements,kaneda2003energy}.

In the sections that follow, the main results
of the Kolmogorov 1941 theory of turbulence are going to be introduced.
These results follow from the basic hypothesis together
with manipulations of the Navier-Stokes equations.

\subsection{The Energy Budget Equation}

From the Navier-Stokes equations, an equation which describes
the flow of energy can be written: The energy budget equation.
This equation provides a quantitative aspect to the cascade picture, in which energy is transported in an inviscid manner from the large to the small scales. It can be obtained by taking a scalar
product of Eq.~\eqref{eq:ns} with $u_i$. Then, an average of the result produces
the desired equation:
\begin{equation} \label{eq:full-EB}
	\big\langle u_i \partial_t u_i \big\rangle+
	\big\langle u_j u_i \partial_j u_i \big\rangle =
	\nu \big\langle u_i \partial^2 u_i \big\rangle
	- \big\langle u_i \partial_i  p \big\rangle +
	\big\langle f_i u_i \big\rangle \ .
\end{equation}
The angle brackets indicate a spatial average, that is, an integral over the whole domain $\Omega$
of the flow:
\begin{equation}
	\langle f \rangle = \frac{\int_{\Omega} d\mathbf{r} \ f(\mathbf{r})}
	{\int_{\Omega} d\mathbf{r} } \ .
\end{equation}
For this reason, the rule of integration by parts, which will
be used extensively, applies directly to the averaged function.
The first term of Eq.~\eqref{eq:full-EB} is a derivative of the energy density (per unit mass).
This can be seen by rewriting the first term as:
\begin{equation} \label{eq:EB-energy}
	\big\langle u_i \partial_t u_i \big\rangle
	= \frac12 \big\langle \partial_t |u_i|^2 \big\rangle
	= \frac{d}{dt} \mathcal{T}(\mathbf{u}) \ ,
\end{equation}
where $\mathcal{T}(\mathbf{u})$ is the energy density:
\begin{equation}
	\mathcal{T}(\mathbf{u}) = \frac12 \big\langle |\mathbf{u}|^2 \big\rangle \ .
\end{equation}

The second term of Eq.~\eqref{eq:full-EB} vanishes, as can be seen
from an integration by parts. It is assumed that the velocity field is smooth,
so that the integration by parts can be carried out, and that it vanishes at the boundaries
of the domain, thus making the boundary term of the integration by parts identically zero.
Then, the incompressibility condition renders the result null:
\begin{equation}
	\big\langle u_j u_i \partial_j u_i \big\rangle =
	\frac12 \big\langle u_j \partial_j |u_i|^2 \big\rangle =
	- \frac12 \big\langle |u_i|^2 \partial_j u_j \big\rangle = 0 \ .
\end{equation}
For the same reason, the pressure term is seen to be zero
after an integration by parts.

The dissipation term does not vanish, but produces an important contribution:
\begin{equation} \label{eq:EB-dissip}
	\nu \big\langle u_i \partial^2 u_i \big\rangle =
	- \nu \big\langle \left\lvert\partial_k u_i\right\lvert^2 \big\rangle
	= - \nu \mathcal{E}(\mathbf{u}) \ ,
\end{equation}
which is used to define the mean enstrophy:
\begin{equation} \label{eq:enstrophy-1}
	\mathcal{E}(\mathbf{u}) = \big\langle |\partial_k u_i|^2 \big\rangle \ .
\end{equation}
This word comes from the greek \textit{strophy}, which means rotation.
The enstrophy is a conserved quantity in two-dimensional flows and
is connected to the energy dissipation and rotation of the flow in general.
Eq.~\eqref{eq:enstrophy-1} is also equivalent to
\begin{equation}
	\mathcal{E}(\mathbf{u}) = \big\langle |\bm{\omega}|^2 \big\rangle \ ,
\end{equation}
where $\bm{\omega} = \mathbf{\nabla} \times \mathbf{u}$ is the vorticity vector.

From Eqs.~\eqref{eq:EB-energy} and \eqref{eq:EB-dissip}, the energy budget equation, Eq.~\eqref{eq:full-EB} can be rewritten as:
\begin{equation} \label{eq:EB}
	\frac{d}{dt} \mathcal{T}(\mathbf{u}) = - \nu \mathcal{E}(\mathbf{u})
	+ \big\langle \mathbf{f \cdot u} \big\rangle \ .
\end{equation}
In this form, it can be seen that kinetic energy density
is provided by the large scale force, $\bm{f}$, and dissipated by
the viscous term (which is proportional to the viscosity and the enstrophy).
%\begin{equation}
%	\frac{d}{dt} \frac{\langle u^2 \rangle}{2} =
%	\frac12 \partial_t \langle u^2 \rangle + \langle u_i u_j \partial_j u_i \rangle
%	= + \nu \langle u_i \partial^2 u_i \rangle + \langle u_i f_i \rangle
%\end{equation}

In particular, two interesting regimes can be derived from the energy
budget equation. The first, in which there is no external force, called
decaying turbulence:
\begin{equation} \label{eq:EB-decay}
	\frac{d}{dt} \mathcal{T}(\mathbf{u}) = - \nu \mathcal{E}(\mathbf{u}) \ .
\end{equation}
In this scenario the role of viscosity and enstrophy in dissipating
energy is clear. The instantaneous kinetic energy dissipation rate is
the rate of change of kinetic energy, and, from this equation, it
can be written as:
\begin{equation}
	\varepsilon' = \nu (\partial_j u_i)^2 \ .
\end{equation}
Nevertheless, it is customary to remove the pressure Hessian contribution,
$(\partial_j u_i) (\partial_i u_j)$,
from the definition of the energy dissipation, since it does not
contribute to the mean dissipation rate. Then, what is usually referred to as the
the dissipation rate is actually
\begin{equation} \label{eq:dissip}
	\varepsilon = \frac12 \nu (\partial_i u_j + \partial_j u_i)^2 \ .
\end{equation}

The other regime of interest in Eq.~\eqref{eq:EB} is one
in which the state of the flow does not change, on average:
\begin{equation} \label{eq:EB-stat}
	\nu \mathcal{E}(\boldsymbol{u}) = \big\langle \mathbf{f \cdot u} \big\rangle \ .
\end{equation}
This is called the stationary regime, in which a balance between
energy input at large scales and dissipation at the Kolmogorov scale can
be observed.

Eq.~\eqref{eq:EB} is a global energy budget, and for this reason the
inviscid transfer of energy from large to small scales can be inferred.
But there is no explicit term responsible for the transfer of energy
across scales in this equation. Such an analysis can be done through a similar
reasoning, but taking into account the point-split mean kinetic
energy:
\begin{equation}
	\mathcal{T}_{\ell}(\bm{r}) = \frac12 \big\langle u_i(\bm{r}) u_i(\bm{r+\ell}) \big\rangle \ .
\end{equation}
This quantity includes an explicit dependence on the scale of observation,
and its dynamics is described by the point-split kinetic energy
budget equation:
\begin{equation} \label{eq:EB-split}
	\frac{d}{dt} \mathcal{T}_{\ell}(\mathbf{u}) + \mathcal{I}_{\ell}(\bm{u})
	= - \nu \mathcal{E}_{\ell}(\mathbf{u})
	+ \mathcal{F}_{\ell}(\bm{u}) \ ,
\end{equation}
% Dubrulle Beyond Eq. 6.12 and personal notes 18/09/19,1
where $\mathcal{E}_{\ell}$ and $\mathcal{F}_{\ell}$ are point-split
versions of the enstrophy and external force in Eq.~\eqref{eq:EB}, defined as:
\begin{align}
	\mathcal{E}_{\ell}(\mathbf{u}) &= \big\langle \partial_k u_i(\bm{r}) \ \partial_k u_i(\bm{r+\ell}) \big\rangle \ ,\\
	\mathcal{F}_{\ell}(\mathbf{u}) &= \big\langle f_i(\bm{r+\ell}) u_i(\mathbf{r}) + f_i(\mathbf{r}) u_i(\bm{r+\ell}) \big\rangle \ .
\end{align}
And $\mathcal{I}_{\ell}$ is the inertial transport term, responsible for the
conservative transport of energy across scales, explicitly written as
\begin{equation} \label{eq:inertial-transport}
	\mathcal{I}_{\ell}(\bm{u}) =
	\frac12 \big\langle u_j(\bm{r}) u_i(\bm{r+\ell}) \partial_j u_i(\bm{r})
	+ u_j(\bm{r+\ell}) u_i(\bm{r}) \partial_j u_i(\bm{r+\ell}) \big\rangle \ .
\end{equation}

It has been mentioned that the fields $p$ and $\mathbf{u}$ are assumed to
be smooth such that the integration by parts and derivations can be perfomed.
Nevertheless, from the expression for the energy
dissipation, Eq.~\ref{eq:dissip} and the third hypothesis of Kolmogorov,
an apparent contradiction can be found.
As $\nu$ approaches zero ($\mathrm{Re} \to \infty$),
the scaling properties of the velocity field must be non trivial
in order for the energy dissipation to remain constant and not vanish.
It is currently understood that the velocity field becomes rough
in at least some small regions of the flow, but not globally, and this
heuristically explains why such manipulations are still valid at high Reynolds numbers.
This is also an indication for the strong fluctuations and spatial
inhomogeneities which turbulence displays and K41 does not
account for.
Fluctuations and roughness of the velocity field is the subject
of Onsager's conjecture, which is discussed in section \ref{sec:onsager}.

From the point-split energy budget, Eq.~\eqref{eq:EB-split},
and from other results known at the time, Kolmogorov was able
to derive one of the most important exact results in the
theory of turbulence, the four-fifths law,
published in \textcite{kolmogorov1941dissipation},
%the third article of 1941.

This result provides an exact value for the skewness
of the velocity difference probability distribution function:
\begin{equation} \label{eq:45law}
	\left\langle\left(\delta \bm{u}_{ \|}(r, \ell)\right)^{3}
	\right\rangle=-\frac{4}{5} \varepsilon \ell \ .
\end{equation}
This quantity is also called the third order structure function.
The longitudinal velocity difference is defined as
\begin{equation}
	\delta \bm{u}_{ \|}(r, \ell) = u_i(\bm{r}+\ell \bm{x}_i) - u_i(\bm{r}) \ .
\end{equation}
where $\bm{x_i}$ is the unit vector along direction $i$,
parallel to the chosen velocity component.

\begin{figure}[ht]
	\centering
	\includegraphics[width=0.8\textwidth]{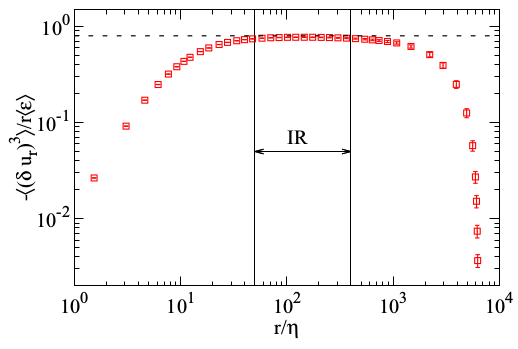}
	\caption[Numerical verification of the four-fifths law]
	{Verification of the four-fifths law in a numerical
	simulation from \textcite{iyer2017reynolds}, with $8192^3$ grid points, and $\mathrm{Re}_{\lambda} = 1300$. The dashed horizontal line corresponds to the theoretical prediction of the 4/5 law, and deviations from this result are used to define the inertial range.}
	\label{fig:k-45-law}
\end{figure}

The four-fifths law reveals that the velocity difference PDF is skewed,
and the skewness is directly proportional to the energy dissipation.
A numerical verification of this law can be seen in
Fig.~\ref{fig:k-45-law}.

Inspired by the second hypothesis of Kolmogorov, Eq.~\eqref{eq:v-scale},
a scaling transformation applied to Eq.~\eqref{eq:45law} reveals
the scaling exponent of Navier-Stokes is $h=1/3$.
If there is only one such scaling exponent, this results in a single functional
form to all structure functions:
\begin{equation} \label{eq:self-sim-zeta}
	S_p(\ell) = \left\langle\left(\delta \mathbf{u}_{ \|}(r, \ell)\right)^{p} \right\rangle
	= C_p (\varepsilon \ell)^{p/3} \ .
\end{equation}
The $C_p$ are constants that do not depend on the Reynolds number,
since the infinite Reynolds limit has already been taken in the derivation of the
four-fifths law.
It was a further hypothesis of Kolmogorov that the constants $C_p$
are universal at small scales, independent of the geometry or the forcing, although later criticism, especially from Landau, was made against the universality hypothesis.
Among the constants, only $C_3$ is universal, and its value
is given by the four-fifths law as $C_3 = -4/5 \varepsilon$. The exponents $p/3$, though, are understood to be universal, but their scaling is not linear as predicted by Kolmogorov.
The critiques against universality are addressed in Section \ref{sec:landau} and the nonlinearity of the exponents, a sign of intermittency, is discussed in the Sections \ref{sec:k62} and \ref{sec:multifractal}.
Despite these objections, $C_2$ has been measured in different settings and its value has been observed to be constant, and equal to $2.0 \pm 0.4$, in different flow conditions \parencite{sreenivasan1995universality}.

\subsection{The Energy Spectrum and the 5/3 Law}

\begin{figure}[t]
	\centering
	\includegraphics[width=.9\textwidth]{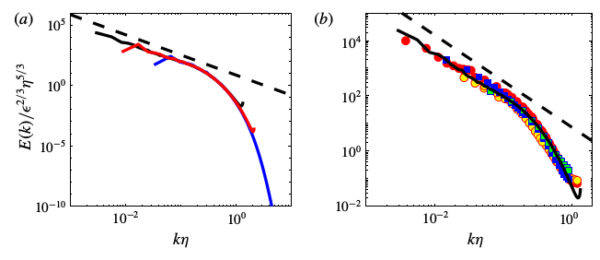}
	\caption[The turbulent energy spectrum]
	{The kinetic energy spectrum is displayed at various values of Reynolds, dif-
	ferent forcing schemes and anisotropy conditions. In both figures, the black
	dotted line is the $k^{-5/3}$ scaling. In (a), results from numerical simulations are shown, while in (b), the black continuous line is from the Johns Hopkins Turbulence Database \parencite{jhtdb}, at $\mathrm{Re}_{\lambda} = 433$ \parencite{li2008}, and the symbols are from several experiments, described in \textcite{debue2018experimental}. The figure was extracted from \textcite{dubrulle2019beyond}.}
	\label{fig:energy-spectrum}
\end{figure}

Another relevant observable for which K41 provides a prediction
is the energy spectrum.
It describes how energy is spread across different energy scales,
and is obtained from the velocity two-point correlation function,
\begin{equation} \label{eq:vv-corr}
	R_{ij}(\mathbf{r}) = \langle u_i (\mathbf{x}) u_j(\mathbf{x}+\mathbf{r}) \rangle \ .
\end{equation}
Its Fourier transform is defined as
\begin{equation}
	\phi_{ij}(\mathbf{k}) = \frac{1}{(2 \pi)^3}
	\int R_{ij}(\mathbf{r}) e^{-i \mathbf{k} \cdot \mathbf{r}} d\mathbf{r} \ ,
\end{equation}
from which we obtain the energy spectrum $E(k)$ as an integral at fixed $k$
over all directions of the Fourier transformed correlation function:
\begin{equation}
	E(k) = \oint \frac12 \phi_{ii}(\mathbf{k}) dS(k) \ .
\end{equation}
In this equation, $dS(k)$ is the surface integration element at a distance $k$
from the origin.
Notice that $E(k)$ only depends on the absolute value of the wavenumber
vector $\mathbf{k}$. Homogeneity and isotropy imply that all the information
contained in the tensor $\phi_{ij}(\mathbf{k})$ can be described by the
scalar $E(k)$ \parencite{pope2000}.

In the inertial range, since there is no characteristic length scale which can
be formed only from $\varepsilon$, Kolmogorov argued that the transfer of energy can
only be self-similar. This means the energy spectrum displays an algebraic
dependence with the scale $k$,
and this algebraic exponent can be found from dimensional analysis.
Given that the kinetic energy dissipation is dimensionally equivalent to
\begin{equation}
	[ \varepsilon ] = [ dU/dt ] = U^2 /T = U^3 /L = U^3 \kappa \ ,
\end{equation}
and the dimension of the energy spectrum is
\begin{equation}
	[ E(k)] = [\phi_{ii} dS(k)] = U^2 \kappa^{-3} \kappa^2 = U^2 \kappa^{-1} \ ,
\end{equation}
we obtain that a self-similar energy spectrum has the functional form:
\begin{equation}
E(k) = C_K \varepsilon^{2/3} k^{-5/3} \ ,
\end{equation}
where $C_K$ is argued to be a universal constant called the Kolmogorov constant.
Its value is approximately 1.5 \parencite{sreenivasan1995universality}.

In the dissipation range, there is a natural length scale provided
viscosity, which hinders the algebraic behavior. Instead, the
energy spectrum decays exponentially in this range.
The first experimental results to show an agreement with the $5/3$ exponents were communicated in \textcite{grant1962turbulence}.
Despite the agreement, in the same year Kolmogorov proposed
a refinement of the 1941 theory, which will be discussed in
the next section.
More recent measurements, both in DNS and in experiments, can be seen in
Fig.~\ref{fig:energy-spectrum}, corroborating the predicted exponent.
One can also see a spiked behavior at small wavenumbers, that
corresponds to the forcing scale, where the energy input is concentrated,
and a fast exponential decay at small spatial scales (large $k$).

\subsection{The Critiques of Landau} \label{sec:landau}

Soon after the publication of the works of 1941, L. D. Landau
expressed critiques regarding the hypothesis of universality.
He argued that fluctuations could spoil universality, which was assumed for the constants $C_p$ and the exponents, in Eq.~\eqref{eq:self-sim-zeta}.
These comments were made at a meeting in Kazan in 1942 and in a footnote
in the first edition of his textbook in fluid mechanics, published
in 1944 \parencite{frisch1995}.

Landau devised an argument to show that, in a flow with
more than one length scale present
(for instance, if the grid generating turbulence has varying grid
spacings), then there are local variations of the kinetic
energy dissipation such that it is not possible to
satisfy Eq.~\eqref{eq:self-sim-zeta} both locally and globally for this flow (the only case in which this is possible is $p=3$).
This version of the argument in terms of a flow with multiple length scales is presented in \textcite{frisch1995}.

As a consequence, whatever is the mechanism used to generate
local variations in $\varepsilon$, universality as proposed in Eq.~\eqref{eq:self-sim-zeta} cannot be true.
But fluctuations in the energy dissipation occur naturally,
and they are quite strong in fully developed turbulence,
even without any external mechanism
to enhance space variations, such as a non-uniform grid.

The comments of Landau were brief, and some of them are known only through recounts in conference proceedings, hence they are still object
of study as to their precise meaning, but substantial credit is given to them in \textcite{kolmogorov1962refinement}, the extension of the 1941 theory which takes fluctuations into account.

\section{Onsager's Conjecture} \label{sec:onsager}

A deeper connection between the solutions of the Euler
and Navier-Stokes equations (at vanishing viscosity)
is subject of a discussion originally stated in \textcite{onsager1949statistical}.
This discussion concerns the properties of weak solutions of the Euler equation.

Weak solutions are functions for which all derivatives may not exist, but which
are still considered solutions of the respective differential equation.
In contrast to them, standard solutions are also called strong solutions.
This formulation was first developed in \textcite{leray1934}, where it
was demonstrated that the Navier-Stokes equations
possess weak solutions on the whole space, $\mathbb{R}^d$, with $d \geq 2$.
In \textcite{hopf1950uber} this demonstration was expanded to limited domains.
For other systems of differential equations, weak solutions are often an intermediate step to a general proof of regularity in the strong solutions, but this path has not been completed for the equations of fluid dynamics, either viscous
or inviscid.

For the Navier-Stokes equations, Eq.~\eqref{eq:ns}, the formal definition
of weak solutions are fields $p(\mathbf{r},t)$ and $\mathbf{u}(\mathbf{r},t)$
that satisfy the following equations \parencite{bernard2000}:
\begin{equation}
\begin{split}
&\int_{\mathbb{R}^3 \times \mathbb{R}} \Big( u_i \partial_t + u_i u_j \partial_j + \nu u_i \partial^2 + p \partial_i + f_i \Big) \phi_i \ d\mathbf{r} \ dt = 0 \ , \\
&\int_{\mathbb{R}^3 \times \mathbb{R}} u_i \partial_i \psi \ d\mathbf{r} \ dt = 0 \ ,
\end{split}
\end{equation}
where $\bm{\phi}(\mathbf{r},t)$ and $\psi(\mathbf{r},t)$ are smooth functions of compact support, with the further constraint of $\partial_i \phi_i = 0$.
A similar definition, without the viscous term, holds for the weak solutions of the
Euler equation.
% other possible references on explicit forms: Buckmaster Vicol Annals of Math 2019, Tao Weak Solutions (includes Leray projection, clear to follow),

Onsager, then, noticed that, while the Euler equation is a conservative system,
its weak solutions, which may display rough and irregular behavior, need not
conserve energy. That is,
consider a weak solution of the Euler equation satisfying
\begin{equation}
	\left|\mathbf{u}(\mathbf{r}, t)-\mathbf{u}(\mathbf{r}^{\prime}, t )\right| \leq C\left|\mathbf{r}-\mathbf{r}^{\prime}\right|^{\theta}
\end{equation}
everywhere, where $C$ is independent of $\mathbf{r},\mathbf{r}^{\prime}$ and $t$.
This condition is called Hölder continuity with an exponent $\theta > 0$, which measures the roughness of the velocity field.
If $\theta \geq 1$, the velocity field is differentiable, whereas if $\theta < 1$, it is a continuous, but nowhere
differentiable function, similar to ideal Brownian paths.

Onsager's conjecture, regarding the weak solution $\mathbf{u}(\mathbf{r}, t)$, states that:
\begin{enumerate}
	\item If $\theta > 1/3$, then this solution conserves energy;
	\item If $\theta \leq 1/3$, there exist weak solutions that do not conserve energy.
\end{enumerate}
The first part was proved in \textcite{constantin1994}.
For the second part, dissipative solutions were first built by \textcite{scheffer1993,shnirelman1997}, and a proof for the open interval $\theta < 1/3$ was
exhibited in \textcite{isett2018} for $d \geq 3$, building upon an argument from \textcite{delellis2007}. Still, the proof
for the case $\theta = 1/3$ remains open.

The connection between the value $1/3$ in this conjecture and the value for the scaling
exponent of the velocity field in Kolmogorov's theory is seen immediately,
but this fact was only explaned in \textcite{duchon2000}.
In this article, it is demonstrated that the term responsible for
dissipation in the weak solutions of the Euler equation
is similar to the inertial transport term
in the Navier-Stokes equation, $\mathcal{I}_{\ell}(\bm{u})$,
Eq.~\eqref{eq:inertial-transport}.
For this reason, the phenomenon of dissipative solutions in the Euler equations was called inertial
dissipation, first described by Onsager:
\begin{displayquote}
	It is of some interest to note that in principle, turbulent dissipation
	as described could take place just as readily wihout the final
	assistance by viscosity. In the absence of viscosity, the standard
	proof of the conservation of energy does not apply, because the velocity
	field does not remain differentiable!
	\parencite{onsager1949statistical}
\end{displayquote}

For this reason, it is also hypothesized that weak solutions
of the Navier-Stokes equations at
vanishing viscosity are equivalent to dissipative weak solutions of the Euler
equation \parencite{eyink2006}.
For
a deeper discussion on this topic, the reader is referred to
\textcite{delellis2019,tao2019onsager} and references therein.

\section{The Theory of 1962 and Intermittency} \label{sec:k62}

In 1962, the first experimental evidences of the success of K41 were
still appearing,
with new measurement techniques being developed.
Nevertheless, there was already theoretical
controversy on the limitations of this theory, such as the critiques
of K41 by Landau. The role of fluctuations
can be seen as crucial in these critiques. In this same year, Kolmogorov
and Alexander Obukhov, a student of Kolmogorov, proposed an extension of the K41 theory,
which accounts for large fluctuations, but which preserved universality at the small
scales from K41.
This theory is commonly called K62.

Among the phenomena considered in K62, there are
deviations from the self-similar exponents of Eq.~\eqref{eq:self-sim-zeta}.
Instead of linear exponents, scaling behavior with arbitrary exponents $\zeta_p$
is expected in the inertial range:
\begin{equation} \label{eq:anomalous-zeta}
    S_p(\ell) = \left\langle\left(\delta \mathbf{u}_{ \|}(r, \ell)\right)^{p} \right\rangle
    = C_p (\varepsilon \ell)^{\zeta_p} \ .
\end{equation}
Strong evidence of this discrepancy was only reported much later,
in \textcite{anselmet1984high}, and has been reinforced ever since.
%although it is
%still hard to measure high order exponents with accuracy and precision.
Several models have been proposed since 1962 to explain such deviations
from the self-similar exponents, beginning with K62,
but such models are still phenomenological and it is still difficult to answer
precisely which of them describes the data the most accurately.

\begin{figure}[h]
    \centering
    \includegraphics[width=1.0\textwidth]{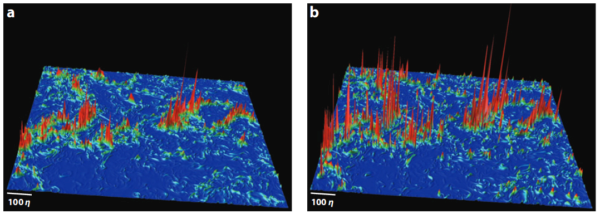}
    \caption[Fluctuations in energy dissipation and enstrophy]
    {This figure, from the large scale numerical simulations of \textcite{ishihara2009},
    shows
    strong spatial fluctuations in energy dissipation (a) and in enstrophy (b),
    }
    \label{fig:space-fluctuations}
\end{figure}

An observation from numerical simulations that strengthens the need
for a statistical description of fluctuations can be seen in
Fig.~\ref{fig:space-fluctuations}, where the values of kinetic
energy dissipation and enstrophy are displayed,
in a single instant, for a cross-section of a three-dimensional flow.
These observables exhibit fluctuations which
are very intense, in marked contrast with Gaussian noise, which would
be evenly spread and of almost uniform intensity.
Instead, there are turbulent spots
where very large fluctuations can be observed.

This phenomenom is called intermittency.
Intermittent fields display scale-dependent fluctuations, in which
intense bursts are observed in small intervals (in space or time).
A velocity field which displays such strong fluctuations, thus,
cannot be accurately described by its mean value, requiring
an understanding of its higher order statistics.

But measuring intermittent fluctuations and the structure
function exponents from Eq.~\eqref{eq:anomalous-zeta}
is a challenge, because capturing large fluctuations requires
very long time series, whether in numerical simulations
or in experiments.

To take fluctuations into account,
a new theory was developed in
\textcite{kolmogorov1961precision,kolmogorov1962refinement,
obukhov1962}.
Its main hypothesis is that,
instead of using the global mean energy dissipation, $\varepsilon$,
its local scale averaged value should be considered.
The kinetic energy dissipation averaged over a scale $\ell$ is
\begin{equation} \label{eq:dissip-coarse}
    \varepsilon_{\ell}(\bm{x},t) = \frac{3}{4 \pi \ell^3}
    \int_{\mathcal{B}_{\ell}(\mathbf{x})} \varepsilon (\bm{x+r},t) \ d \bm{r} \ ,
\end{equation}
in which $\varepsilon$ is integrated over a ball
$\mathcal{B}_{\ell}(\mathbf{x})$ of radius $\ell$ and center $\mathbf{x}$.
This hypothesis was called \textit{local universality} by Kolmogorov and Obukhov, and under this assumption, all statistical properties measured at scale $\ell$ in the inertial range, can only depend on $\varepsilon_{\ell}$, instead of the global quantity $\varepsilon$.

This addresses the critique of Landau: Universal scaling exponents, universal constants $C_p$ and non-homogeneities in the energy dissipation field cannot coexist under the linear scaling defined by Eq.~\eqref{eq:self-sim-zeta}, and not even for the nonlinear scaling of Eq.~\eqref{eq:anomalous-zeta}, for the reasons laid out in Sec.~\ref{sec:landau}. Instead, if the scaling of the structure functions (or any other observable) at some scale $\ell$ is defined in terms of the local energy dissipation, as
\begin{equation}
    S_{p}(\ell) = C_{p}(L)(\varepsilon_{\ell} \ell)^{\zeta_{p}} \ ,
\end{equation}
no such problem exists. The local universality hypothesis warrants the replacement of $\varepsilon$ with $\varepsilon_{\ell}$, and the constants $C_p(L)$ are no longer universal, instead depending on the large scale. In this manner, the universality of the scaling exponents and the non-homogeneities of the energy dissipation field are maintained.

Along with the local universality hypothesis, a conjecture for
the statistics of the energy dissipation was also made. It was
established that $\varepsilon_{\ell}$ displays log-normal fluctuations,
specified by
\begin{equation} \label{eq:lognormal}
\begin{split}
    \langle \varepsilon_{\ell} \rangle &= \varepsilon \ , \\
    \mathrm{Var}[\ln \varepsilon_{\ell}] &= C - \mu \ln (\ell / L) \ .
\end{split}
\end{equation}
In this equation, $C$ is a non-universal constant that depends on the large scales
and $\mu$ is a universal constant, called the intermittency parameter,
which defines the strength of fluctuations. The higher
its value, more intense is intermittency.
Another element which is present in these equations is the
integral length $L$. It is responsible for breaking the self-similar
behavior (scale invariance).

The hypothesis of lognormal fluctuations can be justified with the
Richardson cascade picture, in which the energy is transferred locally from a scale $\ell$ to a scale $a \ell$, with $a \leq 1$, in a self-similar way. The ratio between the energy dissipation at two nearby scales, $\varepsilon_{\ell} / \varepsilon_{a \ell}$, can be thought of as a random variable, its probability distribution depending only on the scale ratio $a$. Then, the whole cascade can be described as the product of several random factors which only depend on the same ratio, as
\begin{equation} \label{eq:epsilon-cascade}
    \frac{\varepsilon_L}{\varepsilon_{a^N L}} =
    \frac{\varepsilon_L}{\varepsilon_{a L}} \frac{\varepsilon_{a L}}{\varepsilon_{a^2 L}} \cdots
    \frac{\varepsilon_{a^{N-2} L}}{\varepsilon_{a^{N-1} L}}
    \frac{\varepsilon_{a^{N-1} L}}{\varepsilon_{a^N L}} \ ,
\end{equation}
for any $\ell = a^N L$. From this expression, $\ln \varepsilon_{\ell}$ is
the sum of several variables with identical distributions
of finite variance. If the
ratios $\varepsilon_{a^n L} / \varepsilon_{a^{n+1} L}$ can be considered
independent, then, as $N$ approaches infinity, the probability distribution of
$\ln \varepsilon_{\ell}$ approaches a normal distribution,
from the Central Limit Theorem. The probability distribution of
$\varepsilon_{\ell}$ is correspondingly a lognormal.

The exponents $\zeta_p$ can be calculated from
the distribution for the scale averaged energy dissipation.
If the probability distribution of $\varepsilon$ is a lognormal
with mean and variance given by Eq.~\eqref{eq:lognormal},
then any moment of $\varepsilon_{\ell}$ is given
by
\begin{equation}
    \langle \varepsilon_{\ell}^{p} \rangle \sim \varepsilon^{p} e^{-p(1-p) C/2}\left(\frac{\ell}{L}\right)^{\mu p(1-p)}.
\end{equation}
Then, using the local universality hypothesis, it is observed that
velocity fluctuations, in the inertial range and
at scale $\ell$, only depend on the energy dissipation at this scale, $\varepsilon_{\ell}$,
and on the scale itself. This means that $\delta u_{\ell}$ has
the same scaling behavior as $(\ell \varepsilon_{\ell})^{1/3}$,
from which the scaling of the structure functions is obtained:
\begin{equation}
    \langle \left(\delta u_{\ell}\right)^{p} \rangle \sim
    \langle \left(\ell \varepsilon_{\ell}\right)^{p / 3} \rangle
    \propto
    (\ell \varepsilon)^{p / 3}\left(\frac{\ell}{L}\right)^{\mu (1-p/3) p/3} \ .
\end{equation}
The anomalous exponents are calculated directly from this expression:
\begin{equation} \label{eq:zeta}
    \zeta_p = \frac{p}{3} \left( 1 + \mu \left( 1-\frac{p}{3} \right) \right) \ .
\end{equation}

Several evidences for deviations from
linear (self-similar) behavior have been reported.
For numerical and experimental results, the reader is referred to
\textcite{ishihara2009,benzi2010inertial,sinhuber2017dissipative,iyer2017reynolds,reinke2018universal}.

The 1962 model is a rich and useful framework for dealing with
large fluctuations. It is known that some
properties of the exponents in Eq.~\eqref{eq:zeta}
at high orders
conflict with mathematical properties expected in general for
them \parencite{frisch1995}.
For this reason, more general models for fluctuations in
turbulence have been proposed, which are discussed in the next section.

\section{The Multifractal Model} \label{sec:multifractal}

The self-similar theory of 1941 is marked by a single scaling
exponent, $h=1/3$. Yet, scale invariance is broken by the
intermittent fluctuations, as can be seen in the K62 model.
Another proposal which employs the language of scale invariance
and includes fluctuations is called multifractality.
A random velocity field with a single scaling exponent is also called
a fractal, in general, because this scaling exponent is in direct correspondence
with the fractal dimension of the random field.
The multifractal approach, instead, states that
a range of scaling exponents is possible, corresponding to
a field with multiple fractal dimensions simultaneously.

This approach began in \textcite{mandelbrot1974intermittent},
where lognormal fluctuations of the energy dissipation
are already treated under a multifractal view.
Further developments were carried out in
\textcite{frisch1985singularity,meneveau1987multifractal, meneveau1991multifractal}.

In the multifractal approach, it is supposed that there is a set
$F \in \mathbb{R}^3$ of fractal dimension
$D_F < 3$
where energy dissipation events and intense fluctuations concentrate.
The complement of this region is made of regular velocity fields,
which can be linearized. In this region, the scaling exponent of
the velocity field is $h \geq 1$,
such that all velocity gradients remain small in this region.
The set $F$ is a multifractal if it is a superposition of subsets $\mathcal{S}_h$,
such that the velocity field scales with an exponent in the range
$[h,h+dh]$ inside $\mathcal{S}_h$.

\begin{figure}[h]
    \centering
    \includegraphics[width=0.7\textwidth]{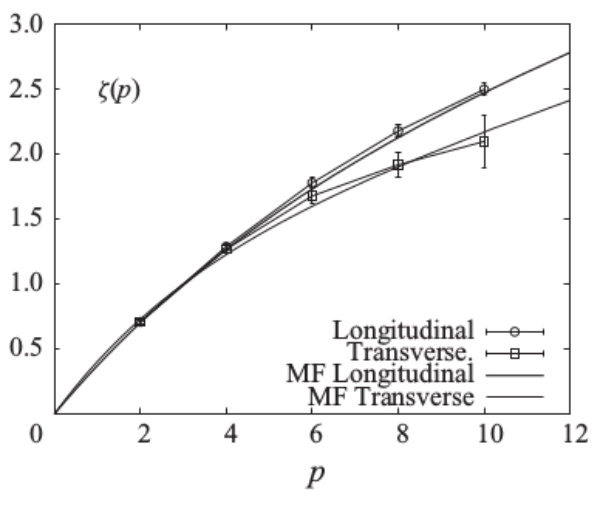}
    \caption
    [Measure of high-order structure functions in numerical
    simulations display deviations from self-similar behavior]
    {Structure functions from the numerical
    simulations of \textcite{benzi2010inertial} display clear deviations from
    the linear prediction of K41. But exponents of order larger
    than eight are still hard to measure with precision due
    to the enourmous amount of data that is needed to capture large fluctuations.
    }
    \label{fig:zeta}
\end{figure}

%the data of Meneveau and Sreenivasan show quite
%convincingly that there are multifractal exponents

In this context, the probability of sampling a singularity
exponent close to $h$ is proportional to the size of the corresponding
set $\mathcal{S}_h$, which is
measured by the fractal dimension $D(h)$.
This probability, then, is
\begin{equation}
    \mathcal{P}_{\ell}(h) dh = \left(\frac{\ell}{L}\right)^{3-D(h)} \rho(h) dh \ .
\end{equation}
The function $\rho(h)$ describes the distribution of values
of $h$ in this velocity field, irrespective of the scale,
it is thus a smooth function of $h$, independent of $\ell$.
Then, the structure functions are given by
\begin{equation} \label{eq:multif-veldif}
    \int \left(\delta \mathbf{u}_{ \|}(\bm{r}, \ell)\right)^{p} \mathcal{P}_{\ell}(h) dh
    \sim \int \left(\ell \varepsilon_{\ell}\right)^{h p}
    \left(\frac{\ell}{L}\right)^{3-D(h)} \rho(h) dh \ .
\end{equation}

Using a saddle-point aproximation for the structure functions,
the structure function exponents are obtained as
\begin{equation} \label{eq:multif-zeta}
    \zeta_p = \underset{h}{\mbox{min}} [ h p + 3 - D(h) ] \ .
\end{equation}
Direct measures of the structure functions
from numerical simulations can be seen in Fig.~\ref{fig:zeta}.
Deviations from the linear exponents of K41 are
clearly seen.

The multifractal model is a general framework for
anomalous scaling, but it does not provide a
quantitative prediction for the exponents,
which depend on an explicit $D(h)$ function.
But relationships between exponents of different observables
which can be predicted from the theory,
and be used to test it. For instance, scaling exponents for
the energy dissipation \parencite{mandelbrot1974intermittent}
or the velocity gradient \parencite{nelkin1990multifractal} can be obtained
as Legendre transforms of the $D(h)$ function,
hence these exponents must be connected to those calculated in
Eq.~\eqref{eq:multif-zeta}.

The former sections have demonstrated the relevance of fluctuations
and probabilities in turbulence.  The next chapter discusses
techniques of stochastic calculus and statistical mechanics which
are used in the works developed in this dissertation.

\end{chapter}

\begin{chapter}{Brief Review of Stochastic Methods}
\label{chap:stoc}

\hspace{5 mm} 
In 1827, the famous botanist Robert Brown observed that pollen immersed
in water followed a strange jiggling motion,
reported in \textcite{brown1828}. Interested
in investigating if this effect was a manifestation
of life, he repeated the experiment with other suspensions
of fine particles: Glass, minerals, and even a fragment of
the sphinx. In all of these situations, the same jiggling motion
was observed, thus ruling out any organic origin
\parencite{gardiner2009}.
% Brown was born in 1773, thus he was already famous in 1827,
% he was also responsible for one of the first descriptions of the
% cell nucleus, on plant pollination and fertilisation,
% he was one of the first to recognize the difference between
% gymnosperms and angyosperms, he also made several
% contributions to plant taxonomy

The origin of this mysterious motion was only resolved in 1905 by Albert Einstein,
who also introduced the probabilistic description of physical
phenomena in his solution \parencite{einstein1905}.
This theory relied on the molecular nature
of matter and the random motion of its molecules, agitated by
thermal fluctuations. He predicted that
the mean square displacement of a suspended particle would be
\begin{equation} \label{eq:displacement}
    \langle x^2 \rangle = 2 D t \ .
\end{equation}
Furthermore, the suspended particle is in thermal equilibrium,
thus its kinetic energy is given by the equipartition law,
which sets the value of the constant $D$ to
\begin{equation}
    D = \frac{k_B T}{6 \pi \mu a} \ .
\end{equation}
In this equation, $k_B$ is the Boltzmann constant and $\mu$ is the
dynamic viscosity of the fluid, it arises because of the Stokes drag in the
suspended particle, assumed spherical of diameter $a$.
Also, $\langle \ \bm\cdot \ \rangle$ represents an expectation value,
calculated with respect to some probability distribution function.

With the theoretical framework developed by Einstein, the
experimental physicist Jean Perrin was able to verify
such random motion and calculate Avogadro's constant,
obtaining the surprisingly accurate value of
$6.4 \times 10^{23} \ \mathrm{mol}^{-1}$ \parencite{perrinnobel}.
This value was in close agreement
with known results of the time, including other independent
measures from Perrin, using several different experimental techniques.
At the time, these experiments were seen as strong evidence
of the existence of atoms and molecules and
Perrin was awarded the Nobel Prize in Physics of 1926
\enquote{for his work on the discontinuous structure of matter}
\parencite{perrinAIP}.

A few years after Einstein,
the French physicist Paul Langevin was able to arrive at the same conclusions
by proposing an equation of motion for the suspended particle
\parencite{langevin1908theorie}:
\begin{equation} \label{eq:origin-langevin}
    m \frac{d^2 x}{dt^2} = - 6 \pi \mu a \frac{dx}{dt} + \eta_t \ .
\end{equation}
In this equation, it is supposed that the particle of mass $m$
is subject to
viscous drag (Stokes drag) by the medium
and to a random external force $\eta_t$, which represents the
incessant impact of the molecules of the liquid on the particle.
He further assumed that this force would be positive and negative
with equal probabilities.
Eq.~\eqref{eq:origin-langevin} was the first example of a stochastic differential equation
used in physics,
and from it
the mean square displacement law can be derived as well,
with the same constant $D$ that Einstein had obtained.
% maybe add a derivation of mean square displacement from
% the langevin equation, but only maybe
% also explain why the equipartition theorem plays a role here,
% something which I never quite understood

Earlier equivalent models for random motion were developed
in the study of time series \parencite{thiele1880} and in stock
markets \parencite{bachelier1900}, although they remained unacknowledged
for a long time (see \textcite{jarrow2004} for a historical review).

The random description of Brownian motion inspired the mathematicians
Norbert Wiener, Raymond Paley and Antoni Zygmund
to develop a rigorous mathematical
explanation for the theory of Einstein.
The construction which obeys Eq.~\eqref{eq:displacement}
was demonstrated in 1923 and is called the
Wiener process, or Wiener measure,
which is informally treated as a synonym for Brownian motion.
%who demonstrated in 1923 the mathematical properties of
%These mathematical constructions relied on new ideas in measure
%theory of the early twentieth century (Daniell, 1913).
% what new ideas? who is Daniell? this quote is probably from Jarrow and Protter
The Wiener process is the continuous-time stochastic process, $W_t$,
which observes the following properties:
\begin{enumerate}
    \item $W_0 = 0$;
    \item $W$ has independent increments;
    \item $W$ has Gaussian increments: $W_{t+s} - W_t$ is distributed with mean $0$
    and variance $s$. This is represented by $W_{t+s} - W_t \overset{d}{=} \mathcal{N}(0,s)$;
    \item $W$ has continuous paths. This means $W_t$ is almost surely continuous in $t$.
\end{enumerate}
As before, the symbol $\overset{d}{=}$ means equality in distribution.
Due to its simple properties, the Wiener process
is used as the basis for numerous other stochastic processes and
in applications to real systems.

The mathematical grounding for the theory of stochastic differential
equations such as Eq.~\eqref{eq:origin-langevin},
though, was only developed years later by the Japanese
mathematician Kiyosi Itô \parencite{ito1944}.
To make mathematical sense of such an equation,
he developed the concept of the
Itô stochastic integral, defined as the limit
\begin{equation} \label{eq:ito-integral}
    \int_{t_0}^t G(t') dW_{t'} =
    \lim_{n \to \infty} \left[ \sum_{i=i}^n G(t_{i-1}) ( W_{t_i} - W_{t_{i-1}} ) \right]
    \ .
\end{equation}
%above, the mean square limit is considered,
%but this is weird, there is no averaging done.
Itô demonstrated that this integral
converges in probability to a well defined random variable.

A well known figure who was also paramount to the theory of
probability and stochastic processes was Andrey Kolmogorov, who
established the axioms
of probability theory and developed a theory for Markov processes.
His work on Markov processes, which elicited the role of the drift and diffusion coefficients, inspired Itô in building a theory of
stochastic calculus. His most famous contribution is
discussed in the next section.

\section{Itô's Lemma}

A generalized Langevin equation is usually written in mathematical
notation as
\begin{equation} \label{eq:langevin}
    du = b[u] dt + g[u] d W_t \ ,
\end{equation}
where the Wiener process is the driving random
contribution. The velocity
$u$ is a vector in $\mathbb{R}^d$,
$b[u]$ is a functional called the drift coefficient and $g[u]$
is the diffusion coefficient.
This equation is similar to the one proposed by Langevin,
and is often written in the physics literature as
%Eq.~\eqref{eq:langevin} written as
%\textcite{zinn2002}
\begin{equation} \label{eq:diff-langevin}
    \dot u = b[u] + g[u] \eta_t \ ,
\end{equation}
where $\eta_t$ is called Gaussian white noise.
Such noise source is equivalent to a time-derivative of the
Wiener process, or, formally, its Radon-Nikodym derivative.
Gaussian white noise can be characterized entirely
by its mean and its two-point correlation function:
\begin{equation} \label{eq:gauss-noise}
	\begin{split}
		&\langle \eta(t) \rangle = 0 \ , \\
		&\langle \eta(t) \eta(t') \rangle = \delta(t-t') \ .
	\end{split}
\end{equation}

%The Langevin equation generates a time dependent probability distribution
%$P(u,t)$ for the solution $u(t)$, which can be formally written as

%This equation is invariant under time translations
%(where can I see it?)

One of the main results in the theory of stochastic processes,
fundamental to the mathematical manipulation of
equations such as Eq.~ \eqref{eq:diff-langevin} is called
Itô's lemma. It is the formula for a change of variables in stochastic processes,
hence it is an expression analogous to the chain rule of
standard calculus.
In the formula for the stochastic integral, Eq.~\eqref{eq:ito-integral},
it can be seen
that the prescription for a derivative has to be carefully
specified. One of the consequences of this is that the
formulation of the chain rule depends on the exact prescription.
The Itô prescription for the derivative is the following:
\begin{equation}
\frac{du}{dt} = \lim_{\epsilon \to 0} \frac{u(t+\epsilon) - u(t)}{\epsilon} \ .
\end{equation}
% i would still have to make this definition
% of the derivative and the formula for the stochastic integral
% compatible, which they are not, currently

If $u$ is governed by Eq.~\eqref{eq:langevin},
the stochastic differential equation which $f(u)$ obeys can be found using
the Itô prescription:
\begin{equation}
\begin{split}
    \frac{df}{dt} &= \lim_{\epsilon \to 0} \frac{f(u(t+\epsilon)) - f(u(t))}{\epsilon} \\
    &= \lim_{\epsilon \to 0} \frac{f(u(t)+\epsilon \dot u(t) + \ldots) - f(u(t))}{\epsilon} \ .
\end{split}
\end{equation}
Then, $f$ can be expanded to second order in a Taylor series,
\begin{equation}
\begin{split}
    \frac{df}{dt} &\approx \lim_{\epsilon \to 0} \frac{1}{\epsilon} \left[ f(u(t)) + \epsilon \dot u(t) \frac{\partial f}{\partial u}
    +\frac12 \epsilon^2 \dot u(t)^2 \frac{\partial^2 f}{\partial u^2} + \ldots - f(u(t)) \right] \\
    &\simeq \lim_{\epsilon \to 0} \left[ \dot u(t) \frac{\partial f}{\partial u}
    +\frac12 \epsilon \dot u(t)^2 \frac{\partial^2 f}{\partial u^2} + O(\epsilon^2) \right] \ .
\end{split}
\end{equation}
With $\dot u^2 = b[u]^2 + g[u]^2 \eta_t^2 + 2 b[u] g[u] \eta_t$
and $\eta_t^2 = O(1/\epsilon)$, the expression for
Itô's lemma can be found:
\begin{equation}
    \frac{df}{dt} = \dot u \frac{\partial f}{\partial u}
    +\frac{g[u]^2}{2} \frac{\partial^2 f}{\partial u^2} \ .
\end{equation}
In the differential formulation, this is written as
\begin{equation}
    df = \left(
    %\frac{\partial f}{\partial t} +
    b[u] \frac{\partial f}{\partial u}
    + \frac{g[u]^2}{2} \frac{\partial^2 f}{\partial u^2} \right) dt +
     g[u] \frac{\partial f}{\partial u} d W_t \ .
\end{equation}
The symbol $O(\ \bm\cdot \ )$ is used for asymptotic big-O notation.

A brief explanation of $\eta_t^2 = O(1/\epsilon)$ is worthy of notice.
From the properties of the Wiener process, it can be seen that
$|W_{t+\epsilon} - W_t|$ is a term of the order of $\sqrt{\epsilon}$. Consequently,
terms of second order in $|W_{t+\epsilon}-W_t|$ cannot be ignored,
since they produce a first order contribution in $\epsilon$.
The order of magnitude of $\eta_t$ can be found from
its defining properties, Eq.~\eqref{eq:gauss-noise} or from
the knowledge that it is a time derivative of $dW_t$,
resulting in $\eta_t^2 = O(1/\epsilon)$.
All of these statements can be proven,
and the reader is referred to \textcite[sec.4.2.5]{gardiner2009} or
\textcite[sec.2.5]{protter2005} on this topic.
% lookup page and section
The former reference is more accessible to physicists in general,
while the latter proof is more mathematically rigorous.

An alternative to the Itô prescription for the derivative is
the Stratonovich one, given by
\begin{equation}
\frac{du}{dt} = \lim_{\epsilon \to 0} \frac{u(t+\epsilon/2) - u(t-\epsilon/2)}{\epsilon} \ .
\end{equation}
An argument equivalent to the previous one, for the Itô prescription,
produces the respective chain rule under the Stratonovich prescription,
which is equal to the chain rule in standard calculus,
\begin{equation}
    \frac{df}{dt} = \dot u \frac{\partial f}{\partial u} \ .
\end{equation}

% lookup difference between Ito and Stratonovich in van Kampen's book

The standard chain rule naively seems to favor the use
of the Stratonovich interpretation, but the differences in
calculation go beyond this point.
Eq.~\eqref{eq:diff-langevin} leaves the interpretation ambiguous
and the issue of which interpretation should be used in a physical
problem depends on the details of the question, such as the source of noise.

Consider the case in which
the variance of the noise $\eta_t$ is a known function with a finite
correlation time $\epsilon$,
instead of Gaussian white noise, which has a singular variance.
It has been demonstrated that, in the limit $\epsilon \to 0$, the equation obtained
is given in the Stratonovich prescription \parencite{mori1975}.
This is usually the case in situations of external noise: When
a random force is added to an otherwise deterministic system.
In these situations, $b[u]$ provides the deterministic
dynamics of the system and the statistical properties of the noise can be studied
independently.

In contrast, in systems where noise is intrinsic (or internal),
it is harder to distinguish the drift and diffusion terms.
This is the case of chaotic systems such as turbulence, where noise
is unavoidable, as discussed in Sec.~\ref{sec:random}.
In many such cases, the Itô prescription is the more appropriate
interpretation because of its non-anticipating nature:
The Itô prescription only requires knowledge of times up to $t$
to calculate derivatives and integrals at time $t$, while the Stratonovich interpretation
requires knowledge of the future of $t$.
This feature makes mathematical demonstrations much harder
in the Stratonovich interpretation and favor the Itô view,
which is employed in all stochastic differential equations
in this dissertation.

Nevertheless, an SDE in the Stratonovich interpretation can be translated
to an equivalent equation in the Ito interpretation,
producing the same physical consequences (for instance, the same probability distribution function).
A Stratonovich SDE is usually written as
\begin{equation}
    du = b[u] dt + g[u] \circ dW_t \ ,
\end{equation}
where the symbol $\circ$ indicates the use
of the Stratonovich rule. This is equivalent to the following
Itô SDE:
\begin{equation}
    du = \left( b[u] + \frac12 \frac{\partial g[u]}{\partial u} g[u] \right) dt + g[u] dW_t \ .
\end{equation}
A proof of this statement can be found in \textcite[sec.6.5.6]{evans2012}.
% https://math.stackexchange.com/questions/2296945/conversion-between-solution-to-stratonovich-sde-and-it%C3%B4-sde/2300426
% https://en.wikipedia.org/wiki/Stratonovich_integral

%It the case of chaotic systems, noise is unavoidable,
%and the
%treatment of turbulence via stochastic systems is an effective
%description.
%These differences are discussed in detail in \textcite[sec.9.5]{vankampen1992}.
The differences between prescriptions were settled
in the discussions of \textcite{vankampen1981,vankampen1992},
and recently reviewed in \textcite{mannella2012}.

At the same time, studying quantum mechanics, physicists developed another method
to deal with the inherent fluctuations of quantum phenomena, namely
the Feynman integral, or functional formalism. In the next section,
the connections between the functional formalism and
stochastic differential equations are developed.

\section{The Functional Formalism}

The theoretical framework of functional methods
were first translated from quantum fields to the
equilibrium statistical mechanics of systems near phase transitions.
In this regime, the correlation length of these systems diverges,
which makes the microscopic details irrelevant, and they
can be described by a continuous limit, that is, a field theory
\parencite{zinn2002,amit2005,mussardo2010}.

Similarly to critical systems, turbulence is a phenomenom involving
several length scales which displays universal behavior at small scales as well.
The relation between these properties and the physics of statistical
systems close to phase transitions was noticed quite early,
and several researchers contributed to the understanding of
both of these problems, turbulence, and critical systems,
such as Landau and Onsager.
But turbulence is already naturally described by continuous variables,
the velocity and pressure, and no fine tuning
is required to observe the scaling behavior of K41,
unlike in phase transitions, where external parameters such
as pressure and temperature have to be carefully
chosen so the system displays critical behavior.
The presence of exponents (that is, scaling behavior)
has also drawn attention
to the possibility of applying renormalization group methods,
which were very successful in the study of critical systems to
understand their scaling exponents and other universal quantities.

The use of functional methods in fluid dynamics began with
\textcite{hopf1952statistical},
where a generating function for multi-point correlation
functions of the velocity was investigated, without success.
%such as the Dyson equation, and not in the framework
%of functional integrals.
This approach spanned several derivative works
\parencite{kraichnan1958higher,lewis1962space,wyld1961formulation,
tatarskii1962application,kraichnan1961dynamics},
though they fail to account for intermittent fluctuations.

A change in paradigm was the development of field
theoretical methods applied to classical stochastic systems.
This was first done in \textcite{martin1973},
with canonical quantization techniques of quantum field theories.
Later, they were extended to the language of functional
integrals in \textcite{janssen1976,dominicis1976}.
These developments are currently known as the
Martin-Siggia-Rose-Janssen-de Dominicis (MSRJD) formalism,
an established technique in the study of classical stochastic systems.
% synonym of technique
%Functional integrals for stochastic dynamics were introduced in
%\textcite{phythian1977functional,langouche1979functional,jensen1981functional}.

Other field theoretical approaches have been developed since the advent
of the MSRJD formalism. As an example, methods in conformal field
theory have been applied to turbulence in
\textcite{polyakov1993theory,falkovich19932d,falkovich1994universal,benzi1995conformal}
and applications of the non-perturbative renormalization group have been pursued in \textcite{mejia2012,canet2016fully}
%,canet2017spatiotemporal,,tarpin2018stationary}.
% any other work I could cite?

The MSRJD approach is valid for any well posed stochastic differential equation (that is, given a smooth initial condition, it generates a unique solution for all times, for each realization of the noise).
The following discussion, though, is limited to SDEs driven by additive Gaussian noise, but the method can be properly extended to other sources of noise through the careful evaluation of a Jacobian.
Additive noise means that the diffusion term does not depend on the $u$ field, which is thus governed by the equation
\begin{equation} \label{eq:additive-langevin}
    \dot u = b[u] + g \eta \ .
\end{equation}
The noise $\eta$ is Gaussian of zero mean and known
two-point correlation, given by
\begin{equation}
    \langle \eta(t) \eta(t') \rangle = \chi(t-t') \ .
\end{equation}
The probability of any single realization of the noise is
then given by the continuous limit of a multivariate normal
distribution,
\begin{equation}
    P[\eta] \propto \exp\left\{-\int dt \ \langle \eta, \chi^{-1} * \eta \rangle / 2\right\} \ .
\end{equation}
In the above expression, a convolution is represented by $*$ and its explicit formulation is
\begin{equation}
    (f*g)(t) = \int_{-\infty}^{\infty} f(t-t') g(t') dt' \ ,
\end{equation}
and $\langle \ \bm\cdot \ , \ \bm\cdot \ \rangle$ represents a
suitable inner product, which depends on the dimensionality of the $u$ field.

This probability is the basis for calculating the expectation
value of any observable $\mathcal{O}[u]$ which depends
on the field $u$. For instance, some useful observables
are $\mathcal{O}[u] = e^{i \lambda u(t)}$ and
$\mathcal{O}[u] = u(t_a) u(t_b)$. The expectation value of the former
is called the characteristic function, which is a Fourier transform of the probability distribution
function, while the expectation value of the latter is the two-point
correlation function.

Given any realization of the noise, $\eta$, the corresponding
(unique) solution of the SPDE is represented by $u_{\eta}$
Then, the expectation value of any observable is calculated as
\begin{equation}
    \langle \mathcal{O}[u] \rangle =
    \int D \eta \ \mathcal{O}[u_{\eta}] P[\eta] \ .
\end{equation}
The functional integral is necessary because an integral is performed
over all possible realizations of the noise, which is a time-dependent
field. This can be understood by
partitioning the time evolution of the system into small intervals,
then this integral would be the limit of
\begin{equation}
    \langle \mathcal{O}[u] \rangle =
    \lim_{n \to \infty}
    \int d \eta_{t_1} d\eta_{t_2} \cdots d\eta_{t_{n-1}}
    \ \mathcal{O}[u_{\eta}] P[\eta] \ .
\end{equation}
This equation can be equivalently written with an extra
functional integral over all possible $u$ fields:
\begin{equation}
    \langle \mathcal{O}[u] \rangle =
    \int Du D\eta \ \mathcal{O}[u] \delta[u - u_{\eta}] P[\eta] \ .
\end{equation}

$\delta[u]$ is a functional Dirac-delta, which
filters the value of the integral over the $u$ field
at each instant of time.
Since $u_{\eta}$ is a solution of Eq.~\eqref{eq:additive-langevin},
a change of variables can be perfomed to replace it
with the stochastic equation itself:
\begin{equation}
    \langle \mathcal{O}[u] \rangle =
    \int Du D\eta \ \mathcal{J} \ \mathcal{O}[u] \delta[\dot u - b[u] - g \eta] P[\eta] \ .
\end{equation}
There is a Jacobian which arises from the change of variables, $\mathcal{J}$,
but it does not depend on $u$ for the case of additive noise
(it is also important that Eq.~\eqref{eq:additive-langevin}
is interpreted with the Itô prescription).
A proof of this can be found in \textcite{nakazato1990symmetries}.
In this case, the Jacobian is simply a normalization constant and can be neglected.

Rewriting the functional Dirac delta as a Fourier transform of a constant, an auxiliary field $p$ is introduced. Then,
discarding all normalization constants, the desired expectation
value can be written as
\begin{equation}
    \langle \mathcal{O}[u] \rangle \propto
    \int Du D\eta Dp \  \mathcal{O}[u] \
    e^{-i \int dt \ \langle p, u - b[u] - g \eta \rangle}
    e^{-\frac12 \int dt \langle \eta, \chi^{-1} * \eta \rangle } \ .
\end{equation}

All terms involving the noise can be exactly integrated.
Making the substitution $\eta \to \eta - i \chi * p$:
\begin{equation}
    \int D\eta \
    e^{-\frac12 \int dt \langle \eta, \chi^{-1} * \eta \rangle
    + i g \int dt \langle p, \eta \rangle}
    %\underset{\eta \to \eta - i \chi * p}{\longrightarrow}
    \xrightarrow[\eta \to \eta - i \chi * p]{}
    \int D\eta \
    e^{-\frac{g^2}{2} \int dt \langle \eta, \chi^{-1} * \eta \rangle
    - \frac12 \int dt \langle p, \chi * p \rangle} \ .
\end{equation}
On the right hand side of this expression, the last term
under the integral does not depend on the noise, while the first
is quadratic on $\eta$.
The integral of this term is a Gaussian integral, which
provides a constant contribution.
Thus, the final form of the MSRJD functional is:
\begin{equation} \label{eq:functional-expec}
    \langle \mathcal{O}[u] \rangle \propto \int Du Dp \ \mathcal{O}[u] \
    e^{-i \int dt \ \langle p, \dot u - b[u] \rangle}
    e^{-\frac{g^2}{2} \int dt \langle p, \chi * p \rangle} \ ,
\end{equation}
from which any statistical observable can be calculated, in principle.
It is worth mentioning, though, that technical difficulties on the exact evaluation of Eq.~\eqref{eq:functional-expec} are the standard, as it happens in any nonlinear field theory.

It is also common, in connection with statistical field theory,
to name the (negative) argument of the exponential the \textit{MSRJD action}:
\begin{equation} \label{eq:msrjd}
    S[u,p] =
    i \int dt \ \langle p, \dot u - b[u] \rangle
    + \frac{g^2}{2} \int dt \langle p, \chi * p \rangle \ .
\end{equation}

This is the starting point for any application of the method,
where a variety of field theoretical tools can be applied.
It is important to remark that the above derivation is valid for Langevin equations with additive noise. In the case of multiplicative noise, the contribution of the Jacobian determinant must be carefully examined, and the choice of discretization is relevant for this. A discussion on multiplicative noise in functional methods can be found in \textcite{arenas2010} and a general discretization, which interpolates between the Itô and Stratonovich cases (the $\alpha$-discretization) is found in \textcite{janssen1992}.

\section{The Instanton Method}

The difficulties in dealing with functional integrals such
as Eq.~\eqref{eq:msrjd} are notorious. The great success
of quantum electrodynamics in making experimental predictions
relies on the presence of a naturally perturbative parameter, the
fine structure constant. Similarly, the first applications of
the MSRJD formalism to fluid dynamics also relied on expansions
around the zero field solution \parencite{forster1977}.
Nevertheless, there is no
natural perturbative parameter in the problem of turbulence
and intermittent
fluctuations is one of its main ingredients.
The standard perturbative expansion, developed around the \enquote{vacuum}, thus cannot account for these effects.

Another approach is the method of background field expansion.
% did it begin in field theory? are you sure?
Instead of a perturbative expansion around the zero field, this method
relied on the expansion around classical (or saddle-point) solutions
of the action.
In the context of disordered systems in
solid state physics, the first applications of the background field
method were done in
\textcite{zittartz1966,langer1967,langer1969}.
In fluid dynamics, the background expansion was first perfomed
for the Burgers equation, a one-dimensional
version of Navier-Stokes, in \textcite{gurarie1996}.
In this setting, the authors studied the tails of the velocity gradient PDF.
This problem has been only
partially resolved and is further investigated in Chap.~\ref{chap:burgers}.

The background field expansion was originally developed in the context of
Yang-Mills theory in \textcite{belavin1975}.
The saddle-point solutions of the respective Euclidean field theory, which also
display non-trivial topology, were called instantons.
This name was a reference to solitons,
waves localized in space, while the instantons are fast transitions
between different ground states, hence they are solutions localized in time
\parencite{schafer1998}.

At the heart of the method is the realization that a
functional integral reduces to the semi-classical limit in the presence
of a small parameter. Then, the integral can be approximated
by its saddle point approximation, using the solution
that minimizes the action.
In the literature of stochastic hydrodynamics, these solutions are called
instantons as well, and represent the maximum likelihood
realization of a specific event. The observed events are usually extreme
fluctuations of intermittent observables, such as velocity gradients,
vorticity, local energy dissipation, or circulation.
This approach captures the leading behavior of the PDF of the desired observable,
but it must be supplemented with fluctuations beyond the instanton
to represent the true probability distribution.

For instance, consider the observable
\begin{equation}
    \mathcal{O}[u] = \delta( F[u(x,t=0)] - a ) \ .
\end{equation}
This means that the event in question is the observable $F$ having
the value $a$ at time $t=0$. The system is allowed to evolve for
an infinite time, from $-\infty$ to $0$, to develop the value
$F[u]=a$ at $t=0$.
Large velocity gradients are one common instance of observable of interest, which is associated to $F[u] = u_x \delta(x)$.

In the MSRJD framework, the probability of such an event is
\begin{equation}
\begin{split}
    P(a) &= \langle \delta( F[u(x,t=0)] - a ) \rangle \\
    &= \int D u Dp \ \frac{1}{2\pi} \int_{-\infty}^{\infty}
    d\lambda \ \mathcal{J}[u] \ e^{-S[u,p]} e^{-i \lambda(F[u]\delta(t)-a)} \ .
\end{split}
\end{equation}
In the limit of extreme fluctuations, $|a| \to \infty$, this integral
can be estimated from its stationary contribution.
As the $u$ and $p$ fields develop extreme values, the action $S[u,p]$ also grows,
which justifies the saddle-point approach.
This limit is also equivalent to $|\lambda| \to \infty$ or to $\chi(0) \to 0$,
corresponding to large fluctuations and vanishing intensity of the external
force, respectively.
%Any of these parameters validates the asymptotic expansion.
The instanton configuration corresponds to the most probable
trajectory compatible with the observable.
In the quantum mechanical analogue,
this is the classical trajectory obtained with $\hbar \to 0$.

The functional derivative of the MSRJD action yields equations for
the instanton fields, corresponding to the extremal points of the action
functional:
\begin{align}
    &\frac{\delta S}{\delta p} = i(\dot u -b[u]) + \chi * p  \ , \\
    &\frac{\delta S}{\delta u} = -i \dot{p} - i (\nabla_u b[u])^T p \ .
\end{align}
The contributions from the Lagrange multiplier (the observable) provide
an initial condition to the instanton fields, which satisfy the following equations:
\begin{align}
    \dot u &= b[u] + i \chi * p \ , \\
    \dot{p} &= - (\nabla_u b[u])^T p + i \lambda \nabla_u F[u] \delta(t) \ .
\end{align}

Integrating around $t=0$, the second equation can be used to obtain the initial condition
for the $p$ field. It is assumed that $p(0^+)=0$ for stability reasons,
otherwise $p$ would grow without bounds and the action would diverge.
Furthermore, the time evolution of the system is only considered for $t \leq 0 $.
The equations obtained for $p$ are:
\begin{equation}
\begin{split}
    \dot{p} &= - (\nabla_u b[u])^T p \ , \\
    p(0^-) &= -i \lambda \nabla_u F[u(x,t=0)] \ .
\end{split}
\end{equation}

The instanton approach can likewise be developed in the language of large deviation theory,
the mathematical framework that studies the asymptotic behavior of
probability distributions \parencite{varadhan1966}.
The connection between the methods of statistical field theory and
large deviation theory has been explored in \textcite{ellis2007,touchette2009}.

Nevertheless, the instanton is only the leading contribution to the
probability distribution function of the observable. This approximation
is more accurate in the limit of large fluctuations, but other effects
have to be considered away from this regime. One possible approach is to
consider perturbative fluctuations to the instanton field, with an expansion of
the form
\begin{equation} \label{eq:perturb-instanton}
\begin{split}
    &u = u_s + \delta u \ , \\
    &p = p_s + \delta p \ ,
\end{split}
\end{equation}
where the subscript $s$ indicates the instanton (or saddle-point) fields and
$\delta u$ and $\delta p$ are fluctuating fields.
After this change of variables, the MSRJD action can be split into
instanton and fluctuation contributions,
\begin{equation} \label{eq:perturb-action}
    S[u,p] = S[u_s,p_s] +
    \Delta S[ u_s, p_s, \delta u, \delta p] \ .
\end{equation}
In this equation, the first term is called the instanton action.
Corrections of first order in the instanton fields do
not play a role because the instanton solutions are extremal solutions of
the action, by definition.
Thus, the second term carries the contributions of quadratic and higher order
in the fluctuation fields, as well as a possible dependence on the
instantons themselves.
A detailed treatment of fluctuations in two models of turbulence
is going to be the subject of the next chapters.

The cumulant expansion is used in these particular cases, a standard method in statistical field theory which has been applied to classical stochastic systems since \textcite{langouche1979}.
This is a perturbative method, hence its validity is limited to regimes in which the diagrammatic contributions in Eq.~\eqref{eq:perturb-action} are small relative to the instanton contribution, that is: $\Delta S[ u_s, p_s, \delta u, \delta p] / S[u_s,p_s] \ll 1$. This condition is met either for small fluctuation intensities ($g \ll 1$) or at large fluctuations of the observable being investigated.

\end{chapter}

\begin{chapter}{Instantons and Fluctuations in a Lagrangian Model of Turbulence}
\label{chap:rfd}

\hspace{5 mm} 

Intermittency has been measured in Galilean-invariant observables such as velocity differences and velocity gradients since \textcite{batchelor1949}. 
Such observables have gained considerable interest in the last decades with the introduction of new experimental techniques to measure all nine components of the velocity gradient tensor,
%with nine and twelve-sensor hot wire probes
\parencite{Wallace9,tsinober1992experimental,zeff2003,wallace2009,WallaceVukoslav2010,KatzSheng2010}. These novel precise measurements inspired new attempts in the modeling of the velocity gradient dynamics \parencite{frisch1995,Chev2006,ChevPRL}, which offers a chance of studying the small scale fluctuations decoupled from the dynamics of large eddies, hence their recent theoretical and experimental interest.

The evolution of the Lagrangian velocity gradient $A_{ij} = \partial_j u_i$ can be obtained from a gradient of the Navier-Stokes equations ($\frac{\partial}{\partial x_j} (N.S.)_i$), resulting in
\begin{equation} \label{eq:velgrad-ns}
    \frac{d A_{i j}}{d t}=-A_{i k} A_{k j}-\frac{\partial^{2} p}{\partial x_{i} \partial x_{j}}+\nu \frac{\partial^{2} A_{j j}}{\partial x_{k} \partial x_{k}} \ .
\end{equation}
In this equation, $d/dt$ stands for the Lagrangian material derivative, $d / d t \equiv \partial / \partial t+u_{k} \partial / \partial x_{k}$. The Lagrangian framework describes the changes of the velocity gradient as it moves along the fluid particle trajectories in the flow, hence $A_{ij}(\mathbf{x},t)$ depends only on the initial position of the fluid particle $\mathbf{x}$, and evolves in time. Nevertheless, Eq.~\eqref{eq:velgrad-ns} is not closed in terms of the velocity gradient on a single trajectory $A_{ij}(\mathbf{x},t)$, because of the last two terms in the right hand side, respectively the pressure Hessian and the viscous term. These terms depend on neighboring Lagrangian trajectories as well. Furthermore, the incompressibility condition provides the equation
\begin{equation}
    \nabla^{2} p=-A_{l k} A_{k l} \ ,
\end{equation}
which is a highly nonlocal condition for the pressure field.
Eq.~\eqref{eq:velgrad-ns} is usually rewritten in a form which isolates local and nonlocal contributions, as
\begin{equation}
    \frac{d A_{i j}}{d t}=-\left(A_{i k} A_{k j}-\frac{1}{3} A_{m k} A_{k m} \delta_{i j}\right)+H_{i j}^{P}+H_{i j}^{\nu} \ ,
\end{equation}
where the pressure and viscous contribution are, respectively
\begin{equation}
    H_{i j}^{p}=-\left(\frac{\partial^{2} p}{\partial x_{i} \partial x_{j}}-\frac{1}{3} \nabla^{2} p \delta_{i j}\right) \quad \text { and } \quad H_{i j}^{\nu}=\nu \frac{\partial^{2} A_{i j}}{\partial x_{k} \partial x_{k}} \ .
\end{equation}

The observation that Eq.~\eqref{eq:velgrad-ns} is unclosed has led to the development of several closure formulations, trying to simplify the equations and still capture the phenomena they describe. The simplest closed Lagrangian model is the Restricted Euler Equation, in which the nonlocal and anisotropic contributions $H_{ij}^p$ and $H_{ij}^{\nu}$ are ignored. This model was first considered in \textcite{leorat1975turbulence} and many of its properties were elicited in \textcite{vieillefosse1982,vieillefosse1984}.
%particularly on the classification of turbulent regions based on invariants of the Restricted Euler equation \textcite{TsinoberInformal}
%it would be good to develop on this, lookup the Meneveau article for this and talk about the finite-time singularities of the RE equation

Numerous other models, including linear damping, stochastic forcing and geometric effects have been developed since the Restricted Euler equation (\textcite{Martin98,girimaji90material,Chertkov99,Jeong2003}). An extensive review is available in \textcite{meneveau2011lagrangian}. This chapter focuses on the Recent Fluid Deformation (RFD) approach, presented in \textcite{ChevPRL}.
% a discussion of the model itself is also suited here. The idea is that one can model the pressure hessian and viscous contribution as isotropic in the past, since the fluid, after some time, has lost memory on the geometric features of the past. Other formulations have been done which assume isotropy, but this is not realistic, hence one can assume isotropy in the past and let the Navier-Stokes equations take care of the time evolution of this blob of pressure hessian / strained viscosity

The RFD model was recast in the Martin-Siggia-Rose-Janssen-de Dominicis (MSRJD) formulation in \textcite{moriconi2014}, where it was suggested that noise renormalization is the main physical mechanism to be considered to understand the onset of fat tails in the PDFs of velocity gradients, a point supported in \textcite{grigorio2017instantons}. But the approach of \textcite{moriconi2014} relied on simplifying hypothesis which were not rigorously verified. The work which generated this chapter, \textcite{apolinario2019instantons}, addresses these hypotheses.

\section{The RFD Model} \label{sec:model}

A natural stochastic extension to the velocity gradient dynamics (Eq.~\ref{eq:velgrad-ns}) is
\begin{equation}  \label{eq:model}
 \dot{{\mathbf{A}}} = V[{\mathbf{A}}] + g \mathbf{F} \ , \ 
\end{equation} 
where $V[{\mathbf{A}}]$ is a nonlinear and nonlocal functional of  ${\mathbf{A}}$, defined from Eq.~\eqref{eq:velgrad-ns}.
There is a random external force, $\mathbf{F} = \mathbf{F}(t)$, of null trace and entries given by a Gaussian stochastic process of zero mean and two-point correlation:
\begin{equation}
 \langle F_{ij} (t) F_{kl} (t') \rangle \equiv G_{ijkl} \delta (t-t') \ , \ 
\end{equation}
where
\begin{equation} \label{eq:tensor-g}
 G_{ijkl} = 2 \delta_{ik} \delta_{jl} - \frac12 \delta_{il} \delta_{jk} - \frac12 \delta_{ij} \delta_{kl} \ .
\end{equation}
This is the most general fourth-order isotropic tensor consistent with the symmetries of Eq.~\eqref{eq:model} \parencite{pope2000}.
The stochastic force strength $g$ is proportional to the energy dissipation rate per unit mass, and can be seen as a perturbative coupling constant.
The incompressibility condition, $\partial_i u_i = 0$, is equivalent to
$\mathrm{Tr} \ \mathbf{A} = 0$.

It is interesting to notice that $\mathbf{A}$ and $\mathbf{F}$ only depend on time and not on space, thus making Eq.~\eqref{eq:model} a closed system. This is achieved through the modeling of the pressure Hessian and the viscous term of Eq.~\eqref{eq:velgrad-ns}, which are replaced by local algebraic functions of the velocity gradient tensor. The closure for the RFD model is obtained from the assumption that velocity gradients are only correlated on short time scales. Hence, the nonlocal contributions can be assumed as isotropic at an arbitrary initial instant of the time evolution.
With these assumptions, two time scale parameters are required in the model, $\tau$ and $T$, respectively corresponding to the dissipative and integral scales. These parameters generate an intermittent system corresponding to Lagrangian turbulence of Reynolds number $ Re = f(g) (T/\tau)^2 $, where $f(g)$ is some unknown (probably monotonic) analytical function of the coupling constant $g$.
% discuss why monotonic? if it were a chapter fully written by me, I know I would discuss it

Mathematically, these assumptions are expressed as the following approximation to $V [ {\mathbf{A}} ]$:
\begin{equation} \label{eq:vpotential}
 V({\mathbf{A}}) = - {\mathbf{A}}^2 + \frac{\mathrm{Tr} ({\mathbf{A}}^2)}{\mathrm{Tr} (\mathbf{C}^{-1})} \mathbf{C}^{-1} 
 - \frac{\mathrm{Tr} (\mathbf{C}^{-1})}{3 T} {\mathbf{A}} \ , \
\end{equation}
where $\mathbf{C}$ is the approximate Cauchy-Green tensor,
\begin{equation}
 \mathbf{C} = \exp[ \tau {\mathbf{A}}] \exp[ \tau {\mathbf{A}}^\T] \mbox{,}
\end{equation}
which governs the deformation in time of advected fluid blobs, within dissipative time scales.
% do I have to explain this? following the rule that this is for novices in the field, yes I do, but imagining that novices will only read the first chapters, then no

Since only the ratio $\tau/T$ has physical significance in the model, the value $T=1$ can be used. Furthermore, in numerical simulations, it was observed that a perturbative expansion of the potential $V(\mathbf{A})$ up to $O(\tau^2)$ is enough to capture its quantitative features \parencite{afonso2010recent,moriconi2014}.
This expansion is given by
\begin{equation}\label{vpowers_def}
V({\mathbf{A}}) = \sum_{p=1}^4 V_p ({\mathbf{A}}) \ , \
\end{equation}
where each $V_p({\mathbf{A}})$ collects velocity gradient contributions of $O({\mathbf{A}}^p)$:
\begin{equation} \label{eq:vpowers}
\begin{split}
 V_1 ({\mathbf{A}}) = &- {\mathbf{A}} \mbox{,} \\
 V_2 ({\mathbf{A}}) = &- {\mathbf{A}}^2 + \frac{\mathbbm{1}}{3} \mathrm{Tr} ({\mathbf{A}}^2) \mbox{,} \\
 V_3({\mathbf{A}}) = &- \frac{\tau}{3} \left( {\mathbf{A}} + {\mathbf{A}}^\T - \frac{2 \mathbbm{1}}{3} \mathrm{Tr} ({\mathbf{A}}) \right) \mathrm{Tr} ({\mathbf{A}}^2)
 - \frac{\tau^2}{3} \mathrm{Tr} ({\mathbf{A}}^\T {\mathbf{A}}) {\mathbf{A}}  - \frac{\tau^2}{3} \mathrm{Tr} ({\mathbf{A}}^2) {\mathbf{A}} \mbox{,} \\
 V_4 ({\mathbf{A}}) = &- \frac{\mathbbm{1}}{9} \tau^2 \mathrm{Tr} ({\mathbf{A}}^\T {\mathbf{A}}) \mathrm{Tr} ({\mathbf{A}}^2) 
 - \frac{\mathbbm{1}}{9} \tau^2 [ \mathrm{Tr} ({\mathbf{A}}^2) ]^2
 + \frac{\tau^2}{3} {\mathbf{A}}^\T {\mathbf{A}} \ \mathrm{Tr} ({\mathbf{A}}^2) \\
 &+ \frac{\tau^2}{6} ({\mathbf{A}}^2 + {\mathbf{A}}^{2 \T}) \ \mathrm{Tr} ({\mathbf{A}}^2) \mbox{.} 
\end{split}
\end{equation}

The RFD model is capable of reproducing several of the statistical features of the turbulent fluctuations of the velocity gradient tensor, observed in numerical simulations and experiments. Close to $g=1.0$, the domain of validity of the model is approximately given by the range $0.05 < \tau < 0.2$. Outside of this range, it was observed in \textcite{ChevPRL}, that the velocity gradient PDFs in the RFD model are strongly dissimilar from experimental and numerical results.

\section{Path-Integral Formulation of Stochastic La\-gran\-gian Mo\-dels} \label{sec:path}

The MSRJD formalism, as described in Chapter~\ref{chap:stoc} is used to express the conditional probability density function of finding ${\mathbf{A}} = {\mathbf{A}}_1$ at time $t= 0$, provided that ${\mathbf{A}} = {\mathbf{A}}_0$ at the initial time $t= -\beta$, as
\begin{equation} \label{eq:rhoPDF}
 \rho ( {\mathbf{A}}_1 | {\mathbf{A}}_0, \beta ) \propto \int_{\Sigma} D[\hat {\mathbf{A}}] D[{\mathbf{A}}]  \exp \left\{ -S[\hat {\mathbf{A}}, {\mathbf{A}}] \right\} \mbox{.}
\end{equation}
In this equation: The auxiliary field is denoted by $\hat{\mathbf{A}}$. The boundary conditions for the path integral are represented by $\Sigma = \{ {\mathbf{A}}(-\beta) = {\mathbf{A}}_0, \ {\mathbf{A}}(0) = {\mathbf{A}}_1 \}$. And the MSRJD action is 
\begin{equation} \label{eq:rfd-action}
     S[\hat {\mathbf{A}}, {\mathbf{A}}] \equiv \int_{- \beta}^0 d t \left\{ i \mathrm{Tr}[\hat {\mathbf{A}}^\T ( \dot{\mathbf{A}} - V({\mathbf{A}}) ) ]
     + \frac{g^2}{2} G_{ijkl} \hat A_{ij} \hat A_{kl} \right\} \mbox{.}
\end{equation}

The stationary state solutions of the RFD equations correspond to large asymptotic times, $\beta \rightarrow \infty$. Furthermore, it can be assumed that for such times the dependence on the initial condition ${\mathbf{A}}_0$ has vanished, making it possible to require periodic boundary conditions in the velocity gradient field,
\begin{equation} \label{eq:boundary}
 {\mathbf{A}}(0) = {\mathbf{A}}(- \beta) \equiv \bar {\mathbf{A}} \ .
\end{equation}
This approach was pursued in \textcite{moriconi2014}, leading to an analytical simplification in the saddle-point equations of the MSRJD action.
The PDF for a stationary state configuration of the system is then reached as the limit
\begin{equation} \label{asymptPDF}
 \rho(\bar {\mathbf{A}}) = \lim_{\beta \rightarrow \infty} \rho(\bar {\mathbf{A}} | \bar {\mathbf{A}}, \beta) \mbox{.}
\end{equation}

As addressed in the general discussion on the MSRJD method, the principal features of the PDF in Eq.~\eqref{asymptPDF} are captured by the instanton configuration, the solution to the saddle point equations of the MSRJD action (Eq.~\eqref{eq:rfd-action}).
% well, not in all regimes, add a remark? again?
This solution is called the instanton field, denoted by $\hat {\mathbf{A}}^{sp}$ and ${\mathbf{A}}^{sp}$, where $sp$ stands for \enquote{saddle point}.
The PDF obtained solely from the instanton fields already captures nontrivial (non Gaussian) behavior induced by the nonlinearity and nonlocality of the Navier-Stokes equation, or an approximation such as the RFD equation.
The instanton fields are the solutions of the Euler-Lagrange equations
\begin{equation} \label{eq:rfd-eq-motion}
 \left. \frac{\delta S[\hat {\mathbf{A}}, {\mathbf{A}}]}{\delta A_{ij}} \right|_{\substack{\hat {\mathbf{A}} = \hat {\mathbf{A}}^{sp} \\ {\mathbf{A}} = {\mathbf{A}}^{sp}}} \!\!\!\!\!\! = 0 \ \ \mbox{and} \ \ 
 \left. \frac{\delta S[\hat {\mathbf{A}}, {\mathbf{A}}]}{\delta \hat A_{ij}} \right|_{\substack{\hat {\mathbf{A}} = \hat {\mathbf{A}}^{sp} \\ {\mathbf{A}} = {\mathbf{A}}^{sp}}} \!\!\!\!\!\! = 0  \mbox{,}
\end{equation}
with the periodic boundary condition $ {\mathbf{A}}^{sp}(0) = {\mathbf{A}}^{sp}(- \beta) = \bar {\mathbf{A}} $.

Besides the instanton solution, fluctuations around these configurations are relevant as well in the reproduction of the correct PDFs from the original model. To consider the effect of fluctuations, the velocity gradient fields in the MSRJD action are replaced as
$\hat {\mathbf{A}} \rightarrow \hat {\mathbf{A}}^{sp} + \hat {\mathbf{A}}$ and ${\mathbf{A}} \rightarrow {\mathbf{A}}^{sp} + {\mathbf{A}}$,
where $\hat {\mathbf{A}}$ and ${\mathbf{A}}$ on the right hand side of these transformations refer to the fluctuation contributions. Since the potential $V(\mathbf{A})$ has been approximated by a sum of polynomials in $\mathbf{A}$, the MSRJD action can be split in the form
\begin{equation}
 S[\hat {\mathbf{A}}, {\mathbf{A}}] \rightarrow S[\hat {\mathbf{A}}, {\mathbf{A}}] = S_{sp}[\hat {\mathbf{A}}^{sp}, {\mathbf{A}}^{sp}] + \Delta S[\hat {\mathbf{A}}, {\mathbf{A}}] \label{sp_fluct} \ .
\end{equation}
% slightly changed
The above expression is exact: The first term, $ S_{sp}[\hat {\mathbf{A}}^{sp}, {\mathbf{A}}^{sp}] $ , holds the contributions to the MSRJD action that contain only the instanton fields, while all the additional terms that involve the fluctuations ${\mathbf{A}}$ and $\hat {\mathbf{A}}$ are included in $\Delta S[\hat {\mathbf{A}}, {\mathbf{A}}]$.
The saddle-point action $S_{sp}[\hat {\mathbf{A}}^{sp}, {\mathbf{A}}^{sp}]$ is simply the 
MSRJD action, (\ref{eq:rfd-action}), evaluated on the instanton
fields, $\hat {\mathbf{A}}^{sp}$ and ${\mathbf{A}}^{sp}$.
From (\ref{eq:rhoPDF}), (\ref{asymptPDF}), and (\ref{sp_fluct}), the vgPDF can be correspondingly rewritten as
\begin{equation}
 \rho(\bar {\mathbf{A}}) = \exp \left\{ -S_{sp}[\hat {\mathbf{A}}^ {sp}, {\mathbf{A}}^ {sp}] \right\} \int D[\hat {\mathbf{A}}] D[{\mathbf{A}}] \exp \left\{ -\Delta S[\hat {\mathbf{A}}, {\mathbf{A}}] \right\} \mbox{.}
\end{equation}

The above path integral can be perturbatively computed following a standard procedure of statistical field theory, the cumulant expansion \parencite{amit2005}. 
With this application in mind, the $ \Delta S[\hat {\mathbf{A}}, {\mathbf{A}}]$ term is exactly split as
\begin{equation} \label{deltaS}
 \Delta S[\hat {\mathbf{A}}, {\mathbf{A}}] = \Delta S_0[\hat {\mathbf{A}}, {\mathbf{A}}] + \Delta S_1[\hat {\mathbf{A}}, {\mathbf{A}}] \mbox{,}
\end{equation}
where $ \Delta S_0 [\hat {\mathbf{A}}, {\mathbf{A}}] $ contains only the quadratic contributions, whereas $ \Delta S_1 [\hat {\mathbf{A}}, {\mathbf{A}}] $ holds all the self-interacting terms of the MSRJD action.
Under this separation, the velocity gradient PDF is calculated as 
\begin{equation} \label{eq:pdf}
 \rho(\bar {\mathbf{A}}) = \exp \left\{ -S_{sp}[\hat {\mathbf{A}}^ {sp}, {\mathbf{A}}^ {sp}] \right\}
 \left\langle \exp [ -\Delta S_1 ] \right\rangle_0 ,
\end{equation}
where the expectation value $ \left\langle \exp [ -\Delta S_1 ] \right\rangle_0 $ is computed 
with respect to the model defined by the quadratic action $ \Delta S_0$.
The PDF in Eq.~\eqref{eq:pdf} can be seen as 
\begin{equation} \label{nn_vgPDF}
\rho(\bar {\mathbf{A}}) \propto \exp \left\{ -\Gamma[\hat {\mathbf{A}}^{sp}, {\mathbf{A}}^{sp} ] \right\} , 
\end{equation}
where $\Gamma[\hat{\mathbf{A}}^{sp},\mathbf{A}^{sp}]$ is the \textit{effective action}, taking into account the contributions from fluctuations around the instanton.

The cumulant expansion is a pragmatic way to evaluate the effective action from the statistical moments of $\Delta S_1$. Up to second order in $\Delta S_1$, the result of this expansion is
\begin{equation} \label{eq:cumul-gamma}
\begin{split}
 \Gamma[\hat {\mathbf{A}}^{sp}, {\mathbf{A}}^{sp} ] &= 
 S_{sp}[\hat {\mathbf{A}}^{sp}, {\mathbf{A}}^{sp} ] + \langle \Delta S_1[\hat {\mathbf{A}}^{sp}, {\mathbf{A}}^{sp} ] \rangle_0 \\
 &- \frac{1}{2}
 \left( \langle \Delta S_1^2[\hat {\mathbf{A}}^{sp}, {\mathbf{A}}^{sp} ]  \rangle_0 - \langle \Delta S_1[\hat {\mathbf{A}}^{sp}, {\mathbf{A}}^{sp} ] \rangle_0^2 \right) \mbox{.}
\end{split}
\end{equation}
A derivation of this formula is found in Appendix \ref{app:cumul}.
Each term in the cumulant expansion corresponds to a Feynman diagram with respect to the free theory $\Delta S_0$. These diagrams can be numerous, with a variety of relative weights, requiring a power counting procedure to single out the most relevant diagrams. This discussion is provided in the next section.

% this paragraph and the paragraph below should be changed. They are talking about the instanton approximation.
It should be emphasized that it is in general difficult to find exact solutions of the saddle-point Eqs.~\eqref{eq:rfd-eq-motion}. 
% review the words below, I'm not sure I like their style
This, however, should not be a matter of great concern, if one is able to find reasonable approximations for the instanton fields, since the substitution (\ref{sp_fluct}) and the second order cumulant expansion result (\ref{eq:cumul-gamma}) are always meaningful perturbative procedures in weak coupling regimes. 

Provided that the Reynolds numbers are not too high, the exact instantons can be approximated by closed analytical expressions derived from the quadratic contributions to the MSRJD action. 
It  is important to have in mind that such a simplification would not work to model the far PDF tails, essentially dependent on the exact nonlinear instantons (in cases where the PDF tails decay faster than a simple exponential).

\section{Application to the RFD Model}  \label{sec:Application}

In this section, the MSRJD formalism, discussed in the last section, is applied to the RFD model. The notation below follows the one introduced in Eq.~\eqref{deltaS}.
The quadratic action is given by
\begin{equation} \label{eq:quadratic}
 \Delta S_0[\hat {\mathbf{A}}, {\mathbf{A}}] = \int_{- \beta}^0 d t \left\{ i \mathrm{Tr} [ \hat {\mathbf{A}}^\T (\dot {\mathbf{A}} + {\mathbf{A}})
 + \frac{g^2}{2} G_{ijkl} \hat A_{ij} \hat A_{kl} \right\} \ ,
\end{equation}
and the interaction term by
\begin{equation} \label{eq:deltaS1}
 \begin{split}
  \Delta S_1 [ \hat {\mathbf{A}}, {\mathbf{A}}] &= - i \sum_{p=2}^{4} \int_{- \beta}^0 d t \ 
  \mathrm{Tr}[ ( \hat {\mathbf{A}}^{sp} )^\T ( V_p({\mathbf{A}}) + \Delta V_p ({\mathbf{A}}) ) ] \\
  &+ \mathrm{Tr}
  [ \hat {\mathbf{A}}^\T V_p ({\mathbf{A}}^{sp})] + \mathrm{Tr} [ \hat {\mathbf{A}}^\T ( V_p({\mathbf{A}}) + \Delta V_p ({\mathbf{A}}) ) ] \mbox{,}
 \end{split}
\end{equation}
where the $V_p$ were defined in Eq.~\eqref{eq:vpowers} and $\Delta V_p$ are
\begin{equation}
 \Delta V_p({\mathbf{A}}) = V_p( {\mathbf{A}}^{sp} + {\mathbf{A}} ) - V_p( {\mathbf{A}}^{sp} ) - V_p( {\mathbf{A}} ) \ . \
\end{equation}
Since each $V_p({\mathbf{A}})$ corresponds to a polynomial in the components of $\mathbf{A}$, it can be seen that the $\Delta V_p({\mathbf{A}})$ isolate those contributions that mix the instanton ($\mathbf{A}^{sp}$ and $\hat{\mathbf{A}}^{sp}$) and fluctuating fields ($\mathbf{A}$ and $\hat{\mathbf{A}}$).

The propagators in the free (quadratic) model are calculated as second order functional derivatives of the free generating functional,
\begin{equation}
 Z[{\mathbf{J}},\hat {\mathbf{J}}] = \int D[\hat {\mathbf{A}}] D[{\mathbf{A}}] \exp \left\{ - \Delta S_0[\hat {\mathbf{A}}, {\mathbf{A}}] +i \int_{ - \beta}^0 d t \ \mathrm{Tr} [{\mathbf{J}}^T {\mathbf{A}} + \hat{\mathbf{J}}^T \hat {\mathbf{A}}] \right\} \mbox{,}
\end{equation}
with respect to the external source fields $\hat {\mathbf{J}}$ and ${\mathbf{J}}$, and evaluated at
$\hat {\mathbf{J}} = {\mathbf{J}} = 0$. There are two nonzero propagators in this model, given by
\begin{equation} \label{eq:q_propagator}
%calculated again 28/03,1
\langle A_{ij} (t) \hat A_{kl} (t') \rangle_0 = \left. \frac{\delta^2 \ln(Z[{\mathbf{J}},\hat {\mathbf{J}}])}{\delta J_{ji}(t) \delta \hat J_{lk}(t')} \right |_{\hat {\mathbf{J}} = {\mathbf{J}}=0}= -i \theta (t-t') \exp(t'-t) \delta_{ik} \delta_{jl} \ \ \mbox{and}
\end{equation}
\begin{equation} \label{eq:0_propagator}
 \langle A_{ij} (t) A_{kl} (t') \rangle_0 =
 \left. \frac{\delta^2 \ln(Z[{\mathbf{J}},\hat {\mathbf{J}}])}{\delta J_{ji}(t) \delta J_{lk}(t')} \right |_{\hat {\mathbf{J}} = {\mathbf{J}}=0} = \frac{g^2}{4} \exp(-|t-t'|) G_{ijkl} \mbox{,}
\end{equation}
which are represented as the Feynman diagrams illustrated in Fig. \ref{fig:free-propagator}. The functions in Eq.~\eqref{eq:0_propagator} are measured in the units of the large time scale $T=1$.

\begin{figure}[ht]
 \centering
 \includegraphics[width=.3\textwidth]{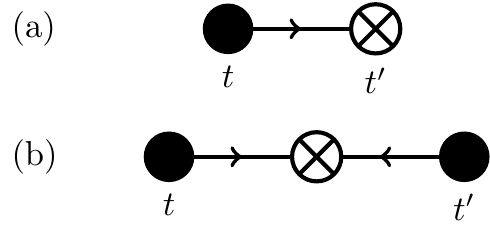}
 \caption
 [RFD model propagators]
 {The unperturbed two-point correlation functions of the RFD model, given by diagrams (a) and (b), respectively related to the time translation invariant expressions (\ref{eq:q_propagator}) and (\ref{eq:0_propagator}).}
 \label{fig:free-propagator}
\end{figure}

The perturbative effective action, Eq.~\eqref{eq:cumul-gamma}, is an algebraic expansion of all the terms in Eq.~\eqref{eq:deltaS1}, producing one hundred and eleven Feynman diagrams. These diagrams are obtained as products of the propagators in Eqs.~\eqref{eq:q_propagator} and ~\eqref{eq:0_propagator} with the application of Wick's theorem \parencite{amit2005,zinn2002}.
As an example, the complete set of fourth-order vertices is depicted in Fig.\ref{fig:Vertices}.

\begin{figure}[ht]
    \centering
    \includegraphics[width=\textwidth]{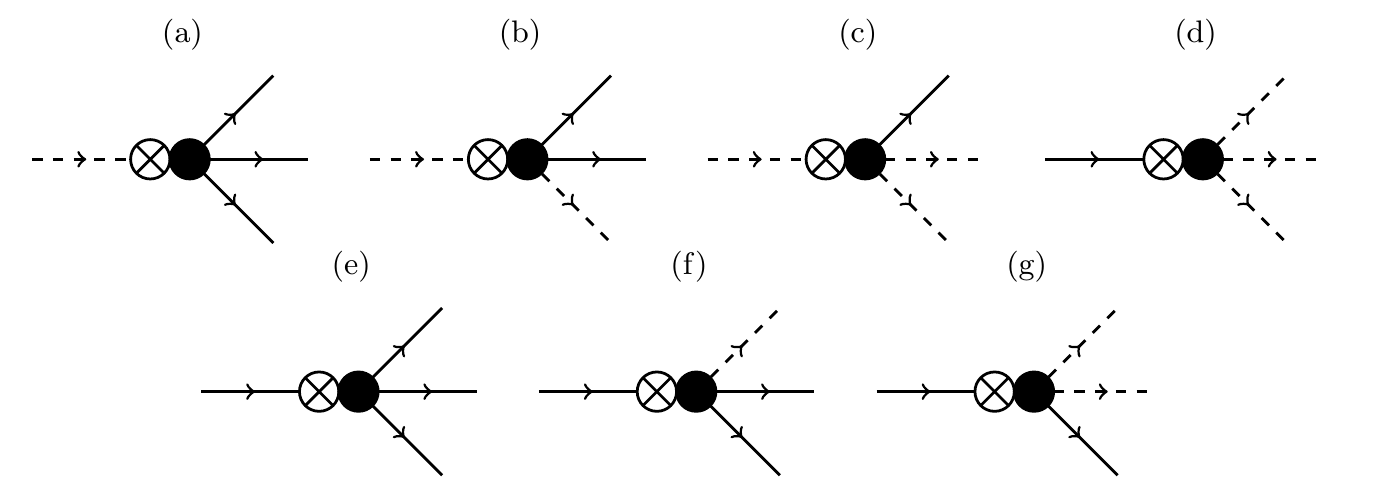}
    \caption
    [Fourth-order vertices in the RFD model]
    {Fourth-order vertices taken from the MSRJD action for the RFD model, Eq. (\ref{eq:deltaS1}). Dashed lines attached to crossed or filled circles, 
    indicate, respectively, the insertion of the instanton fields $ \hat {\mathrm{A}}^{sp}$ (dashed incoming lines) or
    ${\mathrm{A}}^{sp}$ (dashed outgoing lines) in the perturbative vertices. Solid lines have an analogous interpretation, given in
    terms of the fluctuating fields $ \hat {\mathrm{A}}$ and ${\mathrm{A}}$. These vertices are related to the following contributions to the MSRJD action (by ``odd'' or ``even'' parts of traces, we refer to the sum of tensor monomials that contain an odd or even total number of fluctuating fields): 
    (a) $ \mathrm{Tr} [ (\hat {\mathrm{A}}^{sp})^\T V_3({\mathrm{A}}) ] $,
    (b) odd part of $ \mathrm{Tr} [  ( \hat {\mathrm{A}}^{sp})^\T \Delta V_3({\mathrm{A}}) ]$,
    (c) even part of $ \mathrm{Tr} [  ( \hat {\mathrm{A}}^{sp})^\T \Delta V_3({\mathrm{A}}) ]$,
    (d) $ \mathrm{Tr} [ \hat {\mathrm{A}}^\T V_3 ({\mathrm{A}}^{sp})] $,
    (e) $ \mathrm{Tr} [ \hat {\mathrm{A}}^\T V_3({\mathrm{A}}) ] $,
    (f) odd part of $ \mathrm{Tr} [ \hat {\mathrm{A}}^\T \Delta V_3({\mathrm{A}}) ] $ and
    (g) even part of $ \mathrm{Tr} [ \hat {\mathrm{A}}^\T \Delta V_3({\mathrm{A}}) ] $.}
  \label{fig:Vertices}
\end{figure}
% maybe i should review the meaning of these diagrams? it's been so long i can't read them anymore

The evaluation of all Feynman diagrams is possible with the use of computer algebra systems, but the regime of interest is perturbative, where $g$ and $\tau$ are small values, and the vast majority of the diagrammatic contributions can be neglected. The determination of the relevant diagrams has been pursued through a power-counting procedure, taking into account the coupling parameters $g$ and $\tau$, and also the powers of the instanton fields associated to each one of the Feynman diagrams.

Explicitly, the saddle-point equations for the RFD action are:
\begin{equation} \label{eq:rfd-saddle}
    \begin{split}
        &\frac{\delta S[\hat {\mathbf{A}}, {\mathbf{A}}]}{\delta A_{ij}} =
        -i \left( \frac{d \hat A_{ij}}{dt} + \hat A_{kl} \frac{\delta V_{kl}}{\delta A_{ij}} \right) \ , \\
        &\frac{\delta S[\hat {\mathbf{A}}, {\mathbf{A}}]}{\delta \hat A_{ij}} =
        i \left( \frac{d A_{ij}}{dt} - V_{ij} \right) + g^2 G_{ijkl} \hat A_{kl} \ .
    \end{split}
\end{equation}
The roots of the equations above determine the instanton fields. With the second equation, $\hat {\mathbf{A}}^{sp}$ can be expressed in terms of ${\mathbf{A}}^{sp}$, an expression which is used to derive the order of magnitude contribution of each diagram. Graph-theoretical arguments are also important in this derivation.

% this paragraph did not have many changes, it is a difficult paragraph, which I don't understand fully
A diagram is characterized by its number of loops $L$, number of external lines $E$ (representing $\hat {\mathbf{A}}^{sp}$ or ${\mathbf{A}}^{sp}$ fields), and the numbers of vertices $N_3$ and $N_4$. $N_3$ corresponds to vertices of the type $\hat {\mathbf{A}}^{sp}({\mathbf{A}^{sp}})^3$, and $N_4$ to vertices $\hat {\mathbf{A}}^{sp}({\mathbf{A}^{sp}})^4$. The contribution of this diagram is proportional to 
\begin{equation} \label{eq:powers}
 g^{2(L-1)} (1 + a\tau^{N_3})\tau^{N_3 + 2 N_4} f( {\mathbf{A}}^{sp} ) \ .
\end{equation}
In this equation, $f( {\mathbf{A}}^{sp} )$ is a diagram-dependent homogeneous scalar function of ${\mathbf{A}}^{sp}$ with homogeneity degree $E$, that is : $f( \alpha {\mathbf{A}}^{sp}) = \alpha^E f( {\mathbf{A}}^{sp} )$, for any real positive parameter $\alpha$, being $\alpha$ a constant of the order of unity. It is important to note that vertices of type $\hat {\mathbf{A}}^{sp}({\mathbf{A}^{sp}})^2$ do not contribute with factors that depend on their diagrammatic participation number $N_2$, since these diagrams derive from $V_2$ contributions, which do not depend on $\tau$, as it can be seen from the potential (Eq. \ref{eq:vpowers}b). Thus, for each Feynman diagram that takes part in the cumulant expansion, we define, taking into account (\ref{eq:boundary}) and (\ref{eq:powers}), its \textit{power counting coefficient}, as
\begin{equation}
C(g,\tau, A) = g^{2(L-1)} {\hbox{Max($\tau^{N_3}$,$\tau^{2N_3}$)}} \tau^{2 N_4} A^E \label{power_count_coeff} \ , \ 
\end{equation}
where 
\begin{equation}
A \equiv \sqrt{ \mathrm{Tr} [ \bar {\mathbf{A}}^{\T}  \bar {\mathbf{A}}] }
\end{equation}
is a measure of the velocity gradient strength for the velocity gradient tensor $\bar {\mathbf{A}}$ where the vgPDF is evaluated. 

Eq. \eqref{power_count_coeff} is the basis for the ranking of diagrams. In consonance with previous numerical studies \parencite{ChevPRL,moriconi2014,afonso2010recent,grigorio2017instantons}, the values $g=1.0$ and $\tau=0.1$ are taken.
% no changes, starting here
The numerical values of the power counting coefficients are inspected in the interval $0 \leq A \leq 1$, a range where perturbation theory is assumed to hold, a fact we verify \textit{a posteriori} from the computation of PDFs.
It is important that $\tau$ be small enough for the RFD model to yield statistical results which are qualitatively similar to the ones derived from the exact Navier-Stokes equations, as observed in \textcite{ChevPRL}.
The forcing parameter $g$ is our main perturbative parameter, since it controls the intensity of fluctuations. In addition, the velocity gradient strength $A$ must also be small, once we are focused on the evaluation of relevant intermittency corrections near the PDF cores. 
% i have to review this and maybe give the argument of symmetries
% review this again, I'm skipping
The most important five contributions that appear more frequently in that range of velocity gradient strengths are labeled, in ranking order of decreasing importance, with boldface letters from {\textbf{A}} to {\textbf{E}}, and correspond to the following cumulant expansion terms,
\begin{align}
  &\textbf{A:}  \  \langle \mathrm{Tr} \left [ ( \hat {\mathbf{A}}^{sp})^\T V_2({\mathbf{A}}) \right ]_t  \mathrm{Tr} \left [ ( \hat {\mathbf{A}}^{sp})^\T V_2({\mathbf{A}}) \right ]_{t'}   \rangle_0  \sim A^2 \ , \ \label{eqA} \\ 
  &\textbf{B:} \ \langle \mathrm{Tr} \left [ ( \hat {\mathbf{A}}^{sp})^\T V_2({\mathbf{A}}) \right ]_t  \mathrm{Tr} \left [ \hat {\mathbf{A}}^\T \Delta V_2({\mathbf{A}}) \right ]_{t'} \rangle_0 \sim A^2 \ , \ \label{eqB} \\ 
  &\textbf{C:} \ \langle \mathrm{Tr} \left [ ( \hat {\mathbf{A}}^{sp})^\T \Delta V_2({\mathbf{A}}) \right ]_t  \mathrm{Tr} \left [ \hat {\mathbf{A}}^\T V_2({\mathbf{A}}^{sp}) \right ]_{t'} \rangle_0 \sim  A^4/g^2 \ , \ \\
  &\textbf{D:} \ \langle \mathrm{Tr} \left [ ( \hat {\mathbf{A}}^{sp})^\T \Delta V_3({\mathbf{A}}) \right ]_t \mathrm{Tr} \left [ \hat {\mathbf{A}}^\T V_2({\mathbf{A}}^{sp}) \right ]_{t'} \rangle_0 \sim  \tau A^5/g^2 \label{eqD} \ , \ \\
  &\textbf{E:} \ \langle \mathrm{Tr} \left [ ( \hat {\mathbf{A}}^{sp} )^\T V_3({\mathbf{A}}) +  \hat {\mathbf{A}}^\T V_3 ({\mathbf{A}}^{sp}) +  \hat {\mathbf{A}}^\T V_3({\mathbf{A}}) \right ]_t \rangle_0 = {\hbox{const.}}  \label{eqE} \ , \ 
\end{align}
which have their power counting coefficients plotted in Fig. \ref{fig:powers-scaling}a. The histogram analysis of the above top five expectation values is furthermore given in Fig. \ref{fig:powers-scaling}b.
These five diagrams agree with the intuitive notion, brought by
Eq. (\ref{power_count_coeff}), that the more relevant diagrams have smaller values
of $N_3$ and $N_4$, since $\tau = 0.1$. As a matter of fact, we point out that the higher order terms produced from the second order cumulant contributions, and not scrutinized in Eqs. (\ref{eqA} - \ref{eqE}), have prefactors that are proportional to powers of the small parameter $\tau$, and, thus, play a negligible role in the account of perturbative contributions.

\begin{figure}[ht]
 \centering
 \includegraphics[width=\textwidth]{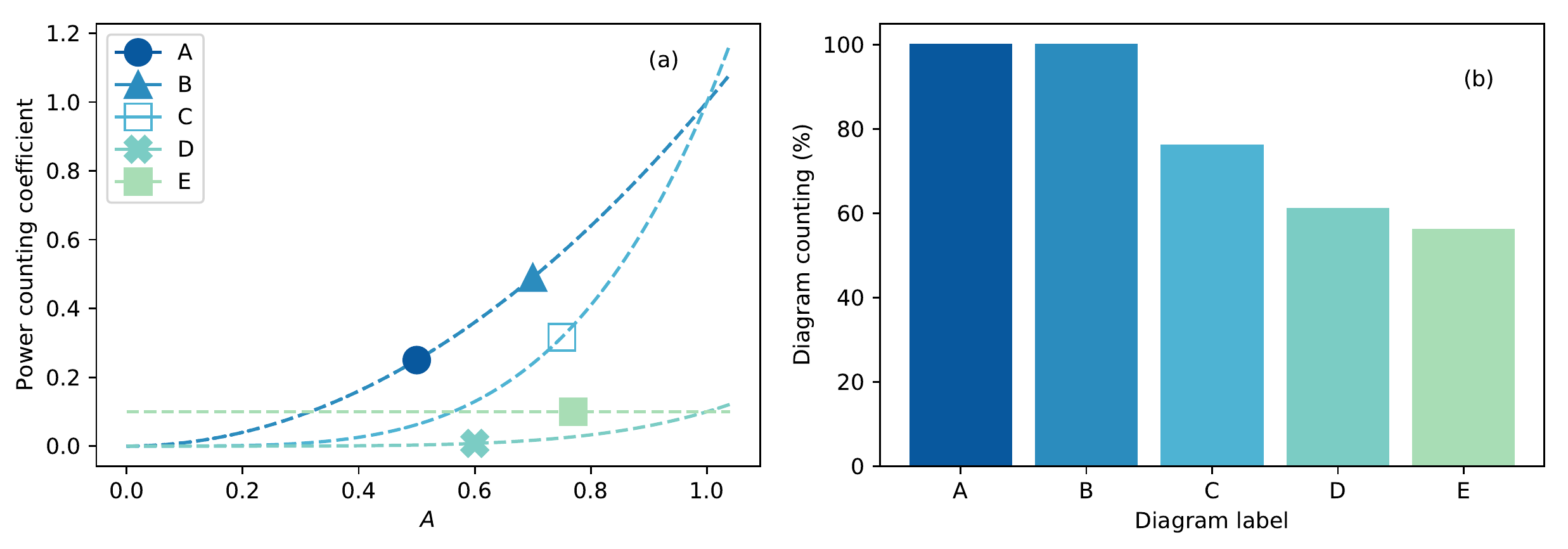}
 \caption
 [Power counting coefficients]
 {(a)  The power counting coefficient as a function of the velocity gradient strength $A$, as defined from the expectation values (\ref{eqA}-\ref{eqE}) taken for $g=1.0$ and $\tau=0.1$. (b) Relative frequencies, within the interval $0 \leq A \leq 1$, of the cases where the power counting coefficients are found to be among the first five largest ones.}
 \label{fig:powers-scaling}
\end{figure}

% did not review this
It turns out that in the considered range of velocity gradient strengths, two contributions, which have exactly the same power counting coefficients,
are clearly dominant over the remaining ones. These are the cumulant corrections {\hbox{\textbf{A}}} and {\hbox{\textbf{B}}}, defined in Eqs. (\ref{eqA}) and (\ref{eqB}). 
Note that the power counting coefficient for the contribution {\hbox{\textbf{E}}}, Eq. (\ref{eqE}), is actually independent of $A$, and, therefore, plays 
no role at all in the evaluation of the PDFs.
Diagram \textbf{E} is only displayed for matters of completeness,
since it casually happens to be larger than many other diagrams. It is important to note that power counting is actually an effective way to identify relevant contributions, provided these are in fact dependent on the velocity gradient tensor - a fact that we check for each one of the selected diagrams.

With respect to the fluctuations, the first order contribution in the expansion around the instantons is null, $ \langle \Delta S_1 [\hat {\mathbf{A}}, {\mathbf{A}}] \rangle_0 = 0$, because the instantons satisfy the Euler-Lagrange equations for the $\Delta S_0$ action. 
From this fact and Eq. (\ref{eq:cumul-gamma}), the MSRJD effective action can be written as
\begin{equation} \label{eq:eff_MSRJD_action}
 \Gamma[\hat {\mathbf{A}}^{sp}, {\mathbf{A}}^{sp} ] = S[\hat {\mathbf{A}}^{sp}, {\mathbf{A}}^{sp} ] + \sum_n C_n [\hat {\mathbf{A}}^{sp}, {\mathbf{A}}^{sp} ] \mbox{,}
\end{equation}
where $n$ labels the several second order cumulant expansion terms $C_n [\hat {\mathbf{A}}^{sp}, {\mathbf{A}}^{sp} ]$, which are dominated by the contributions $\textbf{A}$ and $\textbf{B}$.
Their associated Feynman diagrams, represented in Fig. \ref{fig:oneloop}, are noted to renormalize the noise and propagator kernels in the effective MSRJD action (\ref{eq:eff_MSRJD_action}).
\begin{figure}[ht]
  \centering
  \includegraphics[width=\textwidth]{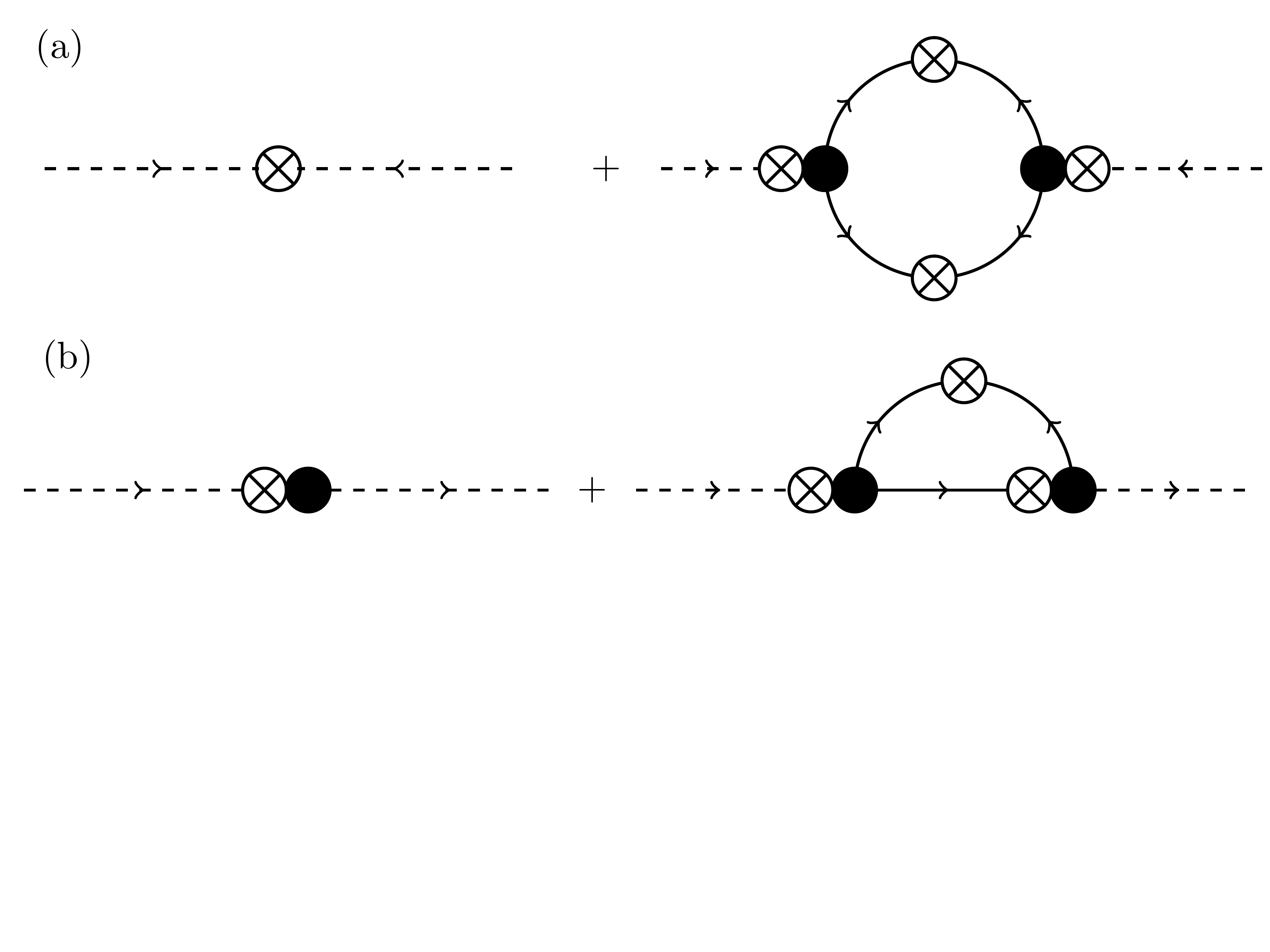}
  \caption
  [Feynman diagrams for noise and potential renormalization]
  {Feynman diagrams for (a) the renormalized noise and (b) the renormalized causal propagator kernels, which take into account the one-loop contributions {\textbf{A}} and {\textbf{B}}, respectively.}
  \label{fig:oneloop}
\end{figure}
The contributions {\textbf{A}} and {\textbf{B}} to the effective action (\ref{eq:eff_MSRJD_action}) can be written, more concretely, as
\begin{equation}\label{eq:diag_a_effective}
 C_{\textbf{A}} [\hat {\mathbf{A}}^{sp}, {\mathbf{A}}^{sp} ]  = \frac12 \int_{-\beta}^0 d t \int_{-\beta}^0 d t'
 \hat A^{sp}_{ij} (t) \hat A^{sp}_{kl} (t') C^A_{ijkl} (t-t')
\end{equation}
and
\begin{equation}\label{eq:diag_b_effective}
 C_{\textbf{B}} [\hat {\mathbf{A}}^{sp}, {\mathbf{A}}^{sp} ]  = \frac12 \int_{-\beta}^0 d t \int_{-\beta}^0 d t'
 \hat A^{sp}_{ij} (t) C^B_{ij} ({\mathbf{A}}^{sp}(t'),t-t') \mbox{,}
\end{equation}
where
\begin{equation}
    C^A_{ijkl} (t-t') = \langle [ V_2 ({\mathbf{A}}(t)) ]_{ij} [V_2 ({\mathbf{A}}(t')) ]_{kl} \rangle_0
\end{equation}
and
\begin{equation}
   C^B_{ij} ({\mathbf{A}}^{sp}(t'),t-t') = \langle [ V_2 ({\mathbf{A}}(t)) ]_{ij} \hat A_{kl} (t') [\Delta V_2 ({\mathbf{A}}(t')) ]_{kl} \rangle_0 \mbox{.}
\end{equation}
These terms correspond to the integration over fluctuating fields.

\subsection{Structure of the MSRJD Effective Action}

The effective action (\ref{eq:eff_MSRJD_action}) can be written, after the introduction of the contributions (\ref{eq:diag_a_effective}) and (\ref{eq:diag_b_effective}), as

\begin{equation}\label{eq:effective_with_integrals}
\begin{split}
\Gamma [ \hat {\mathbf{A}}, {\mathbf{A}}] = i \int_{-\beta}^0 d t \int_{-\beta}^0 d t' \Bigg \{ 
&\mathrm{Tr} \left[ \hat {\mathbf{A}}^\T(t) \left( \frac{d\mathbf{A}}{dt} - V^{\mathrm{ren}}({\mathbf{A}}(t'),t-t') \right)
\right] \\
+ &\frac{g^2}{2} G_{ijkl}^{\mathrm{ren}} (t-t') \hat A_{ij} (t) \hat A_{kl} (t') \Bigg \} \ .
\end{split}
\end{equation}
This action has the same form as the instanton action, $S_{sp}[\hat {\mathbf{A}}^{sp},{\mathbf{A}}^{sp}]$, with two renormalized terms: the noise $G$ and the potential $V$. These renormalizations correspond to the effective role of the fluctuations to the terms in the saddle-point action. Explicitly, these contributions are
\begin{equation}\label{eq:noise_renormalized}
G_{ijkl}^{\mathrm{ren}} (t-t') \equiv  G_{ijkl} \delta(t-t') + C^A_{ijkl}(t-t') 
\end{equation}
and
\begin{equation}\label{eq:potential_renormalized}
V_{ij}^{\mathrm{ren}}({\mathbf{A}}(t'),t-t') \equiv V_{ij}({\mathbf{A}}(t')) \delta(t-t') - C^B_{ij}({\mathbf{A}}(t'),t-t') \ . \ 
\end{equation}
In contrast to the original nonperturbed MSRJD action (\ref{eq:quadratic}), the above renormalized form (\ref{eq:effective_with_integrals}) contains kernels that depend
non-trivially on a pair of time instants $t$ and $t'$. As it is usual (sometimes in an implicit way) in renormalization group
studies \parencite{amit2005,zinn2002,Peskin}
% there used to be references to Barabaki, Kardar and Kogut here, I removed to make the text shorter, since these are almost random citations
, the structure of the renormalized effective action can be simplified in the case of slowly varying fields
(as the instanton fields are assumed to be). This simplification is achieved through the procedure of low-frequency renormalization, which 
in our context consists in replacing the renormalization kernels $C^A_{ijkl}$ and $C^B_{ij}$ by singular ones, 
according to the prescriptions
\begin{subequations}
\begin{align}
&C^A_{ijkl}(t-t') \rightarrow \tilde C^A_{ijkl} \delta(t-t') \ , \ \\
&C^B_{ij}({\mathbf{A}}(t'), t-t') \rightarrow \tilde C^B_{ij}({\mathbf{A}}(t')) \delta(t-t') \ , \ 
\end{align}
\end{subequations}
where
\begin{subequations}
\begin{align}
& \tilde C^A_{ijkl} \equiv  \int_{- \infty}^\infty dt'  C^A_{ijkl}(t- t') \ , \ \label{eq:coeff_diag_a}\\
& \tilde C^B_{ij}({\mathbf{A}}(t)) \equiv \int_{- \infty}^\infty dt'  C^B_{ij}({\mathbf{A}}(t'),t-t') \ . \ \label{eq:coeff_diag_b}
\end{align}
\end{subequations}
Substituting (\ref{eq:coeff_diag_a}) and (\ref{eq:coeff_diag_b}) in (\ref{eq:noise_renormalized}) and 
(\ref{eq:potential_renormalized}), the nonperturbed and the effective MSRJD actions will, then, become
isomorphic to each other, provided that the tensors
$G_{ijkl}$ and $V_{ij}({\mathbf{A}})$ of the nonperturbed action are mapped, respectively, to the tensors
\begin{equation} \label{eq:g_renc}
\tilde G_{ijkl}^{\mathrm{ren}} \equiv  G_{ijkl} + \tilde C^A_{ijkl} 
\end{equation}
and
\begin{equation} \label{eq:v_renc}
\tilde V_{ij}^{\mathrm{ren}}({\mathbf{A}}) \equiv V_{ij}({\mathbf{A}})  - \tilde C^B_{ij}({\mathbf{A}}) 
\end{equation}
that appear in the definition of the effective renormalized action.

It is important to observe, furthermore, that from the traceless property of the stochastic forcing, it follows that $\tilde G_{iikl}^{\mathrm{ren}} 
=\tilde G_{ijkk}^{\mathrm{ren}} = 0$, and we may write, in general, that
\begin{equation}\tilde G_{ijkl}^{\mathrm{ren}} = D_{ijkl} -\frac{1}{3}(x+y) \delta_{ij} \delta_{kl} \ , \
\end{equation}where 
\begin{equation}D_{ijkl} = x \delta_{ik} \delta_{jl} + y \delta_{il} \delta_{jk} \ , \
\end{equation}with $x$ and $y$ being two independent arbitrary parameters. The computation of the noise renormalization diagram, Fig. \ref{fig:oneloop}a,
returns
\begin{equation} \label{eq:g_correction}
 \tilde C^A_{ijkl} = \frac{g^4}{8} \left( 6 \delta_{ik} \delta_{jl}
 - \frac{1}{4} \delta_{il} \delta_{jk} - \frac{23}{12} \delta_{ij} \delta_{kl} \right) \mbox{,}
\end{equation}
and, as a consequence,
\begin{equation} \label{eq:renormalization_xy}
x = 2 + \frac{3}{2} g^2 \ , \  y = -\frac12 - \frac{1}{16} g^2 \mbox{.}
\end{equation}

In its turn, the renormalization of the potential produces the contribution
\begin{equation} \label{eq:v_correction}
    \tilde C^B_{ij} = \frac{g^2}{16} ( 4 A_{ji} - A_{ij} ) .
\end{equation}

The calculation of the renormalization terms in Eqs.~\eqref{eq:g_correction} and \eqref{eq:v_correction} involves the manipulation of several $A_{ij}$ and $G_{ijkl}$ factors in complex tensor contractions. These tensor contractions have been performed with the computer algebra system Mathematica and they are described in Appendix \ref{app:comp}.

Recalling, now, the saddle-point to solve for $\hat {\mathbf{A}}^{sp}$ in terms of ${\mathbf{A}}^{sp}$, the MSRJD effective action can be 
rewritten in a more compact way, up to the same order in perturbation expansion, as a scalar functional uniquely
dependent on the velocity gradient tensor field ${\mathbf{A}}$, namely,
\begin{equation}\label{eq:longGamma}
\begin{split}
\Gamma [{\mathbf{A}}] &= \frac{1}{2 g^2} \int_{-\beta}^0 d t \ 
\left( \frac{dA_{ij}}{dt} - \tilde V^{\mathrm{ren}}_{ij}({\mathbf{A}}) \right)
D^{-1}_{ijkl} 
\left( \frac{dA_{kl}}{dt} - \tilde V^{\mathrm{ren}}_{kl}({\mathbf{A}}) \right)
\\ &= \frac{a}{2 g^2} \int_{-\beta}^0 d t \ 
\left( \frac{dA_{ij}}{dt} - \tilde V^{\mathrm{ren}}_{ij}({\mathbf{A}}) \right)
\left( \frac{dA_{ij}}{dt} - \tilde V^{\mathrm{ren}}_{ij}({\mathbf{A}}) \right)
\\ &+ \frac{b}{2 g^2} \int_{-\beta}^0 d t \ 
\left( \frac{dA_{ij}}{dt} - \tilde V^{\mathrm{ren}}_{ij}({\mathbf{A}}) \right)
\left( \frac{dA_{ji}}{dt} - \tilde V^{\mathrm{ren}}_{ji}({\mathbf{A}}) \right) \ , 
\end{split}
\end{equation}
where
\begin{equation}
    D^{-1}_{ijkl} \equiv a \delta_{ik} \delta_{jl} + b \delta_{il} \delta_{jk} \ , \
\end{equation}
with
\begin{equation} \label{eq:ab_from_xy}
    a = - \frac{x}{y^2 - x^2} \ , \ b = \frac{y}{y^2 - x^2} \ . \     
\end{equation}

In the formulation of Eq.~\ref{eq:longGamma}, it is not necessary anymore to work with a coupled set of saddle-point equations. Instead, the instanton configuration is obtained from a single equation
\begin{equation} \label{eq:euler-lagrange-gamma}
 \left. \frac{\delta \Gamma[{\mathbf{A}}]}{\delta A_{ij}} \right|_{{\mathbf{A}}={\mathbf{A}}^{sp}} = 0 \ . 
\end{equation}
The effective action of Eq.~\ref{eq:longGamma} is usually called the Onsager-Machlup action functional \parencite{onsager1953fluctuations}.

\subsection{Instanton Configurations}

One the working hypothesis made in \textcite{moriconi2014} was the relevance of an approximate analytical instanton, instead of the exact field, solution of Eq.~\ref{eq:rfd-saddle}. In this work, the applicability of this hypothesis was verified. The approximate instantons consist in the saddle-point solutions of a quadratic truncated renormalized effective action,
\begin{equation} \label{eq:gamma_quadratic}
 \Gamma_0[{\mathbf{A}}] \equiv \frac{a}{2 g^2} \int_{-\beta}^0 d t \ \mathrm{Tr}
 \left[\dot {\mathbf{A}}^\T \dot {\mathbf{A}} + {\mathbf{A}}^\T {\mathbf{A}} \right] +
 \frac{b}{2 g^2} \int_{-\beta}^0 d t \ \mathrm{Tr} \left[\dot {\mathbf{A}}^2 + {\mathbf{A}}^2 \right] \ .
\end{equation}
The corresponding approximate saddle-point equation is
\begin{equation}\label{eq:approx_EOM}
 \left. \frac{\delta \Gamma_0[{\mathbf{A}}]}{\delta A_{ij}} \right|_{{\mathbf{A}}={\mathbf{A}}^{sp}} = 0 \Rightarrow  \ddot {\mathbf{A}}^{sp} - {\mathbf{A}}^{sp} = 0
\ ,\ 
\end{equation}
subject to the periodic boundary condition (\ref{eq:boundary}).
Instanton solutions of (\ref{eq:approx_EOM}) have the form
\begin{equation} \label{eq:analytical_instanton}
 {\mathbf{A}}^{sp} (t) = \bar {\mathbf{A}} f(\beta, t) \mbox{,}
\end{equation}
where the function $f(\beta,t)$, defined for $-\beta \leq t \leq 0$, is given by
\begin{equation}
 f(\beta, t) = 2 \frac{\sinh(\beta/2)}{\sinh(\beta)} \cosh(t + \beta/2) \mbox{.}
\end{equation}
Additionally, since the potential functions $V_p({\mathbf{A}})$ are homogeneous functions of degree $p$, their contribution conforms to
\begin{equation}
 V({\mathbf{A}}^{sp}(t)) = \sum_{p = 1}^4 V_p (\bar {\mathbf{A}}) [ f(\beta,t) ]^p \ .
\end{equation}
The above expression together with (\ref{eq:vpowers}), (\ref{eq:g_renc}) and \eqref{eq:v_renc}, leads to the evaluation of the effective action (Eq. \ref{eq:longGamma}) from these following scalar contributions:
% what equation to cite here?
\begin{equation}
\begin{split}
\int_{-\beta}^0 dt \ &\left( \frac{d A_{ij}^{sp}}{dt} - V_{ij}^{\mathrm{ren}}(\mathbf{A}^{sp}) \right) \left( \frac{d A_{ij}^{sp}}{dt} - V_{ij}^{\mathrm{ren}}(\mathbf{A}^{sp}) \right)
= \\ &I_1 (\beta) 
\mathrm{Tr} [ \bar {\mathbf{A}}^\T \bar {\mathbf{A}} ] + \sum_{p=1}^4 \sum_{q=1}^4 I_{p+q}  (\beta) H_{p,q}(\bar {\mathbf{A}}^\T, \bar {\mathbf{A}})
\end{split}
\end{equation}
and
\begin{equation}
\begin{split}
\int_{-\beta}^0 dt \ &\left( \frac{d A_{ij}^{sp}}{dt} - V_{ij}^{\mathrm{ren}}(\mathbf{A}^{sp}) \right) \left( \frac{d A_{ji}^{sp}}{dt} - V_{ji}^{\mathrm{ren}}(\mathbf{A}^{sp}) \right)
= \\ &I_1 (\beta) 
\mathrm{Tr} [ \bar {\mathbf{A}}^2  ] + \sum_{p=1}^4 \sum_{q=1}^4 I_{p+q} (\beta) H_{p,q}(\bar {\mathbf{A}}, \bar {\mathbf{A}}) \ , \    
\end{split}
\end{equation}
where $H_{p,q}(X,Y)$ is an homogeneous scalar function of degrees $p$ and $q$, related, 
respectively, to the matrix variables $X$ and $Y$, and
\begin{equation}
 I_1 (\beta)  \equiv \int_{-\beta}^0 d t [ \dot f(\beta, t) ]^2 \mbox{,}
 \ I_{p+q} (\beta) \equiv \int_{-\beta}^0 d t [ f(\beta, t) ]^{p+q} \mbox{.}
\end{equation}
At asymptotic times, $\beta \rightarrow \infty$, the constants $I_p$ are defined as $I_p = \lim_{\beta \rightarrow \infty } I_p (\beta)$
and their exact values are:
\begin{equation}
\begin{split}
     &I_1 = I_2 = 1 \mbox{,} \  I_3 = 2/3 \mbox{,} \  I_4 = 1/2 \mbox{,}  \nonumber \\
     &I_5 = 2/5 \mbox{,} \  I_6 = 1/3 \mbox{,} \  I_7 = 2/7 \mbox{,} \  I_8 = 1/4 \mbox{.}
\end{split}
\end{equation}
Assembling all the above pieces together, the effective action is
\begin{equation}\label{gamma_A}
\begin{split}
\Gamma[{\mathbf{A}}^{sp}] \equiv \Gamma( \bar {\mathbf{A}}) &= \frac{aI_1}{2 g^2} \mathrm{Tr} [ \bar {\mathbf{A}}^\T \bar {\mathbf{A}} ] + \frac{bI_1}{2 g^2} \mathrm{Tr} [ \bar {\mathbf{A}}^2  ]
 \\ &+ \sum_{p=1}^4 \sum_{q=1}^4 \frac{I_{p+q}}{2g^2} [ a H_{p,q}(\bar {\mathbf{A}}^\T, \bar {\mathbf{A}})
 + b H_{p,q}(\bar {\mathbf{A}}, \bar {\mathbf{A}}) ] \ . \ 
\end{split}
\end{equation}
The normalized vgPDF can now be readily derived from (\ref{nn_vgPDF}) and (\ref{gamma_A}), therefore, as 
\begin{equation}\rho(\bar {\mathbf{A}}) = {\cal{N}} \exp [ - \Gamma(\bar {\mathbf{A}}) ] \ . \  \label{n_vgPDF}
\end{equation}

It is relevant to compare the approximate instanton solutions (Eq. \ref{eq:analytical_instanton}) with accurate numerical solutions in specific cases. As discussed in \textcite{grigorio2017instantons}, diagonal velocity gradient instantons can be obtained from the application of the Chernykh-Stepanov method \parencite{chernykh2001} for the particular boundary conditions,
\begin{equation}
\begin{split} 
&\bar A_{11} = -2 \bar A_{22} = -2 \bar A_{33} \equiv c \ , \ \label{c_boundary} \\
& \bar A_{ij} = 0 {\hbox{  for $i \neq j$}} \ ,  \ 
\end{split}
\end{equation}
where $c$ is an arbitrary constant. The approximate and the numerical instantons for $c=1$, $\tau=0.1$, and $g=0.8$ are both plotted in Fig. \ref{fig:instanton}, for three different values of the $\beta$ parameter. This figure reveals that the approximate instantons are uniformly close to the exact ones, justifying the use of quadratic instantons in the analytical approach. It is also worth pointing out that the regime in question is that of moderate Reynolds numbers and fluctuations only a few standard deviations from the mean, which does not guarantee the same uniform agreement for high Reynolds numbers. The use of approximate instantons also implies, in an exact formulation, that the first order contribution $\langle \Delta S_1 \rangle_0$ is not zero, but these contributions can be safely discarded due to their low relevance in the power counting scheme, and by the small difference between the exact and numerical instantons.

\begin{figure}[ht]
 \centering
 \includegraphics[width=\textwidth]{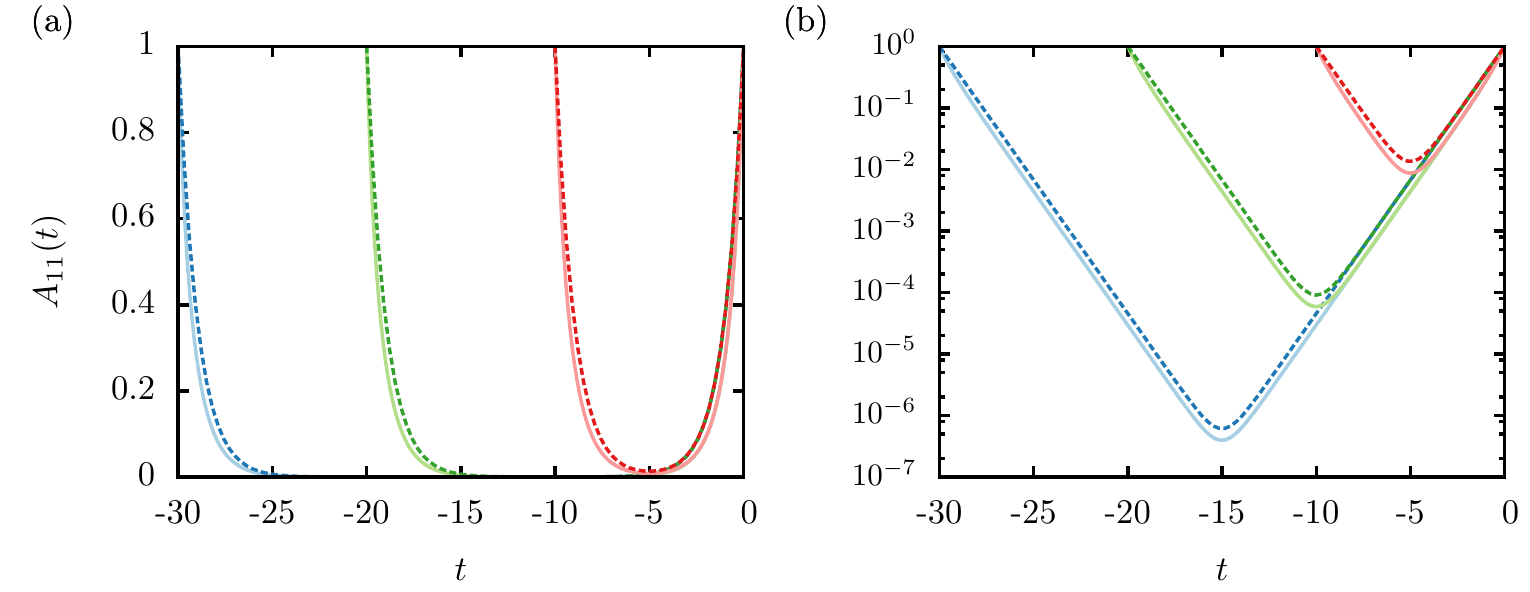}
 \caption
 [Comparison between approximate and numerical instantons]
 {Comparison between approximate (dashed lines) and numerical instantons (solid lines), obtained, respectively from Eq. (\ref{eq:analytical_instanton}) and from the application of the Chernykh-Stepanov method as discussed in \textcite{grigorio2017instantons}, for $c=1$ [that is $\bar A_{11} = 1$, see Eq. (\ref{c_boundary})], $\tau=0.1$, and $g=0.8$ in (a) linear and (b) monolog scales.
 Notice that the approximate and the numerical instantons both refer to the RFD model.
 Blue, green and red curves (left to right) correspond to $\beta =30$, $20$ and $10$.
 }
 \label{fig:instanton}
\end{figure}

%%%%%%%%%%%%%%%%%%%%%%%%%%%%%%%%%%%%%%%%%%%%%%%%%%%%%%%%%%%%%%%%%%%%%%%%%%%%%%%%%%%%%%%%%%%%%%%%%%%%%%%%

\section{Numerical Results} 
\label{sec:Results}

This section discusses the analytical predictions obtained from the renormalized effective action in comparison to the numerical results obtained directly from the RFD model. The observables considered have been previously studied in the context of Lagrangian velocity gradients and reveal many of the non-trivial features of their dynamics.

%\subsection{Marginal velocity gradient PDFs}\label{ssec:vgpdfs}

The PDF in Eq.~\ref{n_vgPDF} is multivariate, it is defined on the domain of nine velocity gradient components. But symmetries between the components yield redundant the individual analysis of all of them. The high-dimensionality of the problem also makes it impractical to consider the whole distribution. For this reason, marginal PDFs have been produced from statistical ensembles of velocity gradients generated with a Monte Carlo procedure. The detailed procedure to sample configurations from the effective action $\Gamma[\hat {\mathbf{A}}^{sp}, {\mathbf{A}}^{sp} ]$ is described in \textcite{moriconi2014}.

The Monte Carlo samples consisted of sets of $ 8 \times 10^6$ velocity gradient tensors, from which were extracted ensembles of $ 24 \times 10^6$ and $ 48 \times 10^6$ diagonal and off-diagonal velocity gradient components, respectively. An illustrative case for the marginal PDFs of the diagonal and off-diagonal components of the velocity gradient tensor is given in Fig. \ref{fig:pdf-single-g}, for controlling parameters $\tau=0.1$ and $g=0.8$.
For this value of $\tau$, the RFD model leads to statistical results similar to the ones observed in realistic turbulent flows \parencite{ChevPRL}.

The PDFs of velocity gradients are compared in four different situations:
\begin{enumerate}[label=(\roman*)]
\item \label{it:DNS} the straightforward numerical simulations of the RFD model, Eq. \eqref{eq:model} with the potential from Eq.~\eqref{eq:vpotential}. This statistical ensemble contains $10^9$ velocity gradient tensors, covering $2 \times 10^5$  integral times scales, 
\item \label{it:NonR} the saddle-point MSRJD action with no renormalization contributions, $S_{sp}[\hat {\mathbf{A}}^{sp}, {\mathbf{A}}^{sp}]$. 
\item \label{it:NoiseR} the {\textit{ partial renormalization}} of the effective MSRJD action, which is renormalized only by the noise contribution, as given by Eqs. (\ref{eq:effective_with_integrals}) and
(\ref{eq:noise_renormalized}), with $\tilde C^B_{ij}=0$ prescribed in (\ref{eq:potential_renormalized}), and
\item \label{it:FullR} the {\textit{full renormalization}} of the effective action, which is renormalized by both the noise and the propagator contributions, as given by Eqs. (\ref{eq:effective_with_integrals}), (\ref{eq:noise_renormalized}), and (\ref{eq:potential_renormalized}).
\end{enumerate}

% review figure captions
\begin{figure}[ht]
 \centering
 \includegraphics[width=\textwidth]{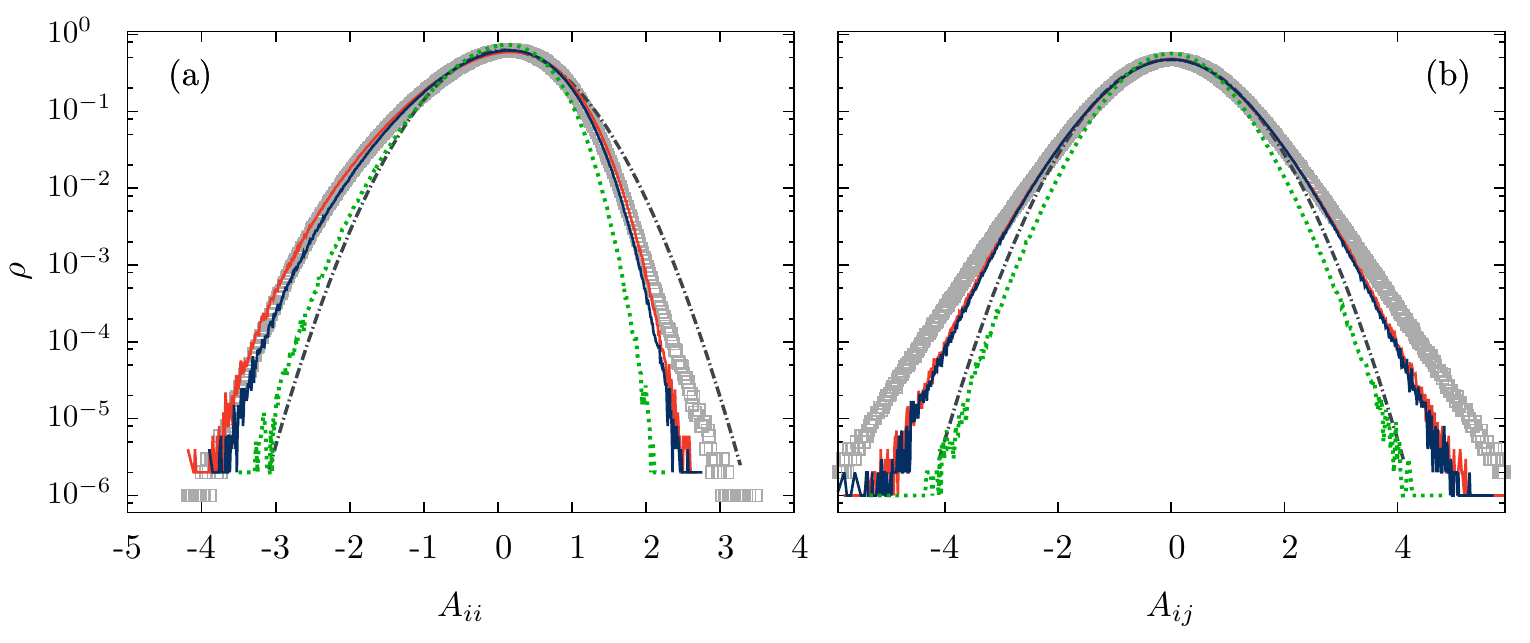}
 \caption
 [PDFs for diagonal and off-diagonal velocity gradient components]
 {PDFs for the (a) diagonal and (b) off-diagonal velocity gradient components, computed for $\tau=0.1$ and $g=0.8$. Open squares refer to the empirical PDFs derived from the numerical solutions of the RFD model, while all the other PDFs follow from analytical expressions obtained at different improvement levels. Green dashed lines correspond to no effective action renormalization, red (light gray) lines to partial renormalization and blue (dark gray) lines to full renormalization.
 The dotted-dashed gray lines correspond to Gaussian fits of the numerical (RFD) data around the PDFs peaks, which clearly show deviations from quadratic behavior both in the partial and full renormalization schemes.
 The diagonal and off-diagonal empirical PDFs have standard deviations and kurtosis given by $\sigma = 0.66$ and $\kappa = 3.3$ and
$\sigma = 0.89$ and $\kappa = 3.7$, respectively.
 }
 \label{fig:pdf-single-g}
\end{figure}

In Fig.~\ref{fig:pdf-single-g}, it can be seen that great improvement is obtained with the use of renormalized actions. Noise renormalization is found to be the leading contribution, and for this reason we may refer to the propagator renormalization contribution as the subleading one.
The results from partial and full renormalization, though, seem essentially equivalent, with small differences observed mainly for the PDF tails of diagonal velocity gradient components. In closer inspection, though, it is found that the core regions of the numerical distribution is more accurately described by the PDF from the complete renormalization procedure.

%One may wonder how the modeled vgPDFs plotted in Fig. \ref{fig:pdf-single-g} (the green, red and blue lines) would change if exact (numerical) instantons \textcite{grigorio2017instantons}, were used instead of the approximated (but analytical) ones considered in this work. It is clear that when renormalization is absent, the use of exact instantons leads in general to better and reasonable vgPDFs for small values of $g$. For larger values of $g$ ($g > 0.4$, roughly), fluctuations become more important and have to be necessarily taken into account for a proper modeling of the vgPDFs \textcite{grigorio2017instantons}.

\begin{figure}[h]
 \centering
 \includegraphics[width=\textwidth]{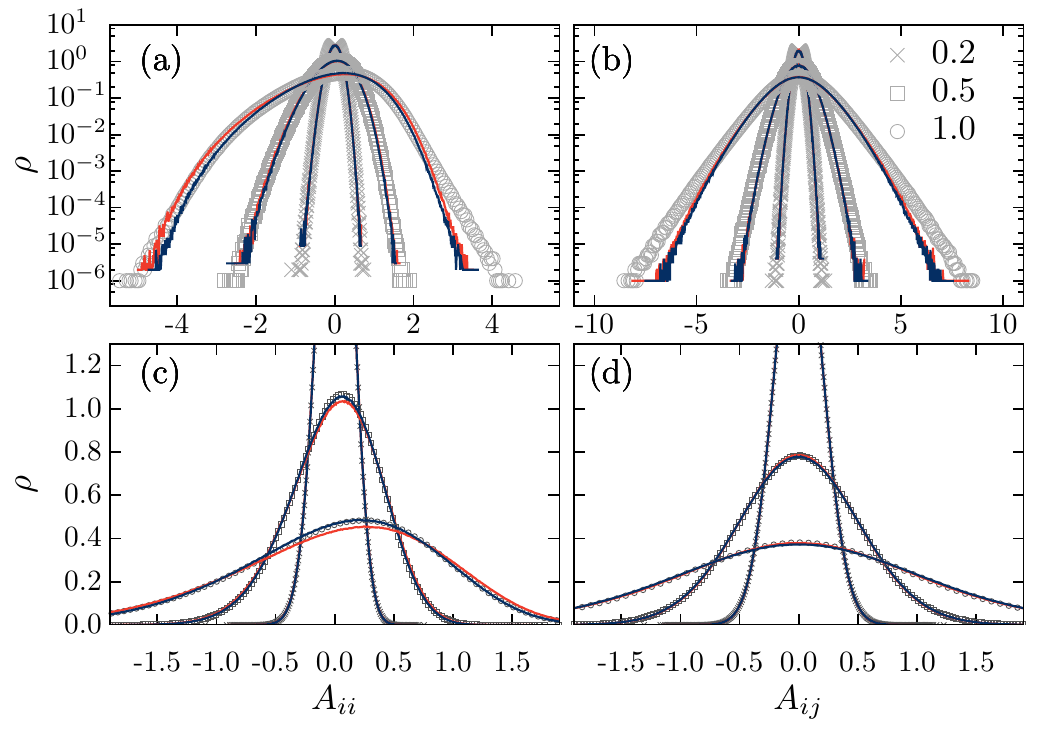}
 \caption
 [PDFs for diagonal and off-diagonal velocity gradient components at several values of the fluctuation intensity $g$]
 {PDFs for the diagonal components of the velocity gradient tensor
 at (a) and (c), and its off-diagonal components, at (b) and (d).
 Figures (a) and (b) are in monolog scale, while figures
 (c) and (d) are in linear scale, and they represent the same
 sets of data.
 Symbols refer to the empirical PDFs derived from the numerical solutions of Eq. (\ref{eq:model}), for different values of the random force
 strength $g$, as indicated in the plots; red (light gray) and blue
 (dark gray) lines refer, respectively, to PDFs obtained from partial
 and fully renormalized effective actions. With better
 visualization in mind, the non-renormalized PDFs have not been plotted.}
 \label{fig:pdf-several-g}
\end{figure}

A more extensive comparison between PDFs is provided in Fig. \ref{fig:pdf-several-g}, where the cores of these distributions are examined with more care. In this figure, three values of $g$ are chosen, up to $g=1.0$, which is close to the limit of validity of perturbation theory. 
This upper limit can be estimated from the perturbative corrections (\ref{eq:renormalization_xy}) and appreciated from the results of simulations: above $g=1.0$ the renormalized PDFs are noted to deviate in a more expressive way from the numerical ones.
It is to be remarked here that the definition of an upper bound for the coupling constant $g$ is by no means a sufficient condition for the 
validity of the perturbative expansion, since it is important that both $g$ and $A$ are not too large for the consistency of the cumulant expansion approach.
There is a clear interplay between these quantities, since the variance of $A$ scales as $g^2$ for small values of $g$, though this yields interesting information only around the PDF peaks. 

\begin{table}[ht]
\centering
%\caption{std}
\begin{tabular}{|c|c|c|c|c|c|c|}
\hline \multicolumn{7}{|c|}{\textbf{Standard Deviations}}
\\ \hline
\textbf{g}  & \multicolumn{2}{c|}{0.2} & \multicolumn{2}{c|}{0.5} & \multicolumn{2}{c|}{1.0} \\ \hline
 {\textbf{Statistical Ensembles}} & D & OD & D & OD & D & OD \\ \hline
Numerical RFD & 0.14 & 0.20 & 0.39 & 0.52 & 0.86 & 1.15 \\ \hline
No Renormalization & 0.14 & 0.20 & 0.35 & 0.48 & 0.68 & 0.88 \\ \hline
Partial Renormalization & 0.14 & 0.20 & 0.39 & 0.52 & 0.90 & 1.13 \\ \hline
Full Renormalization & 0.14 & 0.20 & 0.39 & 0.52 & 0.84 & 1.13 \\ \hline
\end{tabular}
\begin{tabular}{|c|c|c|c|c|c|c|}
\hline \multicolumn{7}{|c|}{\textbf{Kurtoses}}
\\ \hline
\textbf{g}  & \multicolumn{2}{c|}{0.2} & \multicolumn{2}{c|}{0.5} & \multicolumn{2}{c|}{1.0} 
\\ \hline
 {\textbf{Statistical Ensembles}} & D & OD & D & OD & D & OD\\ \hline
Numerical RFD & 3.05 & 3.03 & 3.25 & 3.23 & 3.23 & 3.87 \\ \hline
No Renormalization & 3.03 & 3.01 & 3.12 & 3.09 & 3.19 & 3.33 \\ \hline
Partial Renormalization & 3.03 & 3.01 & 3.14 & 3.12 & 3.14 & 3.51 \\ \hline
Full Renormalization & 3.03 & 3.01 & 3.14 & 3.10 & 3.14 & 3.39 \\ \hline
\end{tabular}
\caption{Standard deviations and kurtoses associated to the vgPDFs shown in Fig. \ref{fig:pdf-several-g}. The labels D and OD stand for statistical ensembles of diagonal and off-diagonal velocity gradient components, respectively. These ensembles are characterized, besides the D/OD classification, from specifying how their associated PDFs are obtained,
according to four alternative schemes, described as the items
\ref{it:DNS} to \ref{it:FullR} in Sec. \ref{sec:Results}:
numerical simulations of the RFD model, non-renormalized saddle-point MSRJD actions, partially renormalized effective actions, and fully renormalized effective actions.
}
\label{tab}
\end{table}

% not changed
The standard deviations and kurtoses of these PDFs are reported in Table \ref{tab}. It is verified that the standard deviations agree remarkably well with those evaluated directly from the RFD model. 
The comparison between kurtoses, on the other hand, has some discrepancies, mainly due to the slow decay of the far tails of the PDFs, which are out of the reach of the present approach.

% To analyze the observed ranges of agreement between the fully renormalized and the empirical vgPDFs in a more quantitative way, we define them as the velocity gradient regions where the calculated perturbative corrections correspond to a given fraction, say $20\%$, of the saddle-point action. The values of $\bar A$ that are obtained from this prescription establish an estimate for the border of validity of perturbation theory, and turn out to be well described by an approximate power-law relation, $ \bar A/g \approx 1.29 \ g^{-0.41} $. Following such a procedure, we find that the limit of applicability of perturbation theory along the lines of the cumulant expansion is actually compatible with the qualitative arguments addressed before.

%\subsection{Joint Statistics of the Velocity Gradient Invariants {\hbox{$Q$}} and {\hbox{$R$}} }

\begin{figure}[ht]
 \centering
 \includegraphics[width=\textwidth]{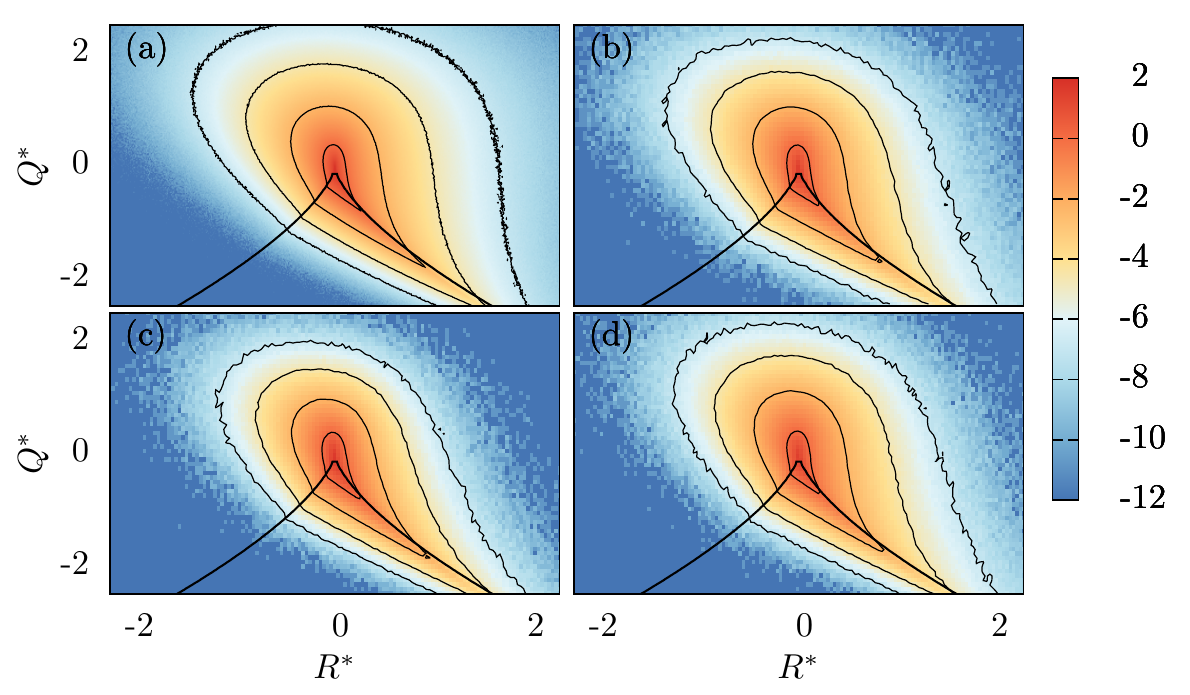}
 \caption
 [Joint PDFs of the velocity gradient invariants $Q^*$ and $R^*$]
 {Joint PDFs (and their level curves) of the velocity gradient invariants $Q^*$ and $R^*$, as obtained from (a) numerical simulations of the RFD model,
 (b) the analytical approach based on the non-renormalized effective MSRJD action, (c) the noise renormalized effective action and (d) the fully renormalized  effective action.
 Solid lines represent the \enquote{Vieillefosse line}, defined by $(27/4)R^2 + Q^3 = 0$ (a constraint which holds for the inviscid evolution of the velocity gradient tensor). The data correspond to the RFD parameters $\tau=0.1$ and $g=0.8$,  and the color bar scale corresponds to powers of 10 in the visualization of the joint probability distribution functions.
 }
 \label{fig:QR}
\end{figure}

The pair of velocity gradient invariants $Q \equiv - \mathrm{Tr}({\mathbf{A}}^2)/2$ and $R \equiv -\mathrm{Tr}({\mathbf{A}}^3)/3$ have been extensively used in the literature as important observables for the investigation of structural aspects of turbulence \parencite{TsinoberInformal}. Turbulent flow regions can be dominated by enstrophy ($Q>0$) or strain ($Q<0$) and, independently, by compression ($R>0$) or stretching ($R<0$) dynamics. It is interesting to work with a dimensionless version of these invariants, 
\begin{equation}
 Q^* = - \frac{\mathrm{Tr}({\mathbf{A}}^2)}{2 \langle \mathrm{Tr}(\mathbf{S}^2) \rangle } \ \ \mbox{and} \ \ 
 R^* = - \frac{\mathrm{Tr}({\mathbf{A}}^3)}{3 \langle \mathrm{Tr}(\mathbf{S}^2)  \rangle^{3/2} } \mbox{,}
\end{equation}
where $\mathbf{S} = ({\mathbf{A}} + {\mathbf{A}}^\T)/2$ is the strain rate tensor, the symmetric part of the velocity-gradient tensor. The joint PDF of $Q^*$ and $R^*$ shows a characteristic teardrop shape, which was first observed in direct numerical simulations of turbulence \parencite{leorat1975turbulence,vieillefosse1982,vieillefosse1984}. This teardrop shape is qualitatively well reproduced by the RFD model \parencite{ChevPRL}.

With the same Monte Carlo ensembles used to produce the marginal PDFs, the joint PDF of $Q^*$ and $R^*$ was produced in the same four regimes, and the relevance of renormalization can be observed for this observable too, as can be seen in Fig.~\ref{fig:QR}. In particular, the difference between the leading order (Fig.~\ref{fig:QR}c) and complete renormalization (Fig.~\ref{fig:QR}d) is noticeable, since the latter is much closer to the numerical results (Fig.~\ref{fig:QR}a) than the former.

\section{Discussion}

The results of \textcite{moriconi2014} motivated this deeper analysis of the field theoretical approach to the RFD approach, a model of Lagrangian intermittency \parencite{ChevPRL}. The work described in this chapter, \textcite{apolinario2019instantons}, pursued a detailed verification of approximation hypotheses which were made in this previous work. 

Notably, two main results stand out: The accuracy of the quadratic instanton, which is used as a proxy for the exact instanton, was verified. And the ranking of the relevance of each diagram was established, advocating for the relevance of the noise renormalization which was empirically observed in \textcite{moriconi2014}.

The cumulant expansion yields accurate results for the core of PDFs at the onset of turbulence, as observed. That is, in regimes of moderate Reynolds numbers, where fat tails start to stand out. In this situation, the expansion is a reliable perturbation technique. At the far PDF tails, on the other hand, the perturbative expansion breaks down, and asymptotically large fluctuations are instead well described purely by the instantons, thus requiring an exact treatment of the saddle-point solutions.

\end{chapter}

\begin{chapter}{The Onset of Intermittency in Stochastic Burgers Hydrodynamics}
\label{chap:burgers}

\hspace{5 mm} 

The Burgers model was introduced in \textcite{burgers1948} as a one-dimensional prototype for Navier-Stokes hydrodynamics. It displays nonlinearities which are similar to those of Navier-Stokes, and became a valuable testing ground for the development of new theoretical and computational approaches in turbulence and hydrodynamics, as foreseen in \textcite{vonneumann1963}. Furthermore, the Burgers equation has been applied to realistic problems in a variety of fields: nonlinear acoustics \parencite{gurbatov1991}, cosmology \parencite{zeldovich1970, gurbatov1984}, critical interface growth \parencite{kardar1986}, traffic flow dynamics \parencite{musha1978,chowdhury2000}, and biological invasion \parencite{petrovskii2005}.
% can i find modern examples? all examples i cited are from the last century, with two exceptions from 2000 and 2005

The stochastic Burgers model determines the evolution of the velocity field $u(x,t)$ which is forced by a random external force $\phi(x,t)$:
\begin{equation}
u_t + u u_x = \nu u_{xx} + \phi \ , \  \label{eq:nu-burgers}
\end{equation}
As usual, $\nu$ is the kinematic viscosity.
And similarly to Navier-Stokes, observables such as velocity gradients, $\xi = \partial_x u$, and velocity differences, $\delta u_{\ell}(x) = u(x+\ell) - u(x)$, display intermittent fluctuations in turbulent steady state solutions.
%\cite{polyakov1995, gurarie1996, khanin1997, balkovsky1997, bec2001, chernykh2001, moriconi2009, grafke2015relevance, friedrich2018}.

Starting from a smooth initial condition, this model leads to the formation of smooth ramps and sudden negative shocks, which can be seen in Fig.~\ref{fig:burgers-shock}a. The probability distribution of the velocity gradient, seen in Fig.~\ref{fig:burgers-shock}b, can also be understood from the picture of ramps and shocks: Positive velocity gradients are small and with little fluctuations, while negative gradients are extremely intermittent, with the largest negative fluctuations forming visible shocks.

\begin{figure}
    \centering
    \includegraphics[width=\textwidth]{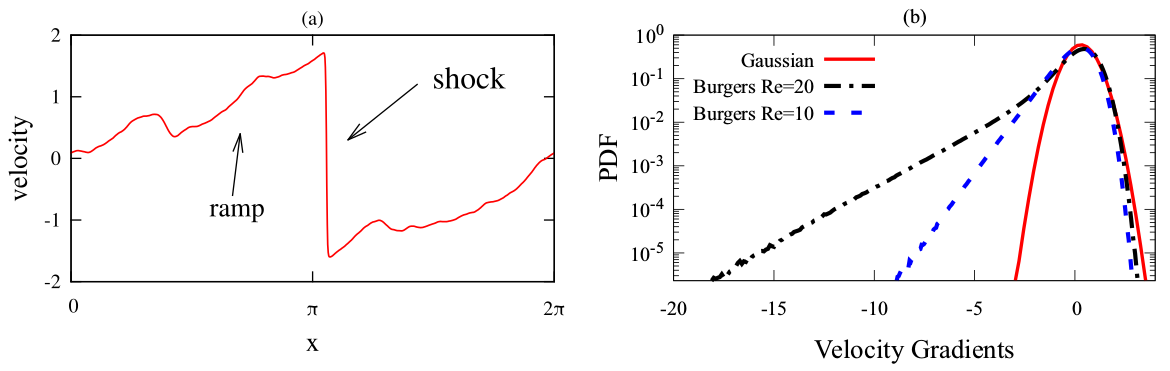}
    \caption
    [Example of shock in Burgers dynamics and its associated velocity gradient PDF]
    {Burgers hydrodynamics is marked by the formation of sudden negative shocks interleaved by smooth positive ramps. A ramp and a shock are identified in (a). This dynamics leads to the PDF observed in (b), where positive gradients decay fast, while negative gradients display intense fluctuations, which grow with higher Reynolds numbers. A comparison between these PDFs and a Gaussian distribution is also shown. This figure was extracted from \textcite{margazoglou2019}.}
    \label{fig:burgers-shock}
\end{figure}

The instanton approach was applied to the Burgers model in \textcite{gurarie1996}, where it was shown that the velocity difference PDFs followed
\begin{equation} \label{eq:pdf_general}
    \rho_g(\delta u) = \exp [ - S(\delta u) ] \ ,
\end{equation}
and that the asymptotic form of the action $S(\xi)$ was
\begin{equation}
    S(\delta u) \sim |\delta u|^{3} \ .
\end{equation}
This result confirms that fluctuations in the positive tails are effectively irrelevant, since their distribution is sub-Gaussian. As a consequence, the instanton PDF is essentially exact.
The same result had been obtained before by different methods in \textcite{polyakov1995}.
% what about feigelman1980, what did he do?
Furthermore, the exponent $3$ is equally observed for positive velocity gradients, a result demonstrated in \textcite{gotoh1998}.
%Vicosity can be neglected in the right tail, which made an analytical result possible and it was obtained that the PDF, if written as

Negative fluctuations, though, due to their intermittent fluctuations, are still subject of research in this field.
%\parencite{khanin1997, balkovsky1997, bec2001, chernykh2001, moriconi2009, grafke2015relevance}
The first result on negative gradients with the instanton method was obtained in \textcite{balkovsky1997}, where it was demonstrated that the left tail of the velocity gradient probability distribution could be written as Eq.~\eqref{eq:pdf_general}, but instead depending on $\xi$. The asymptotic behavior of the PDF tails is given by
\begin{equation} \label{eq:left-asymp}
    S(\xi) \sim |\xi|^\frac{3}{2} \ ,
\end{equation}
which corresponds to a fat tailed distribution.
The hypotheses made in these calculations and the general result were verified in \textcite{chernykh2001}, although, as \textcite{grafke2015relevance} point out, the asymptotic form of $S(\xi)$ is still beyond the reach of direct numerical simulations. Still, the authors observe an agreement between the velocity gradient PDF tails and those provided by the instanton configurations. This agreement, however, relies on an arbitrary multiplication of the random force strength parameter by a Reynolds number dependent adjustment factor. A detailed analytical investigation on why such an empirical \enquote{noise renormalization procedure} works is the central aim of the work developed in \textcite{apolinario2019onset}, work on which this chapter is based.

%The discussion in this chapter applies field theoretical techniques previously described: The MSRJD formalism and the cumulant expansion of fluctuations around the instanton fields. It is natural to expect that corrections to the instanton evaluations of PDF tails have to be supplemented by subdominant fluctuation contributions, but numerical studies of such fluctuations have only been performed very recently through hybrid Monte Carlo techniques \parencite{margazoglou2019, ebener2019}.

\section{Field Theoretical Setup}

A dimensionless version of Burgers equation is obtained by equivalent methods to those of Section~\ref{sec:similarity}, by measuring length in units of $L$, the forcing scale, time in units of $L^2/\nu$ and velocity in units of $\nu/L$. As a result:
\begin{equation}
u_t + u u_x - u_{xx} = g \phi \ , \  \label{burgers_eq}
\end{equation}
where $\phi = \phi(x,t)$ is a zero-mean Gaussian random field used to model large-scale forcing, with correlator
\begin{equation}
\langle \phi(x,t) \phi (x',t') \rangle = \chi(x-x') \delta (t -t') \ . \  \label{phi-c}
\end{equation}
The correlation function $\chi$ is peaked at wavenumber $k=0$ and broadened in Fourier space within a region of size $\Delta k \sim 1$.
In other words, $L \equiv 1$ ($ \sim \Delta k^{-1}$) is taken to be the random force correlation length, defined as the largest relevant length scale in the flow. While most of the considerations in this section are general, a specific form of the correlation function is adopted, as it has been addressed in former works \parencite{balkovsky1997, gotoh1998, grafke2015relevance},
\begin{equation}
\chi(x) = (1-x^2) \exp \left (  - \frac{x^2}{2} \right ) = - \partial_ x^2
\exp \left ( - \frac{x^2}{2} \right ) \ . \
\end{equation}
% cite work which studies universality
This is a case study of particular interest, due to its simple formulation and good analytical properties.

Note that the intensity of forcing is given by the noise strength parameter $g$. This is the only dimensionless parameter in the problem, and can be connected to the Reynolds number through the phenomenology of Section~\ref{sec:k41}. The amplitude $\chi_0 = \chi(0)$ of the external force correlation function is in direct correspondence with the kinetic energy dissipation rate, $\varepsilon$. Furthermore from the K41 theory, $\varepsilon \propto U^3/L$, where $U$ is a typical velocity. From these relations, the following dimensionally correct relation is obtained:
\begin{equation}
    U = (\chi_0 L)^{1/3} \ .
\end{equation}
This equation is used to write the Reynolds number, $\mathrm{Re} = U L / \nu$ in terms of the dimensional parameters of the problem. Finally, the force strength can also be written in terms of the same parameters, as $g = \chi_0^{1/2} L^2 \nu^{-3/2}$, from which it is obtained that
\begin{equation}
    g = \mathrm{Re}^{3/2} \ .
\end{equation}

% I don't fully understand this argument of Luca, but it is based on microscale fluctuations, instead of large scale K41
%Furthermore, once the viscosity $\nu$ and the integral length scale $L$ are normalized to unit in Eq. (\ref{burgers_eq}), by defining the Reynolds number as $Re = L^{4/3}\sqrt[3]{\langle (\partial_x u)^2 \rangle / \nu^2}$ we get $Re = \sqrt[3]{g^2/2}$.

With the Burgers model in mind, velocity gradients are going to be computed in the MSRJD formalism as path-integrations over the velocity field
$u(x,t)$ and its conjugate auxiliary field $p(x,t)$, combined with an ordinary integration over a Lagrange multiplier
variable $\lambda$ as
\begin{equation}
\rho_g(\xi)  = \langle \delta(u_x|_0 - \xi) \rangle = {\cal{N}}^{-1} \int Dp Du \int_{- \infty}^{ \infty} d \lambda \exp \{ -S[u,p,\lambda ; g] \} \ , \
\label{vgPDF}
\end{equation}
where $\cal{N}$ is an unimportant normalization constant (to be suppressed from now on, in order to simplify notation), $u_x|_0$ is the velocity gradient taken at $(x,t) = (0,0)$, and $S[u,p,\lambda ; g]$ is the MSRJD action:
\begin{equation} \begin{split}
&S[u,p,\lambda ; g] = \frac{g^2}{2} \int d t d x ~ p (\chi * p ) + \\
&+i \int d t d x ~ p(u_t+u u_x - u_{xx}) - i\lambda (u_x |_0 - \xi ) \ , \ \label{msr_action}
\end{split} \end{equation}
with $\chi * p \equiv \int dx'\chi(x-x') p(x',t)$, which is a convolution in space, equivalent to the one defined in Sec.~\ref{chap:stoc}.

% here we use the saddle-point for the tails of PDFs while in the previous work we used it for the core of PDFs, do I need to dwell on the differences?
The saddle-point method is adopted to find the asymptotic form of velocity gradient PDF tails. The method provides meaningful answer in cases where the PDF tails decay faster than a simple exponential, $\exp(-c|\xi|)$ for any arbitrary $c>0$, as observed in numerical studies of Burgers turbulence \textcite{gotoh1998}.
The saddle-point configurations $u^c$, $p^c$, and $\lambda^c$ that extremize the MSRJD action are named instantons and they were first obtained for the Burgers mode in \textcite{gurarie1996, falkovich1996}.
As pointed out in the previous chapters, they are the solutions of the Euler-Lagrange equations, which in the present case are:
\begin{equation}
 \left. \frac{\delta S}{\delta u}\right\vert_{u^c,p^c,\lambda^c} \!\!\!\!\!\!\! =0
 \ \ \ , \ \ \
 \left.\frac{\delta S}{\delta p}\right\vert_{u^c,p^c, \lambda^c} \!\!\!\!\!\!\! =0
  \ \ \mbox{and} \ \
 \left.\frac{\partial S}{\partial \lambda }\right\vert_{u^c,p^c, \lambda^c} \!\!\!\!\!\!\! =0   \ . \ \label{sp_eqs}
\end{equation}
It is convenient to rescale $p(x,t)$ and $\lambda$ as
\begin{equation}
    p \rightarrow \frac{p}{g^2} \ \ \mbox{and} \ \
    \lambda \rightarrow \frac{\lambda}{g^2} \mbox{,} \label{rescaling}
\end{equation}
so that the MSRJD action in (\ref{vgPDF}) is rescaled as
\begin{equation}
S[u,p,\lambda;g] \rightarrow \frac{1}{g^2} S[u, p, \lambda; 1]  \label{s_to_s}
\end{equation}
and the Euler-Lagrange equations stated in (\ref{sp_eqs}) become
\begin{subequations}
\begin{align}
 &u_t^c + u^c u_x^c - u^c_{xx} = i\chi * p^c \ , \ \label{sp1} \\
 &p^c_t+ u^c p^c_x + p^c_{xx} = \lambda^c \delta(t) \delta'(x) \label{sp2} \ , \ \\
 &\xi = u^c_x|_0 \label{sp3} \ . \
\end{align}
\end{subequations}
It is relevant to notice that the noise strength $g$ has been factored out of the action in Eq.~\eqref{s_to_s}. Furthermore, the saddle-point solutions $p^c(x,t)$ and $\lambda^c$ are purely imaginary, since it is natural to expect purely real velocity instantons $u^c(x,t)$.

% change this paragraph
When dealing with instantons, one needs, in general, to worry about the existence of degenerate families of saddle-point solutions, associated with symmetries of the action, like translation or gauge invariance. The Fadeev-Popov method is the usual procedure to eliminate such redundant solutions \textcite{coleman1988,moriconi2009}. However, in the formalism addressed here, the degeneracy issue is bypassed through the explicit assignment of the spacetime point $(x,t)=(0,0)$ as the symmetry center around which the instantons evolve, making Eqs. (\ref{sp1}) and (\ref{sp2}) not translationally invariant.

The sign of the dissipation term in Eqs. \eqref{sp1} and \eqref{sp2} works as an effective time-reversal: The equation for $u^c$ has to be solved forward in time, since it has the same sign as the standard Burgers equation, but the equation for $p^c$ has a reversed sign in the dissipation term, making it similar to a backward in time Burgers equation. The time domain of these solutions is
$- \infty < t \leq 0$, with $u^c(x, - \infty) = p^c (x, - \infty) = 0$, and the additional boundary conditions given by Eq. (\ref{sp3}) and $p^c(x,0^+) = 0$. The last of these is equivalent to $p^c(x,0^-) = -\lambda \delta'(x)$, or, in Fourier space, to $\tilde p^c(k,0^-) = - i \lambda k$ \parencite{gurarie1996}. \label{burgers-boundary}

Making use of this time reversal analogy, a fruitful self-consistent numerical strategy for the above saddle-point equations was proposed in \textcite{chernykh2001}. This is now usually called the Chernykh-Stepanov method. One first neglects the boundary condition \eqref{sp3}, and trades it for an arbitrary fixed value of $\lambda$. The equations, then, are solved through an iterative method which, if convergent, leads to arbitrarily precise solutions.

Beginning with $u_1(x,t) \equiv 0$, Eq.~\eqref{sp2} is solved backwards in time for $p_1(x,t)$. This field, $p_1(x,t)$, is then used in Eq.~\eqref{sp1} to obtain $u_2(x,t)$. This procedure can be carried on indefinitely, and the solutions $u^c(x,t)$ and $p^c(x,t)$ can be found within a previously defined precision metric. The velocity gradient $\xi$ is defined, \textit{a posteriori}, from Eq. (\ref{sp3}), as a derivative of the last iterated velocity field. It has been observed numerically that $\xi$ is a monotonically increasing function of $\lambda$, hence Eq.~\eqref{sp3} has a unique solution in this situation. It is imperative to note, though, that the Chernykh-Stepanov method may require further numerical devices to attain convergence for $|\lambda|$ large enough.

Once $u^c(x,t)$, $p^c(x,t)$ and $\lambda^c$ are available, an expansion around the instantons is performed, introducing the fluctuations $u(x,t)$, $p(x,t)$ and $\lambda$, along the lines of Eq.~\eqref{eq:perturb-instanton}:
\begin{subequations}
\begin{align}
&u(x,t) \rightarrow u^c(x,t) + u(x,t) \ , \ \label{subs_u} \\
&p(x,t) \rightarrow p^c(x,t) + p(x,t) \ , \ \label{subs_p} \\
&\lambda \rightarrow \lambda^c + \lambda \ . \ \label{subs_lambda}
\end{align}
\end{subequations}
This substitution rewrites the MSRJD action, accordingly, as
\begin{equation}
\begin{split}
&S[u,p, \lambda; 1]  \rightarrow S[u^c + u,p^c + p, \lambda^c + \lambda; 1] \\
&\equiv S_c[u^c,p^c] + S_0[u,p] + S_1[u^c,u,p^c,\lambda] \ , \
\end{split}
\end{equation}
where $S_c$, $S_0$, and $S_1$ are, respectively: the saddle-point action; the sum of all quadratic forms in the $u$ and $p$ fields that do not depend on $p^c$ and $u^c$; and finally, $S_1$ is the contribution that collects all terms that have not been included in $S_c$ or $S_0$. Explicitly, these terms are
\begin{equation}
\begin{split}
S_c  &\equiv \frac{1}{2} \int d t d x  ~ p^c (\chi * p^c ) + i \int d t d x ~ p^c(u^c_t+u^c u^c_x - u^c_{xx}) \\
&= ({\hbox{using Eq. (\ref{sp1})}}) = - \frac{1}{2} \int d t d x  \ p^c (\chi * p^c)  \ ; \label{sp_action} \\
S_0  &=  \int d t d x  \left \{ \frac{1}{2}\ p (\chi * p)  + i \ p (u_t - u_{xx}) \right \} \ ; \\
\end{split}
\end{equation}
and, up to second order in the fluctuating fields:
\begin{equation}
S_1 =  i \int d t d x \left\{ p^c u u_x - p_x u^c u  \right\}
- i \lambda u_x|_0 \ . \
\end{equation}
It is clear that $S_c$ is a functional of the instanton fields, which on their turn depend on the velocity gradient $\xi \equiv u^c_x |_0 $. Hence we can write, more synthetically, that $S_c = S_c(\xi)$.

Now, taking into account the instanton solutions, we can reformulate the velocity gradient PDF, Eq. (\ref{vgPDF}), as
\begin{equation}
\begin{split}
\rho_g(\xi) &= \exp \left ( - \frac{1}{g^2} S_c(\xi) \right ) \int Dp Du
\int_{- \infty}^\infty d \lambda
\exp \left [ - \frac{1}{g^2} ( S_0 + S_1 ) \right ] \\
& \propto \exp \left ( -\frac{1}{g^2}
S_c(\xi) \right )
\int_{- \infty}^\infty d \lambda \left \langle \exp \left ( - \frac{1}{g^2}
S_1 \right )
\right \rangle_0 \ , \  \label{vgpdf2}
\end{split}
\end{equation}
where $\langle \ \bm\cdot \ \rangle_0$ stands for an expectation value computed in the linear stochastic model defined by the MSRJD action $S_0$.

A perturbative development of the above expression is possible in cases where the fluctuation-dependent contributions are small relative to the leading saddle-point results, that is,
\begin{equation}
S_c(\xi) \gg g^2 \left | \ln \left [ \int_{- \infty}^\infty d \lambda \left \langle \exp \left ( - \frac{1}{g^2}
S_1 \right )
\right \rangle_0 \right ] \right | \ .
\end{equation}
Therefore, the cumulant expansion is a natural method for evaluating Eq.~\eqref{vgpdf2} in a perturbative manner. Considering contributions up to second order in the instanton fields and $\lambda$, we obtain
\begin{equation}
\left \langle \exp \left ( - \frac{1}{g^2} S_1 \right )
\right \rangle_0 = \exp \left \{ - \frac{1}{g^2} \langle S_1 \rangle_0  +   \frac{1}{2 g^4}
\left [ \langle (S_1)^2 \rangle_0 - \langle S_1  \rangle_0^2 \right ]  \right \} \ ,
\label{c-expansion}
\end{equation}
which is used to establish the \textit{effective MSRJD action},
\begin{equation}
\Gamma \equiv S_c + \langle S_1 \rangle_0  - \frac{1}{2 g^2}
\left [ \langle (S_1)^2 \rangle_0 - \langle S_1  \rangle_0^2 \right ] \ . \
\end{equation}

The basic building blocks needed to evaluate (\ref{c-expansion}) are the correlation functions
\begin{equation} \begin{split}
G_{pu}(x,x',t,t') &\equiv \langle p(x,t) u(x',t') \rangle_0 \\
&= -\frac{i g^2}{2 \sqrt{\pi (t'-t)} }
\exp \left [ - \frac{(x-x')^2}{4 (t'-t)} \right ] \Theta(t'-t) \label{Gpu}
\end{split} \end{equation}
and
\begin{equation} \begin{split}
G_{uu}(x,x',t,t') &\equiv \langle u(x,t) u(x',t') \rangle_0 \\
&= \frac{g^2}{2 \sqrt{1+2|t-t'|}} \exp \left [ - \frac{(x-x')^2}{2(1+ 2|t-t'|)} \right ]
\ . \ \label{Guu}
\end{split} \end{equation}
The first propagator connects the velocity field and the auxiliary field, while the second connects two velocity fields at different positions and instants. Thus, these propagators are equivalent to those described for the RFD model, and illustrated in Fig.~\ref{fig:free-propagator}.

From Eqs. (\ref{Gpu}) and (\ref{Guu}), it can be shown that
$\langle S_1 \rangle_0 = 0$ and
\begin{equation}
\langle (S_1)^2 \rangle_0 = I_1 [p^c,u^c] + I_2 [p^c] - \lambda^2 \langle (u_x|_0)^2 \rangle_0 \ , \ \label{S1squared}
\end{equation}
with
\begin{subequations}
\begin{align}
I_1 [p^c,u^c] &\equiv \int_{t,t'<0} dt dt'dx dx'~ p^c(x,t) u^c(x',t') H_1(x,x',t,t')   \ , \label{I1} \\
I_2 [p^c] &\equiv \int_{t,t'<0} dt dt'dx dx'~ p^c(x,t) p^c(x',t') H_2(x,x',t,t') \ , \label{I2}
\end{align}
\end{subequations}
where
\begin{equation} \begin{split}
&H_1(x,x',t,t') = -2 \partial_x [ G_{uu}(x,x',t,t') \partial_x G_{pu}(x',x,t',t) ] \ , \   \\
&H_2(x,x',t,t') = \frac{1}{2} \partial_x^2 [G_{uu}(x,x',t,t')]^2 \ . \
\end{split} \end{equation}
Note that $I_1[p^c,u^c]$ and $I_2[p^c]$, both of $O (g^4)$, are the one-loop contributions which renormalize, respectively, the heat and noise kernels of the original stochastic Burgers equation, Eq.~(\ref{burgers_eq}). These corrections are also equivalent to those for the RFD model, represented in Fig.~\ref{fig:oneloop}, in the previous chapter.

The overall effect of perturbative contributions can always be conventionally accounted by a redefinition of the noise strength parameter $g$ in the expression for the velocity gradient PDF, $\rho_g( \xi) \propto \exp ( - S_c(\xi) /g^2 )$. In the present context, the $g$-independent coefficient
\begin{equation}
    c(\xi) \equiv - \frac{I_1 [p^c,u^c] + I_2 [p^c]}{2g^4 S_c } \ \label{c-xi}
\end{equation}
is enough.
Using (\ref{vgpdf2}), (\ref{c-expansion}), and integrating over $\lambda$ in the Gaussian approximation given by (\ref{S1squared}), a renormalized expression for the velocity gradient PDF is obtained,
\begin{equation}
\rho_g(\xi) \propto \exp \left ( -  \frac{S_c (\xi) }{g_R^2} \right ) \ , \ \label{eff_vgpdf}
\end{equation}
in which
\begin{equation}
g_R \equiv \frac{g}{ \sqrt{1 + c(\xi)g^2}} \label{eff_noise}
\end{equation}
defines an {\textit{effective noise strength}} parameter, which is, in principle, a velocity-gradient dependent quantity that encodes the effects of fluctuations around the instantons, up to the lowest non-trivial order in the cumulant perturbative expansion.

\section{The Onset of Intermittency}

% text was not changed much, is this a problem?
Eq. (\ref{eff_noise}) suggests, in fact, a simple criterion for the consistency of the perturbative analysis.
The cumulant expansion is meaningful, up to second order, if $|c(\xi)|g^2$ is reasonably smaller than unity. It follows, immediately, that for any fixed velocity gradient $\xi$, the cumulant expansion will break down for $g$ large enough. Similarly, since  (as will be seen) $c(\xi)$ is a positive monotonically increasing function of $|\xi|$, the cumulant expansion framework becomes inadequate for large enough $|\xi|$ at any fixed $g$.

Consequently, the consideration of strong coupling regimes ($g \gg 1$, equivalent to high Reynolds numbers) and/or asymptotically large velocity gradient fluctuations is precluded from the cumulant expansion approach. The perturbative analysis, nevertheless, is actually useful to model the shape of velocity gradient PDF left tails in the non-Gaussian region, where $|\xi| > g$, for not very large $g$.
It is expected that, as the noise strength grows and incipient turbulent fluctuations associated to flow instabilities come into play, the onset of non-Gaussian behavior is captured by dominant instanton contributions \enquote{dressed} by cumulant corrections.

It is important, before proceeding, to comment on the challenging technical difficulties associated to the evaluations of $S_c(\xi)$,  $I_1 [p^c,u^c]$, and $I_2 [p^c]$, given, respectively, by Eqs. (\ref{sp_action}), (\ref{I1}) and (\ref{I2}), the essential ingredients in the derivation of the PDF tails. It turns out that the associated integrations based on the numerical instantons are extremely demanding in terms of computational cost. The numerical convergence of integrals is very slow as the system size increases and the grid resolution gets finer.
Nevertheless, a pragmatic scheme for the computation of the saddle-point action $S_c(\xi)$ is available from the numerical work in \textcite{grafke2015relevance}, where it is pointed out that for large negative velocity gradients and at a given noise strength $g$, $S_c(\xi)$ can be retrieved with good accuracy from the velocity gradient PDF $\rho_g(\xi)$ as
\begin{equation}
S_c(\xi) \simeq - g^2\kappa(g) \ln \left [ \frac{\rho_g(\xi)}{\rho_g(0)} \right ] \ , \ \label{s_Sc}
\end{equation}
where $\kappa(g)$ is a $g-$dependent empirical correction factor.

It follows, now, under the light of Eq. (\ref{eff_vgpdf}), that $\kappa(g)$ is nothing more than $(g_R/g)^2$, and, therefore, it should depend on $\xi$ as well.  From such a perspective, one finds that the relevance of Eq. (\ref{s_Sc}) is fortuitously based on the fact that $c(\xi)$, as defined in (\ref{c-xi}), is in general a slowly varying function of $\xi$. Eq.~\eqref{s_Sc} is then used as a practical way to obtain a reasonable evaluation of the saddle-point action.
To distinguish between the exact saddle-point action and the one approximated by \eqref{s_Sc}, the latter is denoted the \textit{surrogate saddle-point action} and represented by $S_{sc}(\xi)$.

Regarding the evaluation of the perturbative functionals $I_1 [p^c,u^c]$ and $I_2 [p^c]$: The full numerical approach is very slowly convergent, if based on the Chernykh-Stepanov numerical solutions of Eqs. (\ref{sp1}) and (\ref{sp2}). Instead, approximate analytical expressions for $u^c(x,t)$ and $p^c(x,t)$ lead to analytical expressions within the same order in the perturbative expansion. Below, this analytical approximation is discussed, along with its use in the determination focus of $S_{sc}(\xi)$, $I_1 [p^c,u^c]$, and $I_2 [p^c]$.

\subsection{Analytical Approximations for the Instanton Fields}

In the asymptotic limit of small velocity gradients, instantons can be well approximated as the solutions of Eqs. (\ref{sp1}) and (\ref{sp2}) simplified by the suppression of nonlinear terms. Working in Fourier space, where
\begin{align}
& \tilde p(k,t) \equiv \int dx \ p(x,t) \exp(-ikx) \ , \ \label{fp} \\
& \tilde u(k,t) \equiv \int dx \ u(x,t) \exp(-ikx) \ , \ \label{fu}
\end{align}
it is straightforward to find, under the linear approximation, that
\begin{align}
& \tilde u^c(k,t) = \lambda^c \sqrt{\frac{\pi}{2}} k \exp \left [ -k^2\left (|t|+\frac{1}{2} \right ) \right ]  \label{fuc}  \ , \ \\
& \tilde p^c(k,t) = -i  \lambda^c k \exp( k^2 t) \Theta(-t) \equiv \tilde p^{(0)}(k,t) \ . \ \label{fpc}
\end{align}
Taking $\lambda^c \equiv -i \lambda$, the velocity gradient at $(x,t)=(0,0)$ is obtained as $\xi = \lambda /2$. From now on it is assumed, thus, that $\lambda$ is a negative real number, since the velocity gradients in question are also negative.

Note that the exact solution for the instanton response field can be formally written as
\begin{equation}
p^c(x,t) = p^{(0)}(x,t) + \delta p^c(x,t) \ , \ \label{pc_full}
\end{equation}
where $\delta p^c(x,t)$ satisfies the boundary conditions
\begin{equation}
\delta p^c (x,- \infty) = \delta p^c(x,0^-) = 0 \ , \ \label{vbc}
\end{equation}
since $p^{(0)}(x,t)$ saturates the boundary conditions for $p^c(x,t)$, stated on p.~\pageref{burgers-boundary}.

The vanishing boundary conditions (\ref{vbc}) suggest that $\delta p^c(x,t)$ can be taken as a perturbation field. Accordingly, the instanton velocity field can be expanded as a functional Taylor series,
\begin{equation}
u^c(x,t) = u^{(0)}(x,t) + \sum_{n=1}^\infty \int \left [ \prod_{i=1}^n dx'_i dt'_i \delta p^c(x_i',t_i') \right ] F_n(x,t, \{x',t'\}_n )  \ , \ \label{uc_full}
\end{equation}
where
\begin{equation}
\{x',t'\}_n \equiv \{ x_1',x_2',...,x_n', t_1', t_2',...,t_n' \}
\end{equation}
and the many variable kernel $F_n(x,t, \{x',t'\}_n )$ is a functional of $p^{(0)}(x,t)$. Note that $u^{(0)}(x,t)$ is independent
(in the functional sense) of $\delta p^c(x,t)$. An infinite hierarchy of equations is obtained for $F_n(x,t, \{x',t'\}_n )$, when (\ref{pc_full}) and (\ref{uc_full})
are substituted into the saddle-point Eqs. (\ref{sp1}) and (\ref{sp2}). In general, $\partial_t F_n(x,t, \{x',t'\}_n )$ will depend in a nonlinear way on the set of $F_m$ and its derivatives, with $m \leq n$.

Interestingly, a closed analytical solution for $u^{(0)}(x,t)$ can be obtained, as
\begin{equation}
\tilde u^{(0)}(k,t) = \lambda^c \tilde F_0^{(1)} (k,t) + (\lambda^c)^2 \tilde F_0^{(2)} (k,t) \ , \ \label{u0}
\end{equation}
where $\lambda^c \tilde F_0^{(1)}(k,t)$ is exactly the same as (\ref{fuc}),
and
\begin{equation} \begin{split}
\tilde F_0^{(2)} (k,t) &= \frac{i k}{32} \sqrt{\frac{3 \pi k^2}{4}} \exp \left [ k^2 \left ( |t| + \frac{1}{2} \right ) \right ] \Gamma \left (-\frac{1}{2}, \frac{3 k^2}{2} \left ( |t|+ \frac{1}{2} \right ) \right )
\\ &-\frac{i k^3}{32} \sqrt{\frac{\pi}{3 k^2}} \exp \left [ k^2 \left ( |t| + \frac{1}{2} \right ) \right ] \Gamma \left (\frac{1}{2}, \frac{3 k^2}{2} \left ( |t|+ \frac{1}{2} \right ) \right ) \ ,
\label{F02}
\end{split} \end{equation}
a result expressed in terms of the incomplete Gamma function, $\Gamma(x,y) = \int_y^\infty t^{x-1} \exp(-t) dt$.
From Eq. (\ref{u0}) (using, again, $\lambda^c \equiv -i \lambda$), the velocity gradient at $(x,t)=(0,0)$ can be
computed, in the approximation where $u^c(x,t) = u^{(0)}(x,t)$, as
\begin{equation} \begin{split}
\xi \equiv \partial_x u^c(x,0)|_{x=0} &= \frac{i}{2 \pi}  \int dk \ k \ \tilde u^{(0)} (k,0) \\
&=\frac{\lambda}{2} + \frac{3 -2\sqrt{3}}{24} \lambda^2 \ .
\end{split} \end{equation}
which, upon inversion, leads to
\begin{equation}
\lambda = 2 \frac{\sqrt{3}  - \sqrt{3 + 2(3 - 2 \sqrt{3}) \xi }}{2-\sqrt{3}} \ . \ \label{lambda-xi}
\end{equation}
In order to see how accurate is Eq. (\ref{lambda-xi}), the numerical instantons from Eqs. (\ref{sp1}-\ref{sp3}), along the lines of the Chernykh-Stepanov procedure, were computed. The solution was implemented through the pseudo-spectral method for a system with size $200$ (recall that $L=1$), and $2^{10}$ Fourier modes. The time evolution is realized through a second order Adams-Bashfort scheme with time step $\delta t = 10/2048 \simeq 5 \times 10^{-3}$ and total integration time $T = 200$. Since instantons evolve within the typical integral time scale $T_0 \sim 1/|\lambda|$, we have investigated the range $0.5 \leq |\lambda| \leq 20.0$, such that $ \delta t \ll T_0 \ll T$.

In Fig. \ref{lambda_fig}, relation (\ref{lambda-xi}) is compared to a result obtained from the numerical instantons.  Following the Chernykh-Stepanov method, $\lambda$ is defined as an external parameter, and the velocity gradient $\xi$ is calculated as derivative of the numerical velocity field obtained after convergence. It can be seen that the predicted relation is reasonably accurate. What is more, the relation between $\lambda$ and $\xi$ is biunivocal in this range, which validates the last step of the Chernykh-Stepanov method, in which Eq.~\eqref{sp3} is solved \textit{a posteriori}.

\begin{figure}[ht]
\centering
\includegraphics[width=.7\textwidth]{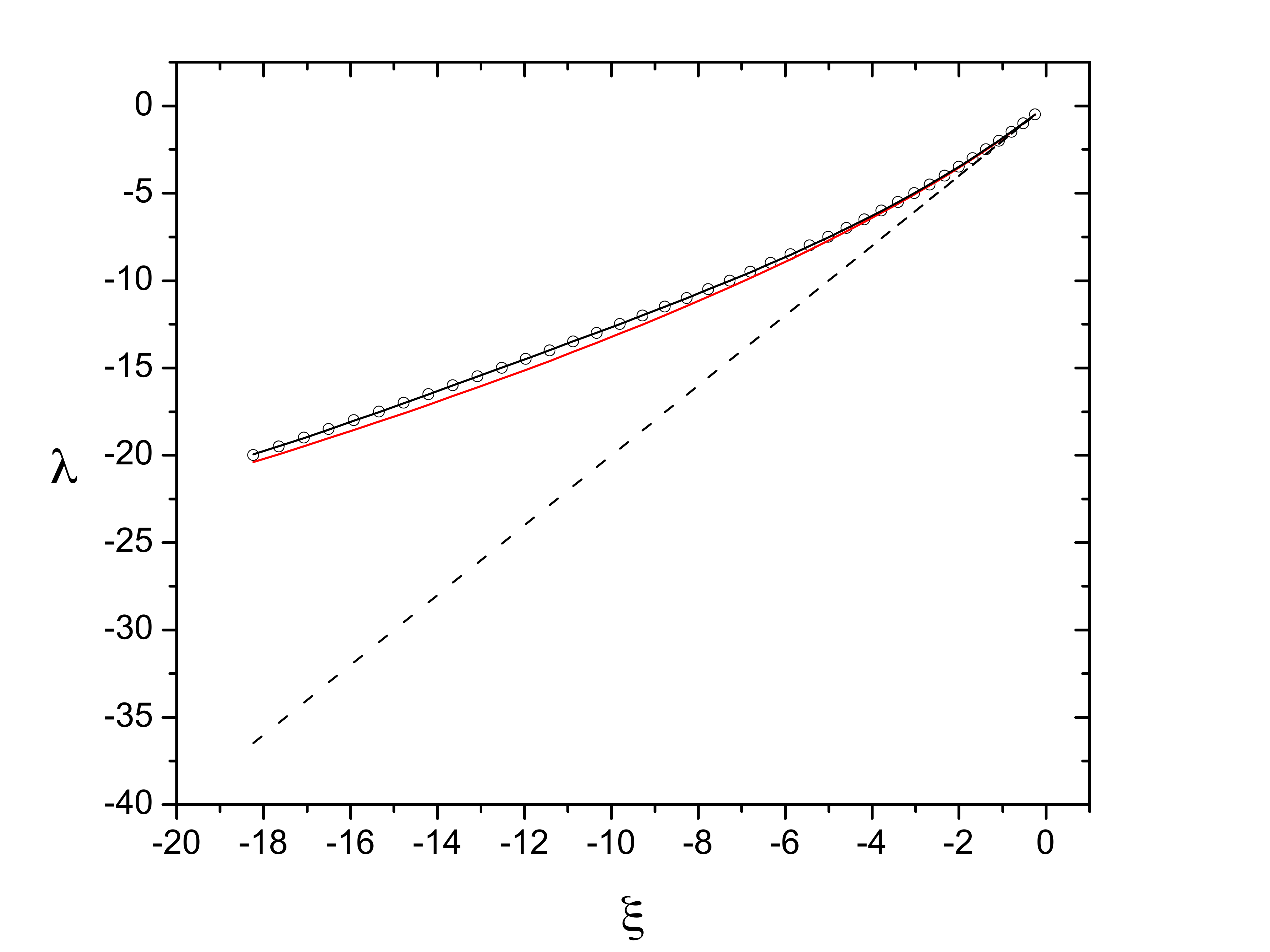}
\caption
[The Lagrange multiplier $\lambda$ is given as a function of the velocity gradient $\xi$]
{The Lagrange multiplier $\lambda$ is given as a function of the velocity gradient $\xi$. Open circles represent values obtained from the numerical solutions of Eqs. (\ref{sp1}-\ref{sp3}) (the black solid line is just a polynomial interpolation of the numerical data); red solid line: approximated instanton relation, Eq. (\ref{lambda-xi}); dashed line: $\lambda = 2 \xi$, which holds for asymptotically small velocity gradients.
}
\label{lambda_fig}
\end{figure}

\subsection{The Surrogate Saddle-Point Action}

The approximate instantons given by Eqs.~(\ref{fpc}) and (\ref{u0}) are useful for the evaluation of $I_1 [p^c,u^c]$ and $I_2 [p^c]$ up to lowest non-trivial order in the functional perturbative expansion around $p^{(0)}(x,t)$. Nevertheless, they are unable to provide the observed dependence of the MSRJD action $S_c(\xi)$ with the velocity gradient $\xi$. In fact, $p^{(0)}(x,t)$ is proportional to $\lambda$, leading, from (\ref{sp_action}), to $S_c(\xi) = \lambda^2/4$, a result that is not supported by the numerical PDFs \textcite{grafke2015relevance}.

\begin{figure}[ht]
    \centering
    \includegraphics[width=.7\textwidth]{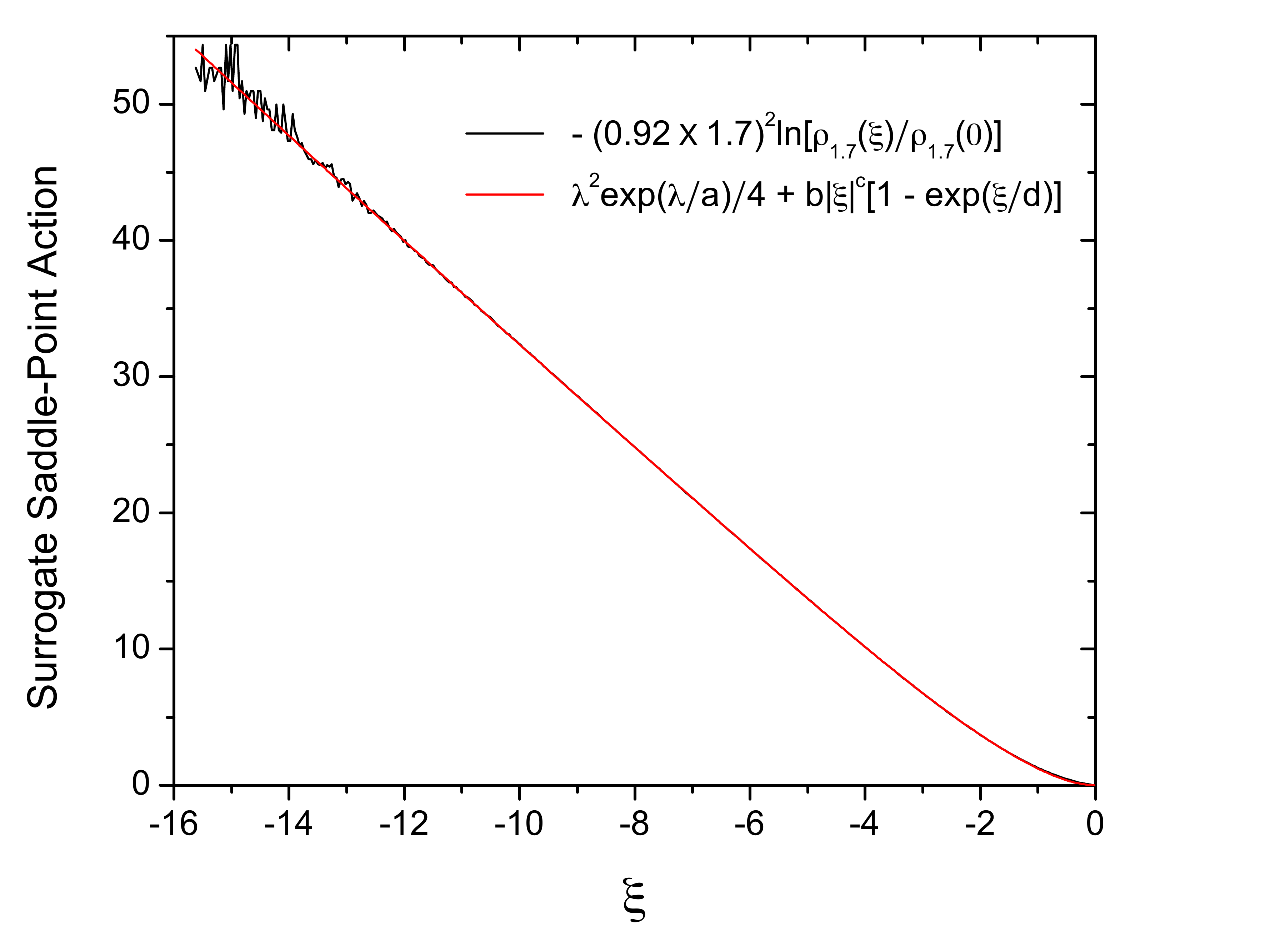}
    \caption
    [Comparison between the surrogate saddle-point action and a fitting function]
    {Comparison between the surrogate saddle-point action, as prescribed by \textcite{grafke2015relevance} for the case of noise strength parameter $g=1.7$ (black solid line), and a four-parameter fitting function (red line) which provides distinct power law asymptotics for domains of small and large velocity gradients.}
    \label{surrogate_action}
\end{figure}

Taking advantage of the results reported in \textcite{grafke2015relevance} for the case of noise strength parameter $g=1.7$, a flow regime close to the onset of intermittency, we set $\kappa(g) = (0.92)^2$ and write down the surrogate saddle-point action (\ref{s_Sc}) as
\begin{equation}
S_{sc}(\xi) \simeq - (0.92 \times 1.7)^2 \ln \left [ \frac{\rho_{1.7}(\xi)}{\rho_{1.7}(0)} \right ] \ . \ \label{s_sc2}
\end{equation}
To obtain the surrogate action and a set of velocity gradient PDFs for several values of $g$, direct numerical simulations were performed. They were used to check (\ref{vgpdf2}) in the approximation given by (\ref{c-expansion}).

The stochastic Burgers equation is solved with a fully dealiased pseudo-spectral method in $N=2048$ collocation points by employing a $2^\text{nd}$ order predictor-corrector time evolution scheme \parencite{canuto2012spectral,kloeden2013}. As in the numerical solution of the instanton fields, the domain size is taken to be $200L$. Velocity gradients are saved every 30 time steps after a suitable transient time, after which the total simulation time is $T \approx 1.2\times 10^7$.

A useful and accurate fitting of the surrogate saddle-point action (\ref{s_sc2}) can be defined as
\begin{equation} \label{interpol}
S_{sc}(\xi) = \frac{\lambda^2}{4} \exp \left ( \frac{\lambda}{a} \right ) + b|\xi|^c \left [1 - \exp \left ( \frac{\xi}{d} \right  ) \right ] \ , \
\end{equation}
where $\lambda$ is given by (\ref{lambda-xi}), and $a=2.046$, $b=2.407$, $c=1.132$, and $d=2.195$ are optimal fitting parameters. This result is shown in Fig. \ref{surrogate_action}.

The interpolation (\ref{interpol}) is actually consistent with the behavior of the local stretching exponent for the saddle-point action. This exponent is defined as
\begin{equation}
    \theta(\xi) = \frac{d \ln S(\xi)}{d \ln |\xi|} \ ,
\end{equation}
and has been used to numerically verify the $3/2$ exponent in the asymptotic expression \eqref{eq:left-asymp}.
At small velocity gradients, the PDF can be approximated by a Gaussian, for which $\theta(\xi) = 2$, whereas at large velocity gradients the exponent transitions to $\theta(\xi) \simeq 1.16$. The expected value, on the basis of the instanton analysis \parencite{balkovsky1997} would be $1.5$, but high resolution numerical simulations have been unable, to this date, to observe this value. Instead, the value $1.16$ has been reported, with hints that even larger fluctuations would have to be investigated to reach the instanton scaling \parencite{gotoh1998,grafke2015relevance}. The main benefit of using (\ref{interpol}) instead of the raw surrogate saddle-point action derived from $\rho_{1.7}(\xi)$ is that it yields a smooth interpolation of data, circumventing error fluctuations that grow at larger values of $|\xi|$.

\subsection{Evaluation of $I_1 [p^c,u^c]$ and $I_2 [p^c]$}

Since $I_1 [p^c,u^c]$ is a linear functional of $u^c(x,t)$, it can be written that, from (\ref{pc_full}) and (\ref{u0}):
\begin{equation} \begin{split} \label{I1I2dp}
I_1 [p^c,u^c] + I_2 [p^c] &= I_1 [p^{(0)},\lambda^c F_0^{(1)}] + \\
&+ I_1 [p^{(0)}, (\lambda^c)^2 F_0^{(2)}] + I_2 [p^{(0)}] + \mathcal{O}[\delta p^c] \ . \
\end{split} \end{equation}
In order to evaluate the first three terms on the RHS of (\ref{I1I2dp}), it is interesting, for the sake of fast numerical convergence,
to write the two-point correlation functions (\ref{Gpu}) and (\ref{Guu}) in Fourier space, viz.,
\begin{subequations}
\begin{align}
\tilde G_{pu} (k,t,t') &= \int d x \ G_{pu}(x,0,t,t') \exp(-ikx) \nonumber \\ &= -i g^2 \exp \left [ - (t' -t)k^2 \right ] \Theta(t'-t) \ , \ \label{fGpu} \\
\tilde G_{uu} (k,t,t') &= \int d x \ G_{uu}(x,0,t,t') \exp(-ikx) \nonumber \\ &= g^2 \sqrt{\frac{\pi}{2}} \exp \left
[ - \left (|t' -t| + \frac{1}{2} \right ) k^2 \right ] \ . \ \label{fGuu}
\end{align}
\end{subequations}
We have, from (\ref{I1}), (\ref{I2}), (\ref{fGpu}), and (\ref{fGuu}),
\begin{subequations}
\begin{align}
I_1[p^{(0)},\lambda^c F_0^{(1)}]  &= \frac{\lambda^c }{2 \pi^2} \int_{t,t'<0} dt dt' \int dk dk' ~  k(k+k') \tilde p^{(0)}(k,t) \times \nonumber \\ &\times \tilde F_0^{(1)} (-k,t') \tilde G_{uu} (k',t,t')   \tilde G_{pu}(k + k',t',t) \nonumber \\
&= \frac{\lambda^2 g^4}{8 \pi} \int dk dk' ~  \frac{k(k+k')}{k^2+k'^2+(k+k')^2}  \exp \left [ -\frac{1}{2} \left (k^2+k'^2 \right ) \right ]  \ , \ \label{I1n}
\\
I_2[p^{(0)}] &= - \frac{1}{2(2 \pi)^2} \int_{t,t'<0} dt dt' \int dk dk' \ k^2 \nonumber \times \\ &\times \tilde p^{(0)}(k,t) \tilde p^{(0)}(-k,t') \tilde G_{uu} (k',t,t')  \tilde G_{uu}(k + k',t,t') \nonumber \\
&= -\frac{\lambda^2 g^4}{16 \pi} \int dk dk' \frac{k^2}{k^2+k'^2+(k+k')^2} \exp \left [ -\frac{1}{2} \left (k'^2 + (k+k')^2 \right ) \right ] \ , \ \label{I2n}
\end{align}
\end{subequations}
implying that
\begin{equation}
I_1[p^{(0)}, \lambda^c F_0^{(1)}] = - I_2[p^{(0)}] = (3-\sqrt{3}) \lambda^2 g^4/24
\end{equation}
and, according to (\ref{I1I2dp}),
\begin{equation}
I_1 [p^c,u^c] + I_2 [p^c] = I_1 [p^{(0)},(\lambda^c)^2 F_0^{(2)}]  + \mathcal{O}[\delta p^c] \ . \ \label{I1dp}
\end{equation}
A straightforward numerical evaluation returns,
\begin{equation}
I_1 [p^{(0)},(\lambda^c)^2 F_0^{(2)}] \simeq 1.6 \times 10^{-3} \lambda^3 g^4 \ . \
\end{equation}
Eqs. (\ref{vgpdf2}), (\ref{c-expansion}), (\ref{S1squared}), and (\ref{I1dp}) provide all the ingredients needed for a perturbative analytical expression for the velocity gradient PDF:
\begin{equation}
\rho_g(\xi) = C(g) \exp \left [- \frac{1}{g^2} S_{sc}(\xi) + \frac{1}{2g^4} I_1 [p^{(0)}, (\lambda^c)^2 F_0^{(2)}] \right ] \ , \ \label{rho_complete}
\end{equation}
where $C(g)$ is a normalization constant that cannot be determined from the instanton approach, since it depends on the detailed shape of the vgPDF for $- \infty < \xi < \infty$, while (\ref{rho_complete}) refers, in principle, to negative velocity gradients which are some standard deviations away from the mean. The relevance of the saddle-point computational strategy (including fluctuations), however, can be assessed from adjustments of $C(g)$ that produce the best matches between the predicted PDFs, Eq. (\ref{rho_complete}), and the empirical ones, obtained from the direct numerical simulations of the stochastic Burgers equation. These numerical fits are carried out in the velocity gradient range $-5g \leq \xi \leq -3g$.
%\begin{equation} \begin{split}
%&\lambda = 1.942 \xi  + 1.135 \times 10^{-1} \xi^2 + \nonumber \\
%&+ 5.707 \times 10^{-3} \xi^3 + 1.115 \times 10^{-4} \xi^4 \ . \
%\end{split} \end{equation}

\begin{figure}[t]
    \centering
    \begin{minipage}{0.5\textwidth}
        \centering
        \includegraphics[trim=20 400 120 0,clip,width=\textwidth]{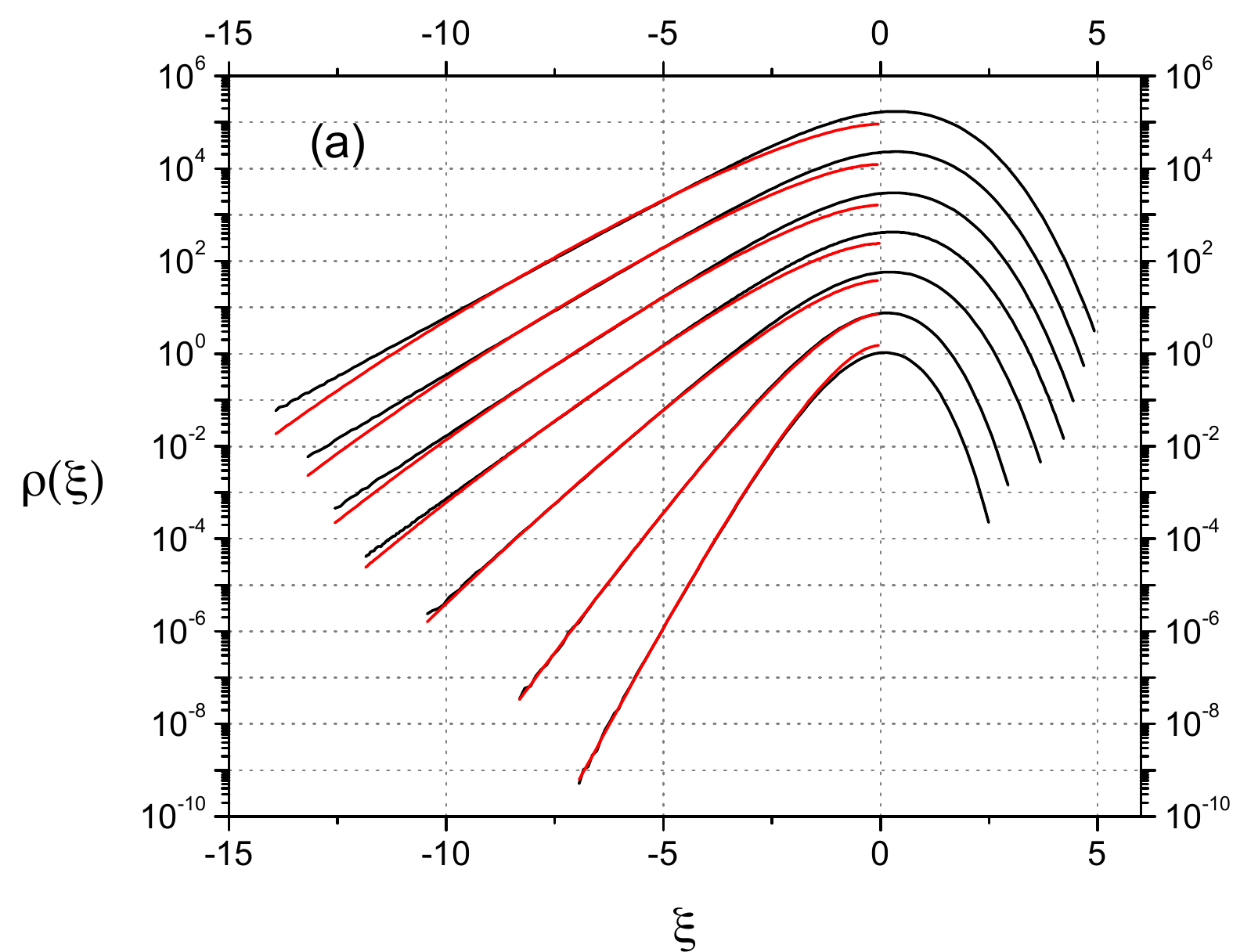} % first figure itself
    \end{minipage}\hfill
    \begin{minipage}{0.5\textwidth}
        \centering
        \includegraphics[trim=20 10 120 370,clip,width=\textwidth]{pdfs.pdf} % second figure itself
    \end{minipage}
    \caption
    [Modeled and empirical velocity gradient PDFs are compared for several values of noise strength]
    {Modeled (red lines) and empirical (black lines) vgPDFs are compared for noise strengths $g=1.0, 1.2, 1.5, 1.7, 1.8, 1.9$, and $2.0$. They have been shifted along the vertical axis to ease visualization, and their associated values of $g$ grow from the bottom to the top in each one of the PDF sets. Figures (a) and (b) give the modeled vgPDFs that include and neglect, respectively, the effects of fluctuations around instantons.}
    \label{vgPDFs}
\end{figure}

Comparisons between the predicted and empirical PDFs are shown in Fig. \ref{vgPDFs}, for $g=1.0$, $1.2$, $1.5$, $1.7$, $1.8$, $1.9$, and $2.0$, with and without the fluctuation correction term proportional to $I_1 [p^{(0)},(\lambda^c)^2 F_0^{(2)}]$, as it appears in (\ref{rho_complete}).
It can be observed in the figure that the surrogate saddle-point action is in fact a very good approximation to the exact one, by inspecting the PDF for $g=1.0$, when the cumulant contribution is almost negligible. As $g$ grows, the relative cumulant contributions grow as well, and become essential for an accurate modeling of velocity gradient PDF tails. For $g=1.7$, as an example, there is an evident intermittent left tail, with an excellent agreement between modeled and empirical PDFs that extends for about four decades.

As it can be seen from Fig. \ref{vgPDFs}, as $g$ grows, the velocity gradient regions where the agreement between the predicted and the empirical vgPDFs is reasonably good shrink in size. This is, of course, expected under general lines, since the cumulant expansion is a perturbative method supposed to break down when the amplitude of saddle-point configurations become large enough, which in our particular case takes place for large negative velocity gradients.

\subsection{Perturbative Domain}

\begin{figure}[h]
    \centering
    \includegraphics[width=.7\textwidth]{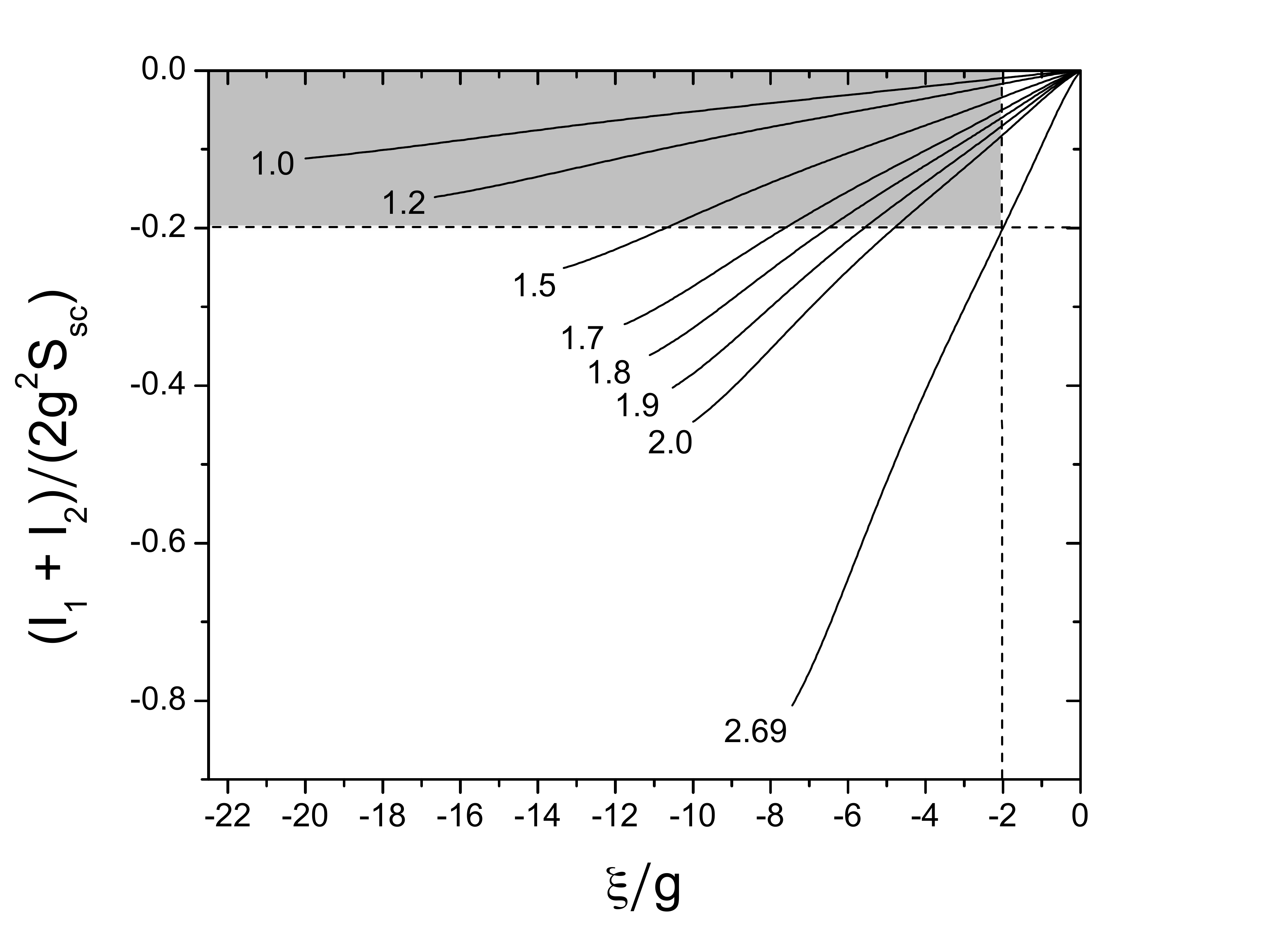}
    \caption
    [The intensity of fluctuations defines the range where the perturbative cumulant expansion is assumed to work]
    {Solid lines, labeled by values of $g$, represent relative corrections to the MSRJD surrogate saddle-point action due to fluctuations around instantons. The intersection points of each one of the solid lines with the vertical and horizontal dashed lines define the range of normalized velocity gradients $\xi/g$ where the perturbative cumulant expansion is assumed to work (highlighted region in the plot).} \label{pert_domain}
\end{figure}

An analysis of the results in Fig.~\ref{vgPDFs} reveals that a fine matching between the predicted and the empirical vgPDFs holds for $|\xi| > 2g$, but starts to lose accuracy when velocity gradients are such that the second order cumulant expansion contributions, $(I_1[p^c,u^c]+I_2[p^c])/2g^4$, are of the order of $20 \%$ (in absolute value) of the dominant saddle-point contributions, $S_{sc}(\xi)/g^2$. In Fig.~\ref{pert_domain}, the ratio between these two quantities is depicted with its dependence on the velocity gradient $\xi$, for the several investigated values of the noise strength parameter $g$, up to $g=2.0$. It can be estimated in this way, then, that $g \simeq 2.7$ is an upper bound for the usefulness of the cumulant expansion method.

\section{Discussion}

The instanton approach to Burgers intermittency was introduced in \textcite{gurarie1996,balkovsky1997} to describe asymptotically large fluctuations, in this case, of the velocity gradient. The representation of its preasymptotic features, though, has remained obscured.

Previous results in the context of Lagrangian turbulence (\textcite{moriconi2014} and \textcite{apolinario2019instantons}, discussed in Chap.~\ref{chap:rfd}), have indicated that non-Gaussian fluctuations can be perturbatively investigated at the onset of intermittency by means of the cumulant expansion technique. The main lesson taken from these studies is that at the onset of intermittency, the MSRJD saddle-point action reaps a renormalization of its noise and heat kernels, as a dynamical effect of fluctuations around instantons.
In this way, accurate comparisons between analytical and empirical velocity gradient PDFs have been achieved.

In this work, reported in \textcite{apolinario2019onset}, a similar approach has been applied to the stochastic Burgers hydrodynamics in order to predict the left tails of its velocity gradient PDF at the onset of intermittency. The results in Fig.~\ref{vgPDFs} show that an account of fluctuations around instantons is necessary to render the instanton approach a meaningful tool for the modeling of Burgers intermittency.

\end{chapter}

\begin{chapter}{Shot Noise Multifractal Model for Turbulent Pseudo-Dissipation}
\label{chap:shotnoise}

The first observations of intermittency in turbulence were reported quite a long time ago \parencite{batchelor1949} and were first addressed theoretically in \textcite{kolmogorov1962refinement,obukhov1962}, as discussed in Sec.~\ref{sec:multifractal}. In these works, scale dependent observables were postulated as the relevant quantities in the study of fluctuations in the turbulent inertial range.
The theory also hypothesized that the kinetic energy dissipation, a positive quantity, follows a lognormal probability distribution, an observation which is remarkably accurate, as reported in experiments and numerical simulations \parencite{yeungpope1989}.
Furthermore, experimental measurements of the kinetic energy dissipation revealed long-range power-law correlations \parencite{gurvitch1963,pond1965}, another key feature of turbulent fields. Multifractal random fields have been a tool to describe and understand turbulent fields with such statistical properties, but their derivation on a first-principle basis is still an open problem.

The origin of the lognormal distribution of the kinetic energy dissipation has been connected to the Richardson energy cascade through several phenomenological models, beginning with discrete cascade models \parencite{novikov1964,yaglom1966,frisch1978}. In summary, these models describe the distribution of energy dissipation across length scales in a turbulent field, from the large energy-injection scale, down to the much smaller dissipation length scales.
The kinetic energy dissipation in an inviscid cascade of $n$ steps, following Eq.~\eqref{eq:epsilon-cascade}, can be written as a random variable
\begin{equation} \label{eq:discrete-cascade}
    \varepsilon_n = W_1 W_2 \cdots W_n \ \varepsilon_0 \ ,
\end{equation}
where the $W_i$ are random variables which determine the energy transferred from one step to the next. For the cascade to be statistically inviscid, it is required that the $W_i$ are positive and have a mean value of one.

% below equally -> identically ?
If the $W_i$ factors in this model are equally and independently distributed, the probability distribution for the small-scale energy dissipation is well approximated by a lognormal, since $\ln \varepsilon_n$ is defined as the sum of several independent random variables of finite mean and variance.
The central limit theorem ensures that the probability distribution of $\ln \varepsilon_n$ approaches a normal distribution. As a consequence, $\varepsilon_n$ itself is a random variable with a lognormal distribution in the limit $n \to \infty$.
%This is a simple way to elicit the relevance of the lognormal distribution and its connection to the energy cascade. Different discrete models rely on this basis, with varying proposals for the way the energy is split at each step, and for the probability distribution of the $W_i$ factors.

These cascade models, though, had the limitation of being discrete and of possessing a special scale ratio between neighboring scales, customarily $\lambda=2$. It was noticed in \textcite{mandelbrot1972} that this special scale ratio should not be present, since turbulent energy dissipation displays multifractal properties for any chosen scale. Instead, a description in which arbitrary values of $\lambda$ are valid and produces multifractal statistics should be preferred and investigated, as pointed out by Mandelbrot.
Furthermore, the discrete models have been able to account for scale-locality of the energy transfer process, but did not account for correlations in time. This means that such fields had no causal structure, making them difficult to connect to the dynamics of turbulent fields described by the Navier-Stokes equations. Some of these issues were addressed by the causal framework of \textcite{pereira2018multifractal}, in which a stationary stochastic process is described and it is analytically verified that it satisfies the same multifractal properties as the small-scale energy pseudo-dissipation of the discrete cascade models.

This chapter describes a causal stochastic process driven by discrete and periodic random jumps, which is used to model the dynamical and multifractal properties of Lagrangian pseudo-dissipation. The observed dynamics is regular at scales below the Kolmogorov length scale, and multifractal at larger scales, demonstrating the possibility to apply discrete (shot) noise in effective models of turbulence.
%But observables which depend on the details of the small-scale dynamics exist, their dynamics is highly dependent on the intermittent fluctuations of turbulent flows, which happens for filaments and surfaces carried by high Reynolds number fluid flows.
Our main motivation here relies on the fact that the time evolution of local Lagrangian observables is sensitive to the existence of spacetime localized perturbations of the turbulent flow, such as vortex tubes. The dynamics of spheroids in turbulent flows has been observed to be dependent on small scale properties \parencite{parsa2012,voth2017}. Their tumbling motion is marked by a regular evolution disrupted by intense jumps, which can be seen in Fig.~\ref{fig:rods}, indicating that a modeling in terms of shot noise might explain their behavior. The main characteristics of a turbulent flow which lead to the preferential alignment of these spheroids is still a problem under investigation, with possible applications in industrial and natural fluid currents.

\begin{figure}[t]
    \centering
    \includegraphics[width=.7\textwidth]{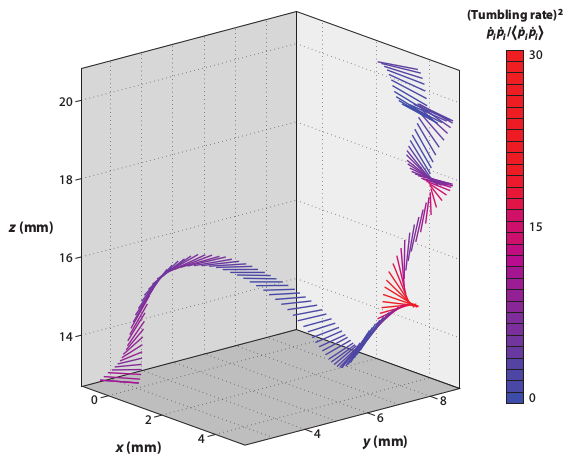}
    \caption
    [Trajectory of a rod in a turbulent flow from experiments]
    {Trajectory of a rod in a turbulent flow from experiments of \textcite{parsa2012}. The color represents the tumbling rate. Figure extracted from \textcite{voth2017}.}
    \label{fig:rods}
\end{figure}

There are infinite ways a stochastic field can be multifractal. In previous models in the literature \parencite{schmitt2003,perpete2011,pereira2018multifractal}, the lognormal description was chosen, due to its interesting statistical features (it is the simplest nontrivial continuous multifractal formulation) and to its history in connection with the statistical theory of turbulence.
Furthermore, the construction of a causal equation for a multifractal field driven by shot noise requires the use of a general version of It{\^o'}s lemma, including the contributions from discontinuities \parencite{protter2005,klebaner2012}. This lemma and its application to the random field in case are discussed in detail.

Focusing on turbulence, it turns out, from experimental evidence \parencite{yeungpope1989}, that several positive-definite observables like the kinetic energy dissipation, kinetic energy pseudo-dissipation, enstrophy, and the absolute value of acceleration can be reasonably well described by lognormal distributions, with a particularly good accuracy being achieved for pseudo-dissipation.
In the work of Yeung and Pope, it is also remarked that the statistical moments of dissipation and enstrophy seem to approach those of the lognormal distributions as the Reynolds number increases, suggesting that further studies are required to settle this issue.
Additionally, it is also known that the lognormal distribution can only be a good approximation to the statistics of dissipation, but not a complete solution which holds close to the dissipative scale or at arbitrarily high orders of the statistical moments. A discussion on consistency requirements for statistical distributions of turbulent observables can be found in \textcite{frisch1995}.

\section{Statistics of Turbulent Energy Dissipation and Pseudo-Dis\-si\-pa\-tion} \label{sec:stat-dissip}

The first theoretical results in the statistical theory of turbulence established the picture of the turbulent cascade on a mathematical ground. This early description of Kolmogorov regards the mean behavior of the inertial range statistics of turbulent velocity fields, but not its fluctuations. Later studies of turbulent fluctuations, leading to the multifractal picture,
revealed that the K41 velocity field is an exactly self-similar field of Hurst exponent $1/3$, that is, a monofractal. This field is homogeneous in space, in contrast to the complex and concentrated structures which form in isotropic flows, revealed by direct numerical simulations and experiments \parencite{ishihara2009,debue2018prf,dubrulle2019beyond}.

Multifractal fields have been proposed as general models to the  turbulent velocity field in the inertial range, although it remains an open problem to fully characterize this multifractal field and its statistical properties.
% blanket = general
For the purpose of modeling a positive-definite scalar field, consider a generic $d$-dimensional multifractal random field $\varepsilon_{\eta}$, which may depend on the spatial variable $\boldsymbol{x}$ and on time $t$, and with a dissipative length scale $\eta$.
The basic statistical properties of this random field are compatible with the features of the discrete cascade models and with experimental and numerical realizations of several observables in turbulence.
The field $\varepsilon_{\eta}$ satisfies, for its statistical moments, the relation
\begin{equation} \label{eq:bare-moments}
\left\langle(\varepsilon_{\eta})^{q}\right\rangle = A(q) \ \eta^{K(q)} \ ,
\end{equation}
where $A(q)$ is a $q$-dependent constant and $K(q)$ is a characteristic function of the multifractal field, connected to how structures at different scales spread across space.
% careful: previously, the microscale eta was represented by a lambda. lambda now is used to represent the scale ratio in discrete (or continuous) cascade models. one has to review this to make sure there are no old lambdas left in the wrong places.
In particular, a lognormal distribution for $\varepsilon_{\eta}$ corresponds exactly to $K(q) = \mu q (q-1) / 2$, where $\mu$ is called the intermittency parameter, which measures the intensity of the fluctuations of this field.
%In the case of Eulerian three-dimensional turbulence, $\mu=0.2$ \parencite{stolovitzky1994,praskovsky1997,chen1997}.
% stolovitzky1992,

The variable $\varepsilon_{\eta}$ is a bare field, since it is defined at the dissipative scale. The multifractal hypothesis makes predictions for the behavior of coarse grainings of $\varepsilon_{\eta}$ as well, which are defined as local averages of the original field at the scale $\ell > \eta$, according to Eq.~\eqref{eq:dissip-coarse}.
In particular, the statistical moments of a coarse-grained multifractal field obey the same statistical behavior as the bare field,
\begin{equation} \label{eq:dressed-moments}
\left\langle(\varepsilon_{\ell})^{q}\right\rangle = A'(q) \ \ell^{K(q)} \ ,
\end{equation}
at scales larger than the bare scale $\eta$ and up to some critical moment $q_{\mathrm{crit}}$, beyond which this scaling becomes linear \parencite{schmitt1994,lashermes2004}.
% schmitt1994 p.98, these references came from PerpeteSchmitt2011
It is vital to know these properties for coarse-grained fields for two main reasons. First, a coarse-grained field is all experimentalists have access to. And second, the features of coarse-grained fields are a fundamental ingredient in Kolmogorov's refined similarity hypothesis, according to which the inertial range statistical properties at scale $\ell$ depend only on $\ell$ itself and on the kinetic energy dissipation coarse-grained at this scale, $\varepsilon_{\ell}$. Thus, the verification that a given set of data does display multifractal statistics compels to the study of its coarse-grained properties.

For the general field $\varepsilon_{\eta}$, given that $X_{\eta} \equiv \ln \varepsilon_{\eta}$, its autocorrelation function decays logarithmically with the distance between the points,
\begin{equation} \label{eq:ln-correlation}
\langle X_{\eta}(0) X_{\eta}(\boldsymbol{r}) \rangle = C - \frac{\sigma^2}{\ln \lambda} \ln |\boldsymbol{r}| \ .
\end{equation}
This property can be easily verified for the the discrete cascade models \parencite{arneodo1998prl,arneodo1998jmathphys,schmitt2003}, in which case $\sigma^2 = \mathrm{Var}[ \ln W ]$ and $C = \langle \ln W \rangle^2 n^2 + \sigma^2 n$, where $n$ is the depth of the cascade and $\lambda$ the scale ratio of the model.
The Fourier transform of this expression corresponds to the power spectrum, which amounts to $1/f$ noise,
\begin{equation}
E_{\eta}(k) \approx k^{-1} \ .
\end{equation}
% can i find other applications of multifractal fields?
This is a common feature of intermittent fields in general and is also valid for coarse-grained fields \parencite{schertzer1987,schertzer1991,bacry2001}.
The properties just presented: Eqs. (\ref{eq:bare-moments}), (\ref{eq:dressed-moments}), and (\ref{eq:ln-correlation}) are the main characteristics of a multifractal field.

To account for fluctuations into account, it was postulated by Kolmogorov and Obu\-khov in 1962 that the kinetic energy dissipation field follows a lognormal distribution with
\begin{equation} \label{eq:kolmogorov-var}
    \mathrm{Var}[\ln \varepsilon_{\ell}] = - \mu \ln (\ell/L) + C \ ,
\end{equation}
% this equation is obtained in schmitt2003
where $L$ is the integral length scale, $\ell$ is the observation scale in the inertial range, $\eta \ll \ell \ll L$, and C is an arbitrary constant. The intermittency parameter $\mu$, the same as in the expression for $K(q)$, was historically introduced in this expression.

It was later realized \parencite{yaglom1966} that the intermittency parameter is also responsible for the power-law correlations of the kinetic energy dissipation, in the form
\begin{equation} \label{eq:gurvitch-corr}
\langle \varepsilon_{\eta}(\boldsymbol{r}) \varepsilon_{\eta}(\boldsymbol{r+\delta r}) \rangle \propto (L/\ell)^{\mu} \ ,
\end{equation}
where $\ell \equiv |\boldsymbol{\delta r}|$ and the parameter $\mu$ is apparent as well.
The cascade models were built to explain these statistical features.

The specific random fields considered in this chapter, as well as \textcite{schmitt2003,perpete2011,pereira2018multifractal} are one-dimensional and depend only on time, since they correspond to some positive-definite observable following a Lagrangian trajectory of the flow.
The Lagrangian view is connected to the space-time structure of the energy dissipation cascade: since eddies are carried by the flow, their statistical distribution is somehow influenced by the transport properties of the turbulent velocity field.
And Lagrangian observables such as velocity differences and velocity gradients have been argued to display scaling and intermittent behavior, following a Lagrangian refined similarity hypothesis, in an equivalent manner to the Eulerian framework. Lagrangian velocity difference structure functions, for instance, are believed to scale as
\begin{equation} \label{eq:lag-k41-scaling}
    \langle \big( \delta u_i(\tau) \big)^n \rangle \propto (\langle \varepsilon_{\tau} \rangle \tau)^{\xi_n}
\end{equation}
in the Lagrangian inertial range, $\teta \ll \tau \ll T$, between the dissipative time scale $\teta$ and the integral time scale $T$.
% Lagrangian integral time is defined in barjona2017
The coarse-grained Lagrangian kinetic energy dissipation, $\varepsilon_{\tau}$, is defined in terms of its bare counterpart, $\varepsilon_{\teta}$, in analogy with Eq.~\eqref{eq:dissip-coarse}:
\begin{equation} \label{eq:dissip-coarse-1dim}
    \varepsilon_{\tau}(t) = \frac{1}{\tau} \int_{t}^{t+\tau} dt'
    \varepsilon_{\teta}(t') \ .
\end{equation}
Its average value $\langle \varepsilon_{\tau} \rangle$ is constant due to the stationarity of the turbulent flow. The dissipative time scale is determined from dimensional analysis as the Lagrangian analogue of the dissipative length scale, and is defined as $\teta = \eta^{2/3} \varepsilon_0^{-1/3}$ and the Lagrangian integral time is defined in terms of the velocity two-point autocorrelation $\rho_L(\tau)$, as $T = \int_0^{\infty} \rho_L(\tau) d\tau$.
In the K41 framework, the scaling exponents of velocity difference structure functions grow linearly, and the equivalent relation in the Lagrangian view is that $\xi_n = n/2$. This can be understood with the framework of \textcite{borgas1993}, which connects Lagrangian and Eulerian self-similarity.
Eq.~\eqref{eq:lag-k41-scaling} has been numerically verified in \textcite{benzi2009,sawfordyeung2011,barjona2017}, and it is notable that finite Reynolds effects are even more pronounced in the Lagrangian frame, making measurements even more difficult \parencite{yeung2002}.

In \textcite{mandelbrot1972}, it was noticed that a random field built as the exponential of a Gaussian field,
\begin{equation} \label{eq:mandelbrot-dissip}
    \varepsilon_{\eta} \propto \exp \{ \sqrt{\mu} X\} \ ,    
\end{equation}
would satisfy these properties. In this equation, the Gausian field $X$ is an It\^o integral over the Wiener measure $dW(t)$, hence $\varepsilon_{\eta}$ can be seen as a continuous product of random factors. 
This is a straightforward, albeit nonrigorous, translation of Eq.~\eqref{eq:discrete-cascade} to the continuum, which has lognormal statistics as well. This construction was mathematically formalized in \textcite{kahane1985}, and the modern understanding of such random fields has led to explicit frameworks in the Eulerian \parencite{pereira2016} and Lagrangian context \parencite{pereira2018multifractal}, which approximate the known statistics of turbulent fields. Since Kahane, this continuous stochastic process with multifractal statistics is called Gaussian Multiplicative Chaos, in connection with the standard additive random process (the Wiener process). For an explanation of the origin of this nomenclature, the reader should consult the footnote in \textcite{rhodes2014}.

The discrete cascades display the same statistical properties as the small scale multifractal field of Eq.~\eqref{eq:mandelbrot-dissip}, but for a single scale ratio.
Another critique of Mandelbrot on the discrete cascades was the absence of a space-time causal structure. The only causal connection in these models is between length scales, a relation which cannot be easily translated to a space-time distribution of turbulent structures or of energy dissipation. A step in this direction was performed in \textcite{schmitt2003}, with the study of a causal one-dimensional stochastic process, formulated in terms of a stochastic differential equation.
In this work, analytical expressions for the statistical moments and two-point correlation functions of this process are proved, in agreement with the multifractal hypothesis and providing a continuous-in-scale extension of the discrete cascade models.
This stochastic process, though, does not generate a stationary state solution, an issue which was resolved in \textcite{pereira2018multifractal}.

The stochastic process of \textcite{pereira2018multifractal} for the evolution of the pseudo-dissipation field $\varphi_P$ is described by the following stochastic differential equation:
\begin{equation} \label{eq:pereira-x}
\begin{split}
    d X_P(t)&=\left[-\frac{1}{T} X_P(t)+\beta_P(t)\right] d t+\frac{1}{\sqrt{\tau_{\eta}}} dW(t) \ , \\
    \beta_P(t)&=-\frac{1}{2} \int_{s=-\infty}^{t} \frac{1}{\left(t-s+\tau_{\eta}\right)^{3 / 2}} dW(s) \ ,
\end{split}
\end{equation}
where the pseudo-dissipation $\varphi_P(t)$ is given by an exponential of the underlying $X_P(t)$ process, explicitly:
\begin{equation}
    \varphi_P(t)=\frac{1}{\tau_{\eta}^{2}} \exp \left\{ \sqrt{\mu} X_P(t)-\frac{\mu}{2} \mathbb{E}\left[X_P^{2}\right] \right\} \ ,
\end{equation}
where $\mu$ is the Lagrangian intermittency exponent, $T$ is the integral time scale, $\teta$ the Kolmogorov dissipative time scale and $W(t)$ is a standard Wiener process.
The stochastic processes $X_P(t)$ and $\varphi_P(t)$ reach a stationary state with lognormal and long-range correlated statistics, in the limit of $\teta \to 0$ (corresponding to infinite Reynolds number). Some of the necessary ingredients in a multifractal process which this equation illustrates are long-term memory and noise correlations, expressed through the $\beta_P(t)$ term, which is driven by the same random noise solution as the main equation, for $\varepsilon_P(t)$.

\section{Stochastic Models of Lagrangian Pseudo-Dissipation} \label{sec:stoc-lag-dissip}

An alternative formulation of multiplicative chaos was done in \textcite{perpete2011}, where a time-discretized causal multifractal process was introduced. This stochastic process satisfies the multifractal properties for its statistical moments and two-point autocorrelation functions, as well as its coarse-grained version, for which the multifractal statistics are verified for any continuous scale ratio (in the limit of infinite Reynolds number).

The stochastic process of \textcite{perpete2011}, with a dissipative timescale $\teta$ and a large timescale corresponding to the integer $N = T/\teta$, is described by
\begin{equation} \label{eq:perpete-x}
    X_D(t) = \frac{1}{\sqrt{\teta}} \sum_{k = 0}^{N-1} (k+1)^{-1/2} \alpha_{t-k} \ ,
\end{equation}
where $\alpha_k$ are independent and identically distributed Gaussian random variables of zero mean and standard deviation $\sqrt{\teta}$. The time, unlike in the previous examples, is only defined for integer $t$. This process also possesses long-term memory over the integral time scale, in connection with Eqs.~\eqref{eq:pereira-x}.
The multifractal process corresponding to Eq.~\eqref{eq:perpete-x} is likewise given by its exponential,
\begin{equation}
    \varphi_D(t) = \varphi_0 \exp \big( \sqrt{\mu} X_D(t) - \mu \mathbb{E}[X_D^2]/2 \big) \ \mbox{,}
\end{equation}
which has lognormal and long-range correlated statistics.
Eq.~\eqref{eq:perpete-x} reaches a stationary state with the following multifractal properties:
\begin{enumerate}[label=(\roman*)]
    \item \label{it:bare}
    Its moments satisfy
    \begin{equation} \label{eq:perpete-moments}
    \mathbb{E}[ \varphi_D^q ] =
    C_q \left( \frac{\teta}{T} \right)^{-K(q)} \ \mbox{,}
    \end{equation}
    with $K(q) = \mu \ q (q-1) / 2$, conforming to the lognormal statistics, and $C_q$ is a factor which can be
    exactly calculated.
    % it'd be interesting to say on which C_q is independent, because this is important, and also to give a name to the K(q) function, maybe
    \item \label{it:dress}
    The coarse-grained moments of $\phi_D$ satisfy a similar relation with the same exponents:
    \begin{equation} \label{eq:perpete-moment}
    \mathbb{E}[ (\varphi_D)_{\tau}^q ] \approx
    c_q \left( \frac{\tau}{T} \right)^{-K(q)} \ \mbox{,}
    \end{equation}
    where the coarse-grained field is defined as a moving average with a window of $\tau$:
    \begin{equation}
    (\varphi_D)_{\tau}(t) = \frac{1}{\tau} \sum_{k=t}^{t+\tau/\teta-1} \varphi_D(k) \ .
    \end{equation}
    Relation (\ref{eq:perpete-moment}) is asymptotic, and is valid in the limit of $T$ going to infinity, with the ratio $\tau/T$ kept fixed. Furthermore, $q$ must be such that $K(q) < q-1$.
    % this resembles the linear growth limit, and Perpete has references on that, I should check it out
    The existence of upper and lower bounds on $c_q$ was demonstrated in \textcite{perpete2011}, while precise values would have to be inspected numerically.
    \item The autocovariance of this process, in the same limit of $T \to \infty$,
    converges to
    \begin{equation} \label{eq:perpete-cov}
    \mathrm{Cov}[\varphi_D(t), \varphi_D(t+\tau)] \approx
    - \mu \ln(\tau/T) \ \mbox{.}
    \end{equation}
\end{enumerate}

Having Eqs.~(\ref{eq:perpete-moments}-\ref{eq:perpete-cov}) in mind, a stochastic differential equation which inherits features from the continuous and discrete instances is described in this chapter. This stochastic field takes into account the small scales in a dynamic manner, such that it follows a smooth time evolution on scales below the Kolmogorov time, but still shows roughness and multifractal behavior on timescales much larger than that. This picture is inspired by the Kolmogorov phenomenology, in which dissipation can be neglected in the inertial range, while it acts in smoothing out the velocity field in the dissipative scale.

Furthermore, the refined similarity hypothesis states that, in the limit of infinite Reynolds numbers, all the statistical properties at scale $\ell$ are uniquely and universally determined by the scale itself and the mean energy dissipation rate coarse-grained at this scale, $\varepsilon_{\ell}$.
By this hypothesis, it is expected that a variety of noise sources generate similar behavior in the inertial range, due to an independence on the details of the dissipative dynamics.
For this reason, several large scale observables of the random field stirred by discrete noise should converge to the same quantities as fields driven by Wiener noise.

%The stochastic process described in this chapter is a model for Lagrangian pseudo-dissipation forced by a discrete noise source, still with a long time memory such as the noise source presented in \textcite{perpete2011}. It is interesting to remark that this stochastic process evolves in continuous time, while being driven by randomness which is periodic in time and only acts in discrete instants. A stationary state arises as solution of this process and its statistical properties are investigated, in comparison with the properties of the multifractal random fields already described in the literature.

Studies of discrete noise (often called shot noise) or a mixture of discrete and continuous noise (or jump-diffusion) have been pursued in others areas of knowledge, such as financial economics \parencite{duffie2000,das2002}, neuronal systems \parencite{patel2008,sacerdote2013}, atomic physics \parencite{funke1993,montalenti1999}, biomedicine \parencite{grenander1994} and image recognition \parencite{srivastava2002}.

Explicitly, consider the stochastic process given by the stationary solution of the following differential equation:
\begin{equation} \label{eq:jump-scalar}
    dX(t) = \left( -\frac{1}{T} X(t^-) + \beta(t) \right) dt + \frac{1}{\sqrt{\teta}} \sum_{\teta \ell \leq t} \alpha_{\ell} \ \delta(t-\teta \ell) \ dt \ .
\end{equation}
The first term on the RHS corresponds to a drift in a usual Langevin equation, and has the same form as the drift term in Eq.~\eqref{eq:pereira-x}. The first contribution in this term is responsible for correlations of the $X(t)$ random field of characteristic time $T$, while the second is in charge of the multifractal correlations in the solution, with a similar role to the long-memory term present in Eqs. (\ref{eq:pereira-x}) and (\ref{eq:perpete-x}). The second term in Eq.~\eqref{eq:jump-scalar} accounts for the discrete random jump contributions. These jumps occur at periodic intervals of length $\teta$ and have an intensity $\alpha_{\ell}$, which is a Gaussian random variable of zero mean and standard deviation of $\sqrt{\teta}$. Each value of $\ell$ corresponds to a jump instant $\ell\teta$, hence the sum is carried for all jump times prior to the observation time $t$.

It is also important to observe the presence of the $t^-$ in the first term which represents an instant infinitesimally preceding the current observation instant. In a stochastic process with jumps, this kind of care is needed, because the current state of the system (at $t$) depends on the continuous evolution up to time $t^-$ and on the value of a jump which may have happened exactly at the instant $t$, and therefore does not affect the previous state of the system, only its future evolution. For this reason, the state $X(t^-)$ and a jump $\alpha_{\ell}$ happening exactly at $\ell \teta = t$ are completely independent events.
In the traditional notation of point processes \parencite{protter2005,klebaner2012}, continuous random fields are taken to be \textit{c\`{a}dl\`{a}g}, a French acronym for \textit{continuous on the right and limit on the left}. This denomination means that jumps occur exactly at the instant $t_{\ell}$, while the left-limit at $t_{\ell}^-$, is not at all influenced by the jump term. Then, for a discontinuous random field $f(t)$ with a random jump happening at $t_{\ell}$, being \textit{c\`{a}dl\`{a}g} is equivalent to
\begin{equation}
    \lim_{t \to t_{\ell}^-} f(t) \neq f(t_{\ell}) \ \ \mbox{and} \ \
    \lim_{t \to t_{\ell}^+} f(t) = f(t_{\ell}) \ .
\end{equation}

The drift term in Eq.~\eqref{eq:jump-scalar} contains a random contribution, $\beta(t)$, in correspondence with the long-term memory random contributions in \textcite{pereira2018multifractal}. The expression for this term is
\begin{equation} \label{eq:beta}
    \beta(t) = -\frac12 \sum_{\teta \ell \leq t}  \ \frac{\alpha_{\ell}}{(t-\teta \ell + \teta)^{3/2}} \ \mbox{,}
\end{equation}
where the $\alpha_{\ell}$ are exactly the same as those already sampled randomly for Eq.~\eqref{eq:jump-scalar}.
The sum is also carried out over all jump times up to the time $t$.

The solution to this equation can be written explicitly in terms of a particular realization of the random jumps:
%\begin{equation}
%X_{\tau_{\eta}}(t)=\int_{s=-\infty}^{t} e^{-\frac{t-s}{T}} \beta_{\tau_{\eta}}(s) d s+\frac{1}{\sqrt{\tau_{\eta}}} \int_{s=-\infty}^{t} e^{-\frac{t-s}{T}} W(d s)
%\end{equation}
\begin{equation}
    \begin{split}
        X(t) &= \int_{s = t-T}^{t} \frac{e^{(s-t)/T}}{\sqrt{t-s+\teta}} \ \sum_{\ell} \alpha_{\ell} \ \delta(s-\teta \ell) \ ds \\
        &+ \frac{1}{\sqrt{T+\teta}} \int_{s=0}^t e^{(s-t)/T} \ \sum_{\ell} \alpha_{\ell} \ \delta(s-\teta \ell+T) \ ds \ ,
    \end{split}
\end{equation}
where, after integrating over the delta functions,
\begin{equation} \label{eq:x-solution}
\begin{split}
        X(t) &= \sum_{t-T < \teta \ell \leq t}
        \frac{e^{(\teta \ell-t)/T}}{\sqrt{t-\teta \ell+\teta}} \  \alpha_{\ell} \\
        &+ \frac{1}{\sqrt{T+\teta}} \sum_{0 < \teta \ell \leq t}
        e^{(\teta \ell-t-T)/T} \ \alpha_{\ell - T/\teta} \ .
\end{split}
\end{equation}
From this solution, several analytical properties of the stationary stochastic field can be calculated and compared to the results of numerical simulations.

Still, the solution in Eq.~\eqref{eq:x-solution} has only Gaussian fluctuations. In analogy to what is done in the discrete and continuous settings, the field with multifractal correlations is, in fact,
\begin{equation} \label{eq:exp-x}
    \varphi(t) = \varphi_0 \ \exp\{\sqrt{\mu} X(t)
    - \mu \mathbb{E}[X(t^-)^2]/2 \} \ ,
\end{equation}
where the mean value of this process is defined as $\varphi_0 = 1/\teta^2$, following the phenomenology of Kolmogorov \parencite{girimaji1990diffusion}.
The variance of the $X(t)$ process, $\mathbb{E}[X^2(t)]$, can be calculated from the analytical solution, Eq.~\eqref{eq:x-solution}:
% this expression: 20/06/19,4
\begin{equation} \label{eq:x2-variance}
\begin{split}
    \mathbb{E}[X^2(t)] &=
    \sum_{t-T \leq \teta\ell \leq t} \teta \frac{e^{2(\teta\ell-t)/T}}{t-\teta \ell + \teta}
    + \frac{\teta}{T + \teta} \sum_{0 \leq \teta\ell \leq t} e^{2(\teta\ell-t)/T-2} \\
    &+ \frac{2 \teta}{\sqrt{T+\teta}} \sum_{t-T \leq \ell \leq t} \frac{e^{2(\teta\ell-t)/T-1}}{\sqrt{t-\teta \ell + \teta}} \ ,
\end{split}
\end{equation}
thus it can be seen simply as a function of time.

Eq.~\eqref{eq:exp-x} also explains the choice of periodic discrete noise with period $\teta$, instead of the common choice of Poisson noise with an equal characteristic time, which is often what is referred to as shot noise \parencite{morgado2016}. The variable $z = \exp{\sqrt{\mu}X}$, where $X$ is a sum of $N$ Gaussian random variables, follows a lognormal probability distribution. In the case of Poisson noise, the amplitudes of the jumps would be given by the normal distribution as well, but the number of jumps would be random, with a mean $N$, and $z$ would not follow a lognormal distribution exactly. In the limit of $N \to \infty$, though, both distributions coincide, by the central limit theorem.

As was done in \textcite{schmitt2003,pereira2018multifractal}, a dynamical equation for the pseudo-dissipation itself can be obtained from the dynamical equation for $X(t)$, Eq.~\eqref{eq:jump-scalar}, and the relation between the $X$ and $\varphi$ variables, Eq.~\eqref{eq:exp-x}.
Consider for a moment the general stochastic differential equation
% is this going to make confusion with any other F and G variables? maybe, there is F down below, in F A_{ij} factor
% is this really t^- or t? because noise is calculated exactly at t, while the G function is probably calculated in the limit
% okay, PROBABLY F and G have to be continuous, or the expression for X(t) - X(t^-) below will be different, and maybe other things will also be different
\begin{equation} \label{eq:x-sde}
    dX(t) = F(t^-, X(t^-)) \ dt + \sum_{0 \leq t_{\ell} \leq t} G(t^-,X(t^-)) \delta(t-t_{\ell}) \alpha_{\ell} \ dt \ ,
\end{equation}
where $F$ and $G$ are arbitrary functions of $t$ and $X(t)$, respectively called the drift and jump terms. This equation does not have any continuous noise term (proportional to a Wiener measure $dW(t)$), because the stochastic differential equation proposed in this work does not possess the Wiener term either.
In addition, an appropriate set of initial conditions for $X(t)$ is provided.
The new variable, $Y$, is obtained from the original variable through an arbitrary continuous function $f$, as
% what hypotheses on the function f?
\begin{equation} \label{eq:f}
    Y(t) = f(t,X(t)) \ .
\end{equation}
A stochastic differential equation for $Y(t)$ is obtained \parencite{protter2005,klebaner2012} with It\^{o}'s lemma for semimartingales, which is the appropriate expression for a change of variables in a stochastic process, equivalent to the chain rule in standard calculus. Semimartingales are generalizations of local martingales: While the latter are represented by continuous stochastic processes, such as the standard Brownian motion, the former may display discontinuous jumps, which are central to the current discussion. The solution $X(t)$ of Eq.~\eqref{eq:x-sde} is thus a semimartingale.

In its semimartingale formulation, It\^{o}'s lemma is expressed as
\begin{equation} \label{eq:ito-original}
    \begin{split}
    &Y(t) = Y(0) +
    \int_0^t \partial f(s^-,X(s^-)) / \partial s \ ds \\
    &+ \int_0^t f'(s^-,X(s^-)) dX(s) + \frac12 \int_0^t f''(s^-,X(s^-)) d[X,X]^c(s) \\
    &+ \sum_{0 \leq t_{\ell} \leq t} \Big( f(t_{\ell},X(t_{\ell})) - f(t_{\ell}^-,X(t_{\ell}^-))
    - f'(X(t_{\ell}^-)) (X(t_{\ell}) - X(t_{\ell}^-)) \Big) \ .
    \end{split}
\end{equation}
The integration interval, from $0$ to $t$, includes several jump instants, denoted by $t_{\ell}$ with an integer index $\ell$ differentiating each jump. Because of the discontinuities, it is important to prescribe that the $X(t)$ process is \textit{c\`{a}dl\`{a}g}, which means that terms of the form $X(s^-)$ should be calculated as the limit
\begin{equation}
    X(s^-) = \lim_{t \to s^-} X(t) \ .
\end{equation}
If $s$ is a jump instant, this limit does not include the contribution from the discontinuous jump, which is only accounted for in $X(s)$. Whereas if $s$ is not a jump instant, $X(s)$ and $f(s,X(s))$ are continuous at this point.
The first four terms in the RHS of Eq.~\eqref{eq:ito-original} are exactly equal to those in It\^{o}'s lemma for continuous processes, with the only difference that the discontinuous jumps require a distinction between left and right limits. As in the traditional It\^{o}'s lemma, the derivatives $f'(t,X(t))$ and $f''(t,X(t))$ are taken with respect to $X(t)$.

The \textit{continuous quadratic variation} $[X,X]^c(t)$ of the Wiener process is simply $t$, concluding the identification with the lemma for local martingales.
In general, the quadratic variation is defined by
\begin{equation} \label{eq:quad-var}
    [X,X]_{t}=\lim _{\delta t \rightarrow 0} \sum_{k=1}^{n}\left(X_{t_{k}}-X_{t_{k-1}}\right)^{2} \ ,
\end{equation}
where time has been partitioned into $n$ intervals of size $\delta t_k = t_k - t_{k-1}$ and $\delta t$ is the maximum size among these partitions \parencite{protter2005,klebaner2012}. The continuous quadratic variation is the continuous part of Eq.~\eqref{eq:quad-var}.
If the stochastic force is purely jump-discontinuous, as is the case in Eq.~\eqref{eq:jump-scalar}, its continuous quadratic variation is zero. Also, using Eq.~\eqref{eq:x-sde}, the discontinuity $X(t_{\ell}) - X(t_{\ell}^-)$ which appears in Eq.~\eqref{eq:ito-original} is equal to $G(t_{\ell}^-,X(t_{\ell}^-)) \alpha_{\ell}$.
Thus, replacing Eq.~\eqref{eq:x-sde} in Eq.~\eqref{eq:ito-original}, one of the terms in $f'(s^-,X(s^-)) dX(s)$ is canceled by $f'(t_{\ell}^-,X(t_{\ell}^-)) (X(t_{\ell}) - X(t_{\ell}^-))$. With this, \textit{It\^{o}'s lemma for pure jump processes} is obtained:
\begin{equation} \label{eq:ito-semi}
\begin{split}
    Y(t) &= Y(0) +
    \int_0^t \frac{\partial f}{\partial s}(s^-) \ ds
    + \int_0^t f'(X(s^-)) F(s^-,X(s^-)) ds \\
    &+ \sum_{\ell} \Big( f(t_{\ell},X(t_{\ell})) - f(t_{\ell}^-,X(t_{\ell}^-)) \Big) \ .
    \end{split}
\end{equation}
In differential notation, this is equivalent to
\begin{equation} \label{eq:ito-jump}
\begin{split}
    d Y(t) &=
    \partial f / \partial t \ dt +
    f'(X(t^-)) F(t^-,X(t^-)) dt \\
    &+ \sum_{\ell} \Big( f(t_{\ell},X(t_{\ell})) - f(t_{\ell}^-,X(t_{\ell}^-)) \Big)
    \delta(t-t_{\ell}) dt \ .
    \end{split}
\end{equation}

At first glance, this definition may look circular, because the variable $Y$ and the variable $X$ appear simultaneously. In fact, only the initial condition for $X(t)$ is needed, which is easily converted to an initial condition for $Y(t)$. All other appearances of $X(t)$ in Eq.~\eqref{eq:ito-jump} are causal, referring to values of $Y(t)$ already calculated, thus $X(t) = f^{-1}(t,Y(t))$. The term $f(t_{\ell},X(t_{\ell}))$, when $t_{\ell}$ is a jump instant, needs the value of $X$ at the current instant, which is simply the left-limit at the jump instant with the random contribution added:
\begin{equation} \label{eq:x-jump-gen}
    X(t_{\ell}) = X(t_{\ell}^-) + G(t_{\ell}^-,X(t_{\ell}^-)) \alpha_{\ell} \ .
\end{equation}
Thus, Eq.~\eqref{eq:ito-jump} is an entirely self-consistent way to determine the time evolution of the random field $Y(t)$.

In the specific model considered in this work, the equation for $X(t)$ is Eq.~\eqref{eq:jump-scalar}. $X(t)$ is a stochastic process with Gaussian fluctuations and its exponential is the variable of interest, with lognormal fluctuations and long-range correlations.
Through It\^{o}'s lemma (Eq.~\ref{eq:ito-semi}), a stochastic differential equation for the pseudo-dissipation field is reached, $\varphi(t) = f(X(t))$, where the specific function $f$ corresponding to Eq.~\eqref{eq:f} depends only on $X$ and is defined by
\begin{equation} \label{eq:exp}
    f(X(t)) = \varphi_0 \ \exp\{\sqrt{\mu} X(t)
    - \mu \mathbb{E}[X^2(t^-)]/2 \} \ \mbox{.}
\end{equation}

The equation for the pseudo-dissipation field obtained through It\^o's lemma is
\begin{equation} \label{eq:phi-x}
\begin{split}
    &d\varphi(t) = \varphi(t^-)
    \Bigg( - \frac{1}{T} \ln \frac{\varphi(t^-)}{\varphi(0)} -     \frac{\mu}{2 T} \mathbb{E}[X^2(t^-)] + \sqrt{\mu} \beta(t) \\ &- \frac{\mu}{2} \frac{\partial \mathbb{E}[X^2(t^-)]/2 }{\partial t} \Bigg) \ dt + \sum_{\ell} \Big( f(\varphi(\teta \ell)) - f(\varphi(\teta \ell^-)) \Big) \delta(t-\teta \ell) dt \ .
        \end{split}
\end{equation}
Together with an initial condition, this stochastic process is then completely well defined.
Since a long-term memory is present, it is necessary to provide $X(s)$ for $s \in ]-T,0]$, corresponding to the past time-evolution of $X$. After a few integral times, the influence of the initial condition vanishes, and the process reaches a stationary state. At the jumping times, a new random jump intensity $\alpha_{\ell}$ is sampled and this updates the value of the $X$ variable as in Eq.~\eqref{eq:x-jump-gen}, with
\begin{equation}
    X(\teta \ell) = X(\teta \ell^-) + \alpha_{\ell} / \sqrt{\teta} \ .
\end{equation}
From the above expression, the pseudo-dissipation is simply obtained as $\varphi(\teta \ell) = f(X(\teta \ell))$.

\section{Numerical Procedure} \label{sec:numerical}

Numerical simulations were performed to verify the statistical properties of the shot noise driven process. The time evolution of Eq.~\eqref{eq:jump-scalar} can be split in a deterministic contribution from the drift term, $(-X/T+\beta)$, and a jump term, proportional to a random jump intensity $\alpha_{\ell}$. There are sophisticated algorithms to obtain the solutions of general jump-diffusion equations, such as those illustrated in \textcite{casella2011,gonccalves2014}, which provide a framework to deal with complex time dependence in the drift or diffusion coefficients, cases where the solution cannot be obtained with a straightforward stochastic Euler algorithm. Instead, the diffusion term in Eq.~\eqref{eq:jump-scalar} does not display any time dependence, and the $\beta(t)$ term has a long-term memory, requiring a simpler algorithm in its implementation. With these considerations in mind, the Euler algorithm described in \textcite{casella2011} was applied in the simulation of the stochastic jump process of Eq.~\eqref{eq:jump-scalar}.

The jumping times are known in advance, since they are periodic, and given simply by $(0,\teta,2 \teta,\ldots)$. For each interval between two jumps, $((\ell-1)\teta,\ell\teta)$, the drift term is simulated with an Euler algorithm, which is used to calculate $X(t_{\ell}^-)$. Then, the jump term is added, with the random sampling of a new random quantity which is added to determine $X(t_{\ell})$.
To setup the initial conditions for the simulation, the jump intensities $\alpha_{\ell}$ are arbitrarily defined for a complete integral time in the interval $[-T,0]$. Eq.~\eqref{eq:beta} depends on the whole time evolution of the system, hence a truncation in the past evolution is required. A complete integral time has been chosen since it provides accurate results in comparison with the theoretical means and standard deviations, as is detailed in the next section. The random jump intensities in this past interval are sampled exactly like the intensities in the core of the simulation, as Gaussian random variables of mean zero and standard deviation $\sqrt{\teta}$. The choice of $X(0)=0$ is made as well.
The time necessary to reach a statistically stationary state in every simulation run is optimized by this choice of initial conditions, and is found to be less than two integral times for all simulations performed.

The above algorithm is used to build a sample path for the stochastic process in Eq.~\eqref{eq:jump-scalar}. This procedure was run for sample paths of thirty integral times in total, and three hundred sample paths were drawn for each value of $\teta$. Thus, an ensemble containing $9 \times 10^3$ integral times is built for each $\teta$, providing the significant statistics used to verify the multifractal properties of the stationary random field.

The chosen values of $\ln(\teta/T)$ range from $-1.0$ to $-6.0$. The more negative values correspond to more intermittency and higher Reynolds number.
The time step for the simulation was chosen to be $2 \times 10^{-3} \teta$ and the Lagrangian intermittency parameter used is $\mu = 0.3$,  which was measured in Lagrangian trajectories from direct numerical simulations in \textcite{huang2014}.

Once the $X(t)$ process is calculated with this algorithm, the pse\-udo-dis\-si\-pa\-tion $\varphi(t)$ is obtained as its exponential, from Eq.~\eqref{eq:exp}. It was verified that the mean and standard deviation of $X(t)$ follow the analytical results (Eq.~\ref{eq:x2-variance}) within error bars. This is particularly important for the evaluation of $\varphi$, which depends on the time periodic function $\mathbb{E}[X^2(t)]$. It is simpler and more precise to apply the analytical expression for this function (Eq.~\ref{eq:x2-variance}) than to store the previous integral times and compute standard deviations on the fly. For our results, the first five integral times were discarded, even though the observed times to reach a stationary state were always smaller than this. These results are reported in the next section.

%%%%%%%%%%%%%%%%%%%%%%%%%%%%%%%%%%%%%%
%%%%%%%%%%%%%%%%%%%%%%%%%%%%%%%%%%%%%%
%%%%%%%%%%%%%%%%%%%%%%%%%%%%%%%%%%%%%%
%%%%%%%%%%%%%%%%%%%%%%%%%%%%%%%%%%%%%%

\section{Numerical Results} \label{sec:results}

A sample trajectory of the shot noise multifractal process governed by Eq.~\eqref{eq:phi-x}
is depicted in Fig.~\ref{fig:phitraj}, along with its mean behavior.
%In the figure the sample trajectory (in blue), shows intense and inhomogeneous positive fluctuations, in yellow, it is depicted the ensemble average, which fluctuates very close to the theoretical mean, the dashed line in black. The discrete random jumps at regular intervals cannot be identified at this scale, hence a closer look at this stochastic process is shown in the inset, where individual jumps can be seen.
Trajectories for this example were generated for $\ln\teta/T = -5.60$, which corresponds to one of the highest Reynolds numbers achieved in these simulations.
For higher Reynolds numbers, even larger ensembles would be needed to display the same agreement between the empirical ensemble averages and theoretical predictions.
This ensemble size is sufficient for other statistical measures, though, such as probability distributions and correlation functions, because averages can be taken over ensembles and time translated samples.

In Fig.~\ref{fig:xtraj}, the same detailed range as the one of the inset of Fig.~\ref{fig:phitraj} is shown, which now contains the corresponding sample path of the $X(t)$ process, along with the empirical ensemble and theoretical means. The individual jumps are noticeable: They are equally likely to be positive or negative, and their intensity does not vary as vigorously as for the $\varphi(t)$ variable. The yellow curve is the ensemble average, and it is very close to the theoretical value for the mean of $X(t)$. The global character of this stochastic process is not shown, but it resembles a standard Gaussian process, since the small time scale and the periodicity of the jumps cannot be resolved if the observation window is closer to the integral scale, $T$.

\begin{figure}[ht]
    \centering
    \includegraphics[width=.7\textwidth]{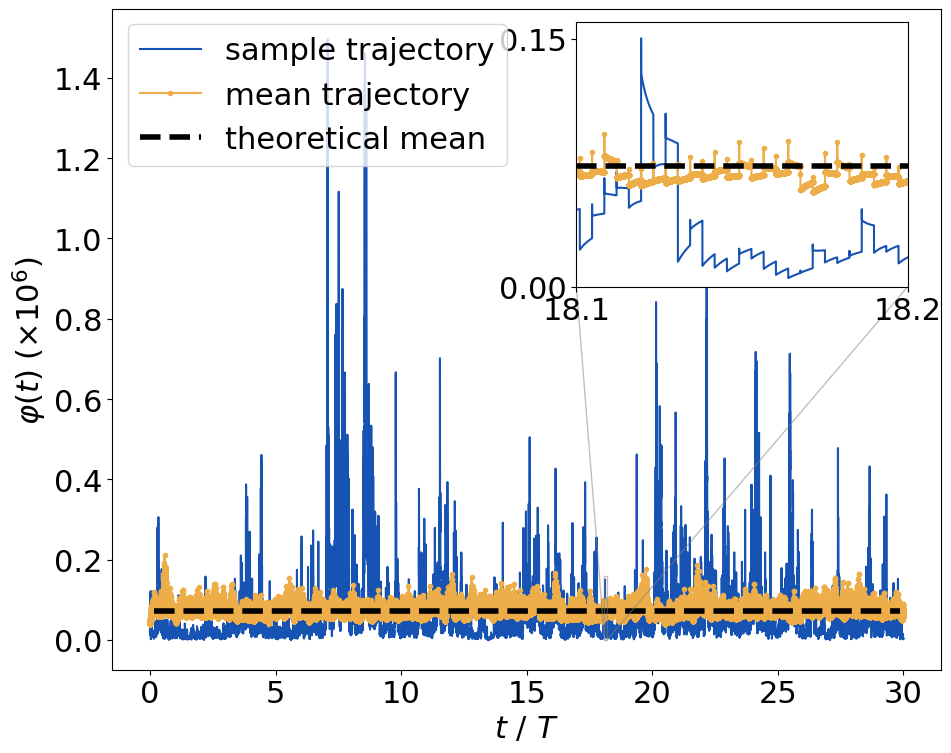}
    \caption
    [A sample path of the shot noise stochastic process]
    {An illustration of the shot noise stochastic process for the energy pseudo-dissipation $\varphi(t)$ (Eq.~\ref{eq:phi-x}). A sample path is drawn (blue), from an ensemble of three hundred paths, and shows strong and non-Gaussian fluctuations, characterized by localized large positive bursts. The ensemble mean (yellow) and the theoretical mean (black, dashed) are shown as well, and it can be seen that the numerical results accurately reproduce the correct average.
    %Another noticeable feature in the ensemble trajectory is how fast the numerical solution reaches the stationary state, starting from the initial condition $\varphi(0)=1/\teta^2$.
    In this picture, $\ln(\teta/T) = -5.60$.
    The inset shows a small stretch of the full time evolution, expanded to show details of the stochastic process at small time scales, where individual jumps can be seen. The inhomogeneity of the fluctuations can also be noticed in this smaller excerpt.
    }
    \label{fig:phitraj}
\end{figure}

\begin{figure}[ht]
    \centering
    \includegraphics[width=.7\textwidth]{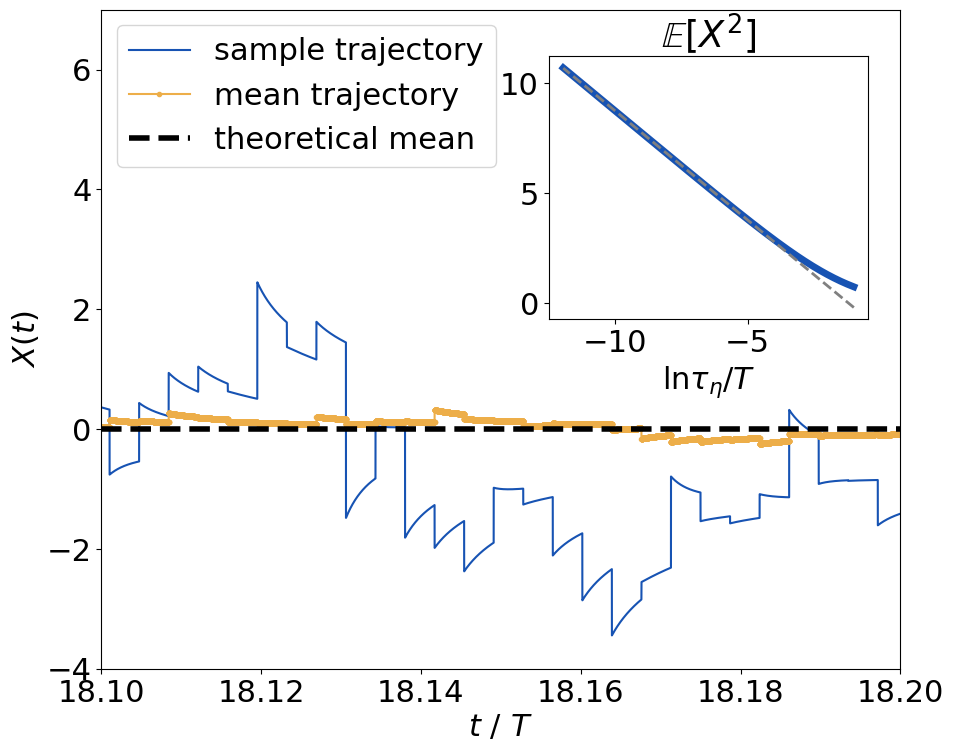}
    \caption
    [Sample path of the underlying Gaussian shot noise stochastic process]
    {
    The interval depicted and the data of this figure are the same as those of the inset in Fig.~\ref{fig:phitraj}. The time evolution of the stochastic Gaussian process $X(t)$ (Eq.~\ref{eq:jump-scalar}) is shown. The individual jumps can be seen at periodic intervals of $\teta$, and the fluctuations are much more regular, since $X$ is a Gaussian process. The colors represent the same data as in the previous figure (with $\ln(\teta/T) = -5.60$): the same sample trajectory in blue, the ensemble average in yellow and the theoretical mean in black, dashed.
    %Again, the ensemble average is consistently close to the theoretical value.
    In the inset, the variance of the process $X(t)$ is shown, with its dependence in $\ln(\teta/T)$. The variance is calculated analytically with Eq.~\eqref{eq:x2-variance} and a clear asymptotic linear behavior as $\teta \to 0$ is indicated by a fit in gray.}
    \label{fig:xtraj}
\end{figure}

In the same figure, in the inset, the asymptotic behavior of the variance of $X(t)$ is shown. For the continuous field in \textcite{pereira2018multifractal}, it was demonstrated that
\begin{equation} \label{eq:lim-varx}
    \mathbb{E}\left[\left(X_P\right)^{2}\right] \underset{\teta \to 0}{\sim} -\ln \left(\frac{\teta}{T}\right) \ .
\end{equation}
The equivalent relation for $X(t)$ is verified in Fig.~\ref{fig:xtraj}.
A linear fit is depicted together with the analytical curve, and the linear coefficient obtained is $0.993$. It has also been observed that this coefficient grows closer to $1.0$, the expected value for the continuous process, as the range of the fit is extended to more negative values of $\ln \teta/T$. This is an important property in the numerical verification that the shot noise driven process indeed displays multifractal statistics.

\begin{figure}[ht]
    \centering
    \includegraphics[width=\textwidth]{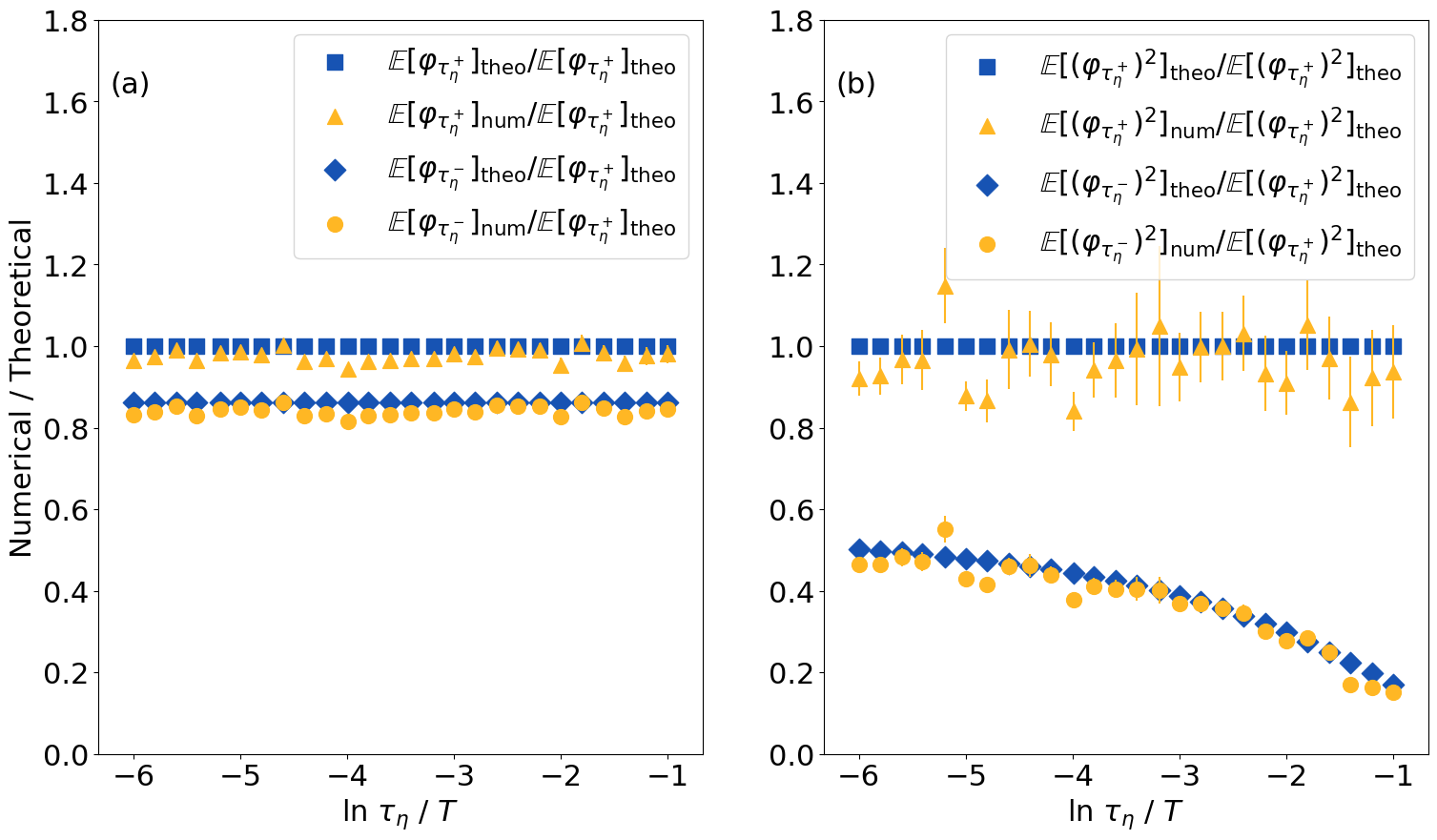}
    \caption
    [Verification of low order one-point statistical properties of shot noise process]
    {Comparison between low-order one-point statistical properties of the numerical solutions of Eq.~\eqref{eq:phi-x} and their exact values. It is a consistency check on the results of the numerical calculations. The ensemble mean (a) and the variance (b) are shown. Both the mean and the variance were calculated at instants immediately before and after the jumps, and these instants are represented respectively by $\teta^-$ and $\teta^+$.
    % Also, to make visualization more clear and the data easier to distinguish, all of the data points are given in units of the respective theoretical values after the jumps.
    Yellow symbols correspond to the numerical results, plotted with error bars in both cases, and blue corresponds to theoretical results.
    %The values of $\ln \teta/T$ range from $-1.0$ to $-6.0$ and display all of the numerical solutions obtained.
    }
    \label{fig:mean_var_match}
\end{figure}

Considering an instant $t$ and all other instants which differ by a multiple of $\teta$ from $t$, these points follow the discrete process described in \textcite{perpete2011} for different initial conditions, and its multifractal properties can be demonstrated analytically. In particular, it is obtained that
\begin{equation} \label{eq:phi-mom-periodic}
    \mathbb{E} [ \varphi^q(t) ]_{\{t \sim t+\ell\teta\}} = \varphi_0^q \exp \left\{ \mathbb{E}[X^2(t)] K(q) \right\} \ ,
\end{equation}
in which the subscript $\{t \sim t+\ell\teta\}$ for the expectation value means that, in addition to the ensemble average, an average over all equivalent instants (separated by a multiple of the dissipative scale $\teta$) is taken as well. From this relation, taking into account Eq.~\eqref{eq:lim-varx}, which has been verified numerically in Fig.~\ref{fig:xtraj}, the multifractal dependence of the statistical moments is obtained:
\begin{equation}
    \mathbb{E} [ \varphi^q(t) ]_{\{t \sim t+\ell\teta\}} = B(t) \ \varphi_0^q \left( \frac{\teta}{T} \right)^{-K(q)} ,
\end{equation}
where $B(t)$ is a function of period $\teta$.
The inset of Fig.~\ref{fig:xtraj}, though, displays the stronger result
\begin{equation} \label{eq:moments-multifrac}
    \mathbb{E} [ \varphi^q ] = \varphi_0^q \left( \frac{\teta}{T} \right)^{-K(q)} \ ,
\end{equation}
where the expectation value is taken over ensemble and time translated samples. This scaling is compatible with the inset of Fig.~\ref{fig:xtraj}, which shows scaling of the time average $\overline{\mathbb{E}[X^2(t)]}$, defined by
\begin{equation} \label{eq:x2-varmean}
    \overline{\mathbb{E}[X^2(t)]} = \frac{1}{\teta} \int_0^{\teta} \mathbb{E}[X^2(t)] \ dt \ ,
\end{equation}
reason for which there is no time dependence.

\begin{figure}[ht]
    \centering
    \includegraphics[width=.7\textwidth]{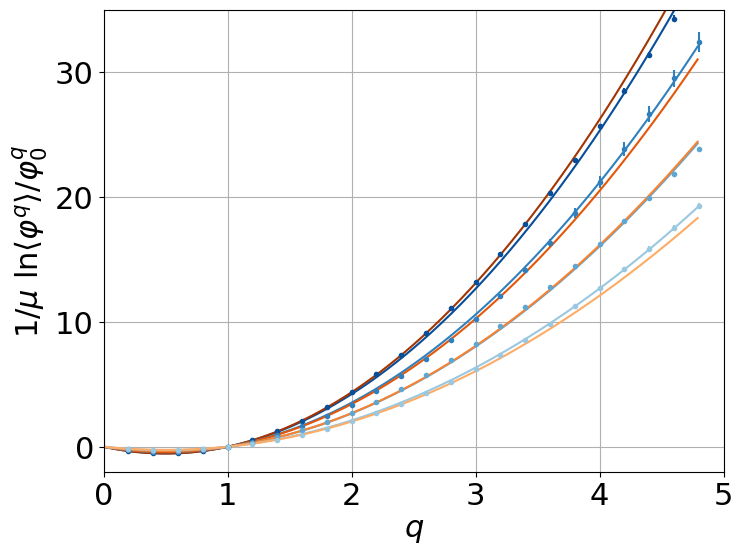}
    \caption
    [Statistical moments of shot noise process]
    {Statistical moments of the $\varphi(t)$ stochastic process, where averages are done over the ensemble and time translated samples. The numerical results correspond to the blue points, which align into a different curve for each value of $\ln \teta/T$, these curves are indicated in blue, calculated with a quadratic fit. All of the blue points include error bars. The values of $\ln\teta/T$ in this figure are $(-3.0,-3.8,-4.6,-5.6)$, with darker colors corresponding to more negative values (higher Reynolds number). In orange, quadratic theoretical curves corresponding to each of these values are displayed.
    %These theoretical curves are quadratic, and follow the blue points and the blue curves closely for most of the calculated moments.
    The curves only deviate from each other for higher moments or higher Reynolds numbers, both regions where a significantly higher statistical ensemble would be needed.}
    \label{fig:parabola}
\end{figure}

Fig.~\ref{fig:mean_var_match} is a consistency test of the numerical solution of Eq.~\eqref{eq:phi-x}, compared with respective analytical results for the mean and variance of $\varphi(t)$ immediately before and after the jumps.
In this figure, the ensemble is larger than in the previous two figures: All independent trajectories were considered, as well as all jumps in a single trajectory. In this fashion, all points immediately before (after) a jump are equivalent in order to calculate the mean and variance of $\varphi(t)$ before (after) jumps, since $\mathbb{E}[X(t)]$ and $\mathbb{E}[X^2(t)]$ vary periodically in time.. The points in yellow correspond to numerical averages while those in blue correspond to theoretical values, and it can be seen that, with little exceptions, the theoretical values are within the error bars of the corresponding numerical data points. Those exceptions are expected to be corrected with a larger statistical ensemble.
%in the time average, can I say that I used all points? and indeed I have not used ALL points, I believe, since there is a stride used to speed up the calculations and test the figures, I don't know if this is the final figure produced.
% There is a problem here: Is this really X or is this observable $\varphi? Because X has mean zero, then how can I divide by Theo[X] and obtain a result that makes sense? This has to be checked!!
%A continuous comparison has been performed and it can be seen that both curves follow the same trends, even if the numerical results show some small fluctuations. In most of the cases it has been observed a strong match between analytical and theoretical results.
%The results are closer for the mean, which is simpler to obtain, but are also quite close for the standard deviation. Only some values deviate from the theoretical ones, and these deviations are consistent across different observables, which indicates that these results would look better with even more data, with larger statistics.

The statistical moments $\mathbb{E}[\varphi^q(t)]$ calculated from the ensemble and time translated samples, are shown in Fig.~\ref{fig:parabola} for several values of $q$ and of $\ln \teta/T$.
This plot verifies relation (\ref{eq:moments-multifrac}), in which all time dependence has been integrated.
The numerical results, in blue points, fall in different quadratic curves according to their value of $\ln\teta/T$, in agreement with
\begin{equation}
    \mathbb{E}[\varphi^q(t)] = \varphi_0^q \exp \left\{ \overline{\mathbb{E}[X^2(t)]} K(q) \right\} \ ,
\end{equation}
with $\overline{\mathbb{E}[X^2(t)]}$ calculated from Eqs.~(\ref{eq:x2-variance}) and (\ref{eq:x2-varmean}).
This value is used to trace the orange theoretical curves in Fig.~\ref{fig:parabola}.
The data points are well approximated by parabolic fits (blue curves) which show reasonable agreement with the theoretical expectations.
Some deviation between the points and the curves are only noticeable for higher Reynolds numbers (more negative values of $\ln \teta/T$), represented by the darker curves, and for the higher moments.
% In both of these cases, it is expected that reliable results follow for larger statistical ensembles.
The blue curves in this figure were obtained with a fit over a quadratic function $K_1(q) = a q (q-1)$, and the agreement with the points and the theoretical curves is remarkable, especially for low order moments. This result is another evidence for the lognormal behavior of the jump stochastic process.

\begin{figure}[ht]
    \centering
    \includegraphics[width=.7\textwidth]{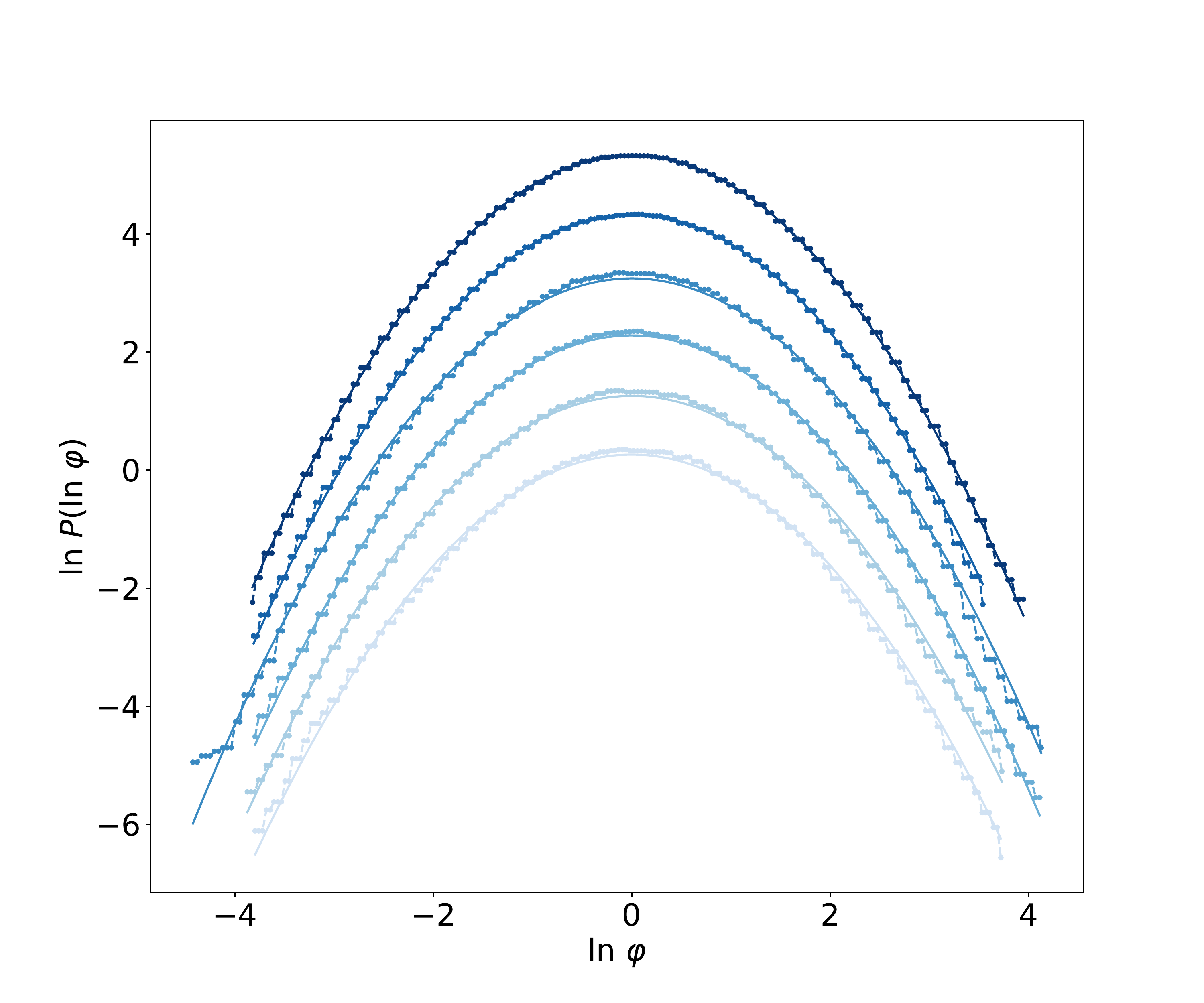}
    \caption
    [Normalized PDFs of shot noise process]
    {Normalized PDFs of $\ln \varphi$ are shown for the following values of $\ln \teta/T: (-6.0, -5.0, -4.2, -3.6, -2.8, -2.0)$, where darker colors correspond to more negative values (higher Reynolds number).
    The curves fall accurately on the continuous curves, which were obtained with a fit through a quadratic curve. This means that the probability distribution of pseudo-dissipation is lognormal for all values of $\teta$. All PDFs have been scaled to a standard Gaussian distribution (mean zero and unit variance), and they have been arbitrarily displaced upwards to simplify visualization.
    %All points were obtained from the ensemble of numerical solutions of Eq.~\eqref{eq:jump-scalar}, and averages over the ensembles and time translated samples have been done.
    }
    \label{fig:pdfs}
\end{figure}

Another form of visualizing the lognormal statistical distribution of the pseudo-dissipation $\varphi$ can be directly implemented from its probability distribution function.
They can be seen in Fig.~\ref{fig:pdfs} for several values of $\ln\teta/T$.
%The blue points correspond to numerically obtained PDFs, from the ensemble of numerical solutions, and the colors follow the same convention as in the other figures, with darker colors representing more negative values of $\ln\teta/T$.
The mean and variance of the pseudo-dissipation have already been verified against their analytical results in Fig.~\ref{fig:mean_var_match}, hence only normalized PDFs (zero mean and unit variance) are shown in Fig.~\ref{fig:pdfs}. In this way, a direct comparison between the PDFs and an exact lognormal distribution can be done. The continuous curves are fits through quadratic functions, revealing that all of the curves fall closely on the expected distribution.

\begin{figure}[ht]
    \centering
    \includegraphics[width=.7\textwidth]{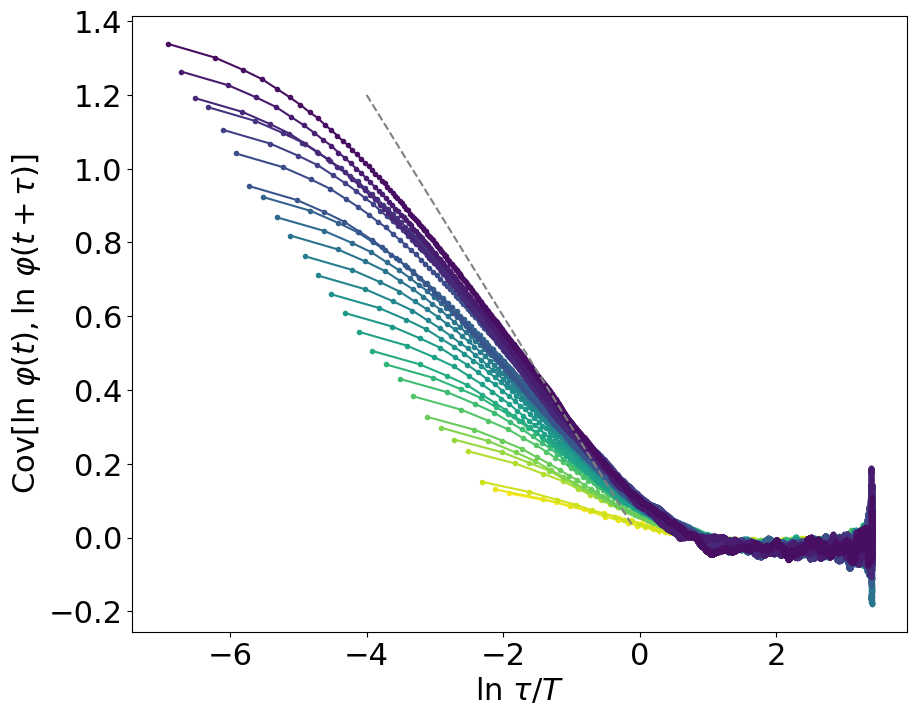}
    \caption
    [Autocovariance of the shot noise process]
    {Numerical results for the autocovariance function of the pseudo-dissipation. $\tau$ is the separation between the points in this function.
    Colors range from yellow to purple, increasing in this order from less to more negative values of $\ln\teta/T$, thus the upper curves, showing a wider scaling region, are those with highest Reynolds numbers.
    The dashed line is the asymptotic relation for autocovariance in the continuous limit, where this function scales linearly with $\ln\tau/T$.
    It can be seen that as the Reynolds number grows, the region where a linear scaling can be seen grows, each curve becomes more closely linear, closer to the theoretical result for the continuous limit.
    % I also have to discuss what are the values of $\ln\teta/T$ which are included here and if they are all values which were simulated (that is, the same values as in the mean_var_match figure)
    }
    \label{fig:cov}
\end{figure}
% change fig. to say "ln" instead of "log"

Besides their lognormal behavior, another of the most relevant features of the dissipation and pseudo-dissipation statistics is their long-range correlations, which the multifractal hypothesis is able to reproduce \parencite{gurvichyaglom1967,meneveausreenivasan1991,sreenivasanantonia1997}. The autocovariance of the pseudo-dissipation field, $\mathrm{Cov}[\ln \varphi(t),\ln \varphi(t+\tau)]$ has been calculated to verify the existence of long-range correlations. The covariance is calculated as
\begin{equation} \label{eq:gen-cov}
    \mathrm{Cov}[X,Y] =
    \mathbb{E}[(X-\langle X \rangle)(Y-\langle Y \rangle)] \ ,
\end{equation}
and the respective numerical results can be observed in Fig.~\ref{fig:cov}. In this figure, $\tau$ is the separation between two data points, where the range of interest lies in $\tau > \teta$. It can be seen that correlations grow for more negative values of $\ln\teta/T$, and as they grow, a larger scaling region can be seen for intermediate values of $\ln\tau/T$. This region is analogous to the inertial range in three-dimensional Navier-Stokes turbulence. In the scaling region, the autocovariance displays a dependence with $\ln\tau/T$, which is very close to linear, a relation which had been observed in \textcite{pereira2018multifractal}. This linear dependence can be understood by noticing the relation
%rewriting the autocovariance of $\ln \varphi(t)$ as
\begin{equation}
    \mathrm{Cov}[\ln \varphi(t),\ln \varphi(t+\tau)]
    = \mu \mathbb{E}[X(t) X(t+\tau)] \ ,
\end{equation}
where a linear dependence in $\mu$ is observed. The second term, the autocorrelation of $X$, is an extension of Eq.~\eqref{eq:lim-varx}, and in the limit $\teta \to 0$, it also displays a linear dependence in $\ln\tau/T$, which leads to
\begin{equation} \label{eq:phi-cov-asymp}
    \mathrm{Cov}[\ln \varphi(t),\ln \varphi(t+\tau)] \underset{\teta \to 0}{\sim} - \mu \ln \left( \frac{\tau}{T} \right) \ .
\end{equation}
The scaling region is a measure of the inertial range and is seen to grow with higher Reynolds. Also, in the gray dashed line, the exact asymptotic relation for the continuous multifractal field, Eq.~\eqref{eq:phi-cov-asymp}, is shown, and it can be observed that the stochastic process with discrete jumps approaches the continuous limit as the intervals between jumps become smaller.

\begin{figure}[ht]
    \centering
    \includegraphics[width=.7\textwidth]{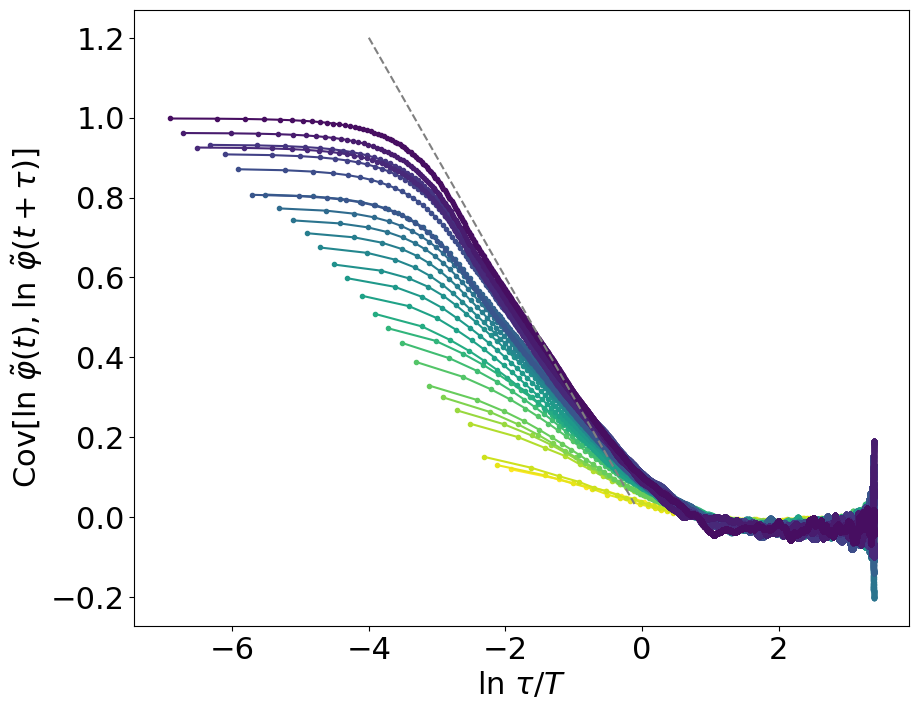}
    \caption
    [Autocovariance of the coarse-grained shot noise process]
    {Autocovariance of the coarse-grained field $\tilde\varphi(t)$, in this figure the local averaging is done over a scale $\teta/2$.
    A clear scaling range, which is much more linear, can be seen in all of the curves, becoming more pronounced as the Reynolds number grows. Also, the slope of these linear curves is much closer to the theoretical value for the continuous limit, which is shown exactly the same as in the previous figure.}
    % i guess i should write a little more ...}
    \label{fig:cov_smooth}
\end{figure}

These statistical properties were also investigated for time averaged fields, denoted by $\tilde \varphi(t)$ and calculated as
\begin{equation} \label{eq:coarse-phi}
    \tilde \varphi(t) = \frac{1}{\tau} \int_{t}^{t+\tau} \varphi(t') \ dt' \ ,
\end{equation}
where $\tau$ is the averaging scale under consideration.
This observable is inspired by the hypothesis of refined similarity in the Lagrangian context as discussed in Section \ref{sec:stat-dissip}.
The one point statistical measures (namely PDF and statistical moments, in Figs.~\ref{fig:parabola} and \ref{fig:pdfs}) of coarse grained fields did not show appreciable difference from their fine grained versions, and for this reason the corresponding figures are not shown.
Yet for the autocovariance, which is a two point statistical observable, a different behavior for the coarse-grained field is noticed, still compatible with the asymptotic description of the continuous field.
In Fig.~\ref{fig:cov_smooth}, the autocovariance of the coarse-grained fields is seen, and the linear behavior observed in Fig.~\ref{fig:cov} for the fine-grained covariance is revealed to be even more pronounced:
The inertial range is more clearly visible, and grows as $\teta\to 0$, and its slope closely approaches the theoretical value in the continuous limit.

\begin{figure}[ht]
    \centering
    \includegraphics[width=.7\textwidth]{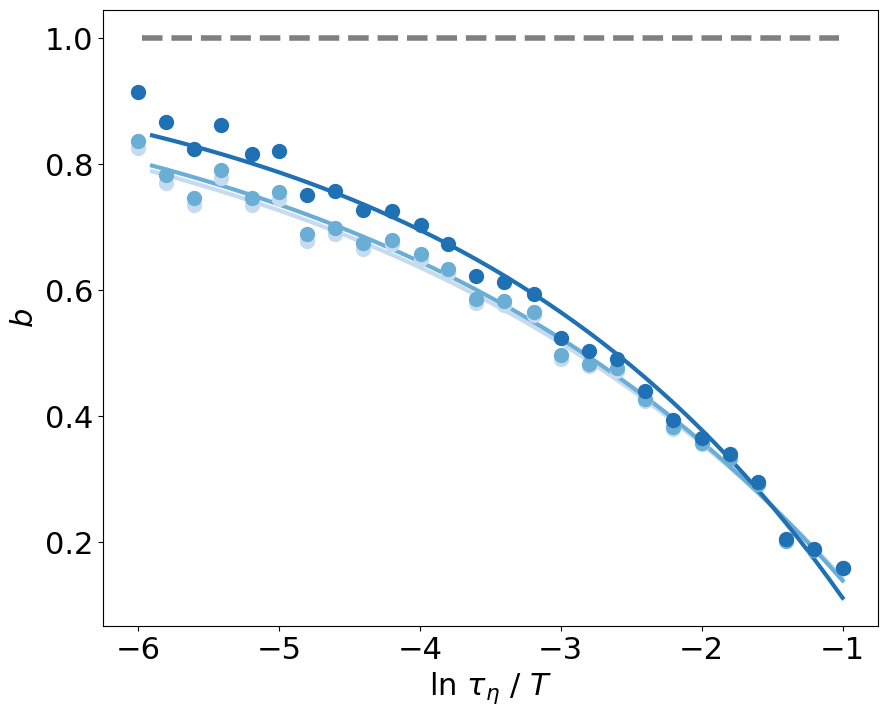}
    \caption
    [Slope of the scaling range of the autocovariance for coarse-grained fields]
    {Each point has been obtained from a numerical fit of the inertial range of the autocovariance, according to Eq.~\eqref{eq:fit_cov}. In this range, the asymptotic scaling Eq.~\eqref{eq:fit_cov} is valid. Each color corresponds to a different coarse-graining scale, where the values shown are $\tau=(\teta/3,\teta/2,\teta)$, higher values are represented in darker colors. An exponential fit, with Eq.~\eqref{eq:b_exp_fit}, through these numerical values was done to demonstrate the tendency of the data to approach the value $b=1$. This exponential fit is shown in the continuous curves. The gray dashed line on the top corresponds to $b=1$, indicating the high Reynolds number limit.}
    \label{fig:cov_irange}
\end{figure}

In order to investigate the convergence to the continuous limit, the autocovariance in the inertial range was fitted to an asymptotic functional form linear in $\ln\tau/T$ with a free parameter:
\begin{equation} \label{eq:fit_cov}
    \mathrm{Cov}[\ln \tilde\varphi(t),\ln \tilde\varphi(t+\tau)]
    = - b \mu \ln \tau/T \ .
\end{equation}
The constant $b$ is a measure of the rate of convergence to the asymptotic continuous behavior, where $b=1$. The evolution of this parameter as the dissipative scale $\teta$ changes can be seen in Fig.~\ref{fig:cov_irange}, where the points correspond to numerical fits over the respective inertial ranges. Each color represents a different coarse-graining scale $\tau$ in Eq.~\eqref{eq:coarse-phi}, and as this scale grows, convergence to the continuous becomes faster. This property was observed in the autocovariance, in Fig.~\ref{fig:cov_smooth}, and is verified in Fig.~\ref{fig:cov_irange}.

Further evidence of the accelerated convergence produced by coarse-graining was obtained with a numerical fit of the curves in Fig.~\ref{fig:cov_irange}. These points slowly approach the asymptotic continuous value, $b=1$, and an exponential fit can make this argument quantitative. The function
\begin{equation} \label{eq:b_exp_fit}
    \chi(\teta) = 1 + \alpha \exp(\beta \ln \teta/T)
\end{equation}
approaches 1 as $\teta \to 0$, and is represented in the figure in continuous curves. The curves serve as a guide to the eye on the evolution of the slope $b$ as the Reynolds number grows, and furthermore show that for the higher values of $\tau$, this convergence is hastened. The exponential shape is only a plausible approximation to a curve which asymptotically approaches a value, hence fluctuations around this curve can be seen in the data. Furthermore, the inertial range is narrow for values of $\ln\teta/T$ closer to zero, which make the fit more delicate in this region. It can also be observed from Fig.~\ref{fig:cov_irange} that an increase of a few percent in the value of $b$ (Eq.~\ref{eq:fit_cov}) would require the smallest $\teta$ to be one or two orders of magnitude lower, corresponding to a significant increase in computational effort.

\section{Discussion}

Positive-definite quantities such as dissipation, pseudo-dissipation and enstrophy have been observed to display nearly lognormal probability distributions and long-range correlations \parencite{yeungpope1989}, and such statistical properties can be understood under the multifractal formulation of turbulent flows, leading to a connection between the statistics and the geometrical properties of the energy cascade.

The stochastic process driven by shot noise discussed in this chapter was verified to display multifractal properties, particularly a lognormal distribution and a power-law long-range correlation. These properties have been verified for fine and coarse grained fields, which is an important feature in the application of such models to real world Lagrangian trajectories.

%New questions that the work brings
%We should look at the statistics of turbulent flows at a finer scale, close to the dissipation range, and investigate means to represent it more accurately as a stochastic process, if this can at all be performed. The connection with Navier-Stokes would still be elusive, but this is a step in the direction of understanding fluctuations in turbulence, particularly Lagrangian fluctuations.

\end{chapter}
\begin{chapter}{Conclusion}
\label{chap:conclusion}

In this dissertation, two main approaches to the theoretical description of large fluctuations in turbulence have been pursued: functional methods and the modelling through non-Markovian stochastic processes, with techniques coming from statistical field theory, random fields, statistical analysis and numerical methods. In this closing chapter, the main results of each previous chapter and directions for further research are laid out.

In Chap.~\ref{chap:rfd}, an analytical technique for the description of PDF cores of non-gaussian observables is described and applied to the RFD model of Lagrangian turbulence. This analysis verified the accuracy of the analytical instanton approximation and determined a hierarchical ordering of the perturbative diagrams according to their relevance. The same technique was also applied to the Burgers model in Chap.~\ref{chap:burgers}, and it was revealed that a provisional noise renormalization procedure, previously observed on the velocity gradient PDFs can be explained as the product of fluctuations around the asymptotic instanton description.

Further questions remain on these matters, nevertheless. Numerical (exact) instantons could generate results of higher precision and a broader regime of validity to the studies of \textcite{apolinario2019instantons,apolinario2019onset}. The exact instantons have been studied in \textcite{grafke2015instanton,grigorio2017instantons,ebener2019}, but not the contribution from their fluctuations.
This analytical approach permits the study of complex statistical observables such as conditional measures without the need to generate huge ensembles for the measurement of the frequency of intense fluctuations.

% A useful aspect of the analytical approach to the velocity gradient PDFs is the possibility of generating large statistical ensembles for conditioned statistics. In this approach, ensembles can be easily produced which are considerably larger than those corresponding to a straightforward numerical solution of the model equations. This is an area of intense current investigation, particularly on algorithms for the preferential sampling of large deviations \parencite{giardina2006,giardina2011,margazoglou2019}.

In the stochastic Burgers model, the negative velocity gradient asymptotic exponent of $3/2$ (Eq.~\ref{eq:left-asymp}), although obtained as a theoretical result in \textcite{balkovsky1997}, has so far not been observed in numerical simulations of this system, instead the value of $1.16$ is reported \parencite{grafke2015relevance}. The investigation of the influence of noise in this problem, though, requires non perturbative techniques to describe the preasymptotic tails of the negative velocity gradient probability distribution.

Further applications of the functional methods described are also possible. For instance, the investigation of circulation statistics in three or two-dimensional Navier-Stokes turbulence \parencite{moriconi2004, smith1997, falkovich2011}, or the transport of passive scalars \parencite{balkovsky1998instanton}.

%Nevertheless, a full characterization of the velocity gradient PDFs in Burgers turbulence remains an open problem. Its behavior in the regime of fully developed turbulence is a challenge which cannot be undertaken with the cumulant expansion, and instead, nonperturbative techniques would be required for the path-integration of fluctuations around the instantons. Improved analytical solutions for the instantons would be required as well.

The work of Chap.~\ref{chap:shotnoise} adds to the effort of building a causal structure to continuous and scale invariant energy cascades. This is done through the modeling of the Lagrangian pseudo-dissipation with a multifractal random field driven by periodic shot noise, which is found to be lognormal and long-range correlated. Nevertheless, futher investigation is required to verify from DNS the statistical properties of Lagrangian dissipation and pseudo-dissipation, in order to understand how well such models represent real energy cascades. In this investigation, understanding the behavior of coarse-grained cascades is crucial, as discussed. 

%It is compelling to note that the analytical advantages of the lognormal formulation make it suitable for applications in several other fields where intermittent fluctuations play a role, besides turbulence, such as in financial economics \parencite{mandelbrot1997,ghashghaie1996,liu1999}, cosmology \parencite{coles1991} and condensed matter systems \parencite{kravtsov1997,serbyn2017}

The understanding of Lagrangian fluctuations is key to the effective modeling of transport properties, either of particles or fields, and to the understanding of the motion of extended structures in turbulence, such as filaments, rods and surfaces. They are heavily influenced by the localized intense bursts of energy dissipation, but still poorly understood theoretically.

\end{chapter}

%%%%%%%%%%%%%%%%%%%%%%%%%%%%%%%%%%%%%%%%%%%%%%%%%%%%%%%%%%%%%%%%%%%%%%%%%

%%%%%%%%%%%%%%%%%%%%%%%%%%% Bibliografia %%%%%%%%%%%%%%%%%%%%%%%%%%%%%%%%

\newpage
\addcontentsline{toc}{chapter}{References}
\printbibliography

%%%%%%%%%%%%%%%%%%%%%%%%%%%%%%%%%%%%%%%%%%%%%%%%%%%%%%%%%%%%%%%%%%%%%%%%%

%%%%%%%%%%%%%%%%%%%%%%%%%%%%% Apendices %%%%%%%%%%%%%%%%%%%%%%%%%%%%%%%%%

\appendix
\begin{chapter}{Computation Resources}
\label{app:comp}

\hspace{5 mm} 

Some of the computational tools developed during the work on this thesis might be of interest to other students, and the purpose of this section is to share these tools.
The tool below was built with the \texttt{Mathematica} software and its code is freely available on Github, with a short link provided.

\section{Tensor Contraction}

This program performs contraction of indices on Euclidean
tensors written with the Einstein summation convention.
There already are specific packages for tensor manipulations,
especially in the context of General Relativity, but they provide
more general functionality at the expense of a more complex
interface.
In the context of the RFD model (Chapter \ref{chap:rfd}), only manipulations of Euclidean tensors are necessary, a case in which there is no difference between covariant and contravariant indices.

The function \texttt{contract} was built in order to carry out tensor contractions in arbitrarily long products of Euclidean tensors. 
For instance, consider the isotropic tensor $G_{ijkl}$ defined in Eq.~\eqref{eq:tensor-g}:
\begin{equation}
G_{ijkl} = 2 \delta_{ik} \delta_{jl} - \frac12 \delta_{il} \delta_{jk} - \frac12 \delta_{ij} \delta_{kl} \ .
\end{equation}
The diagrammatic contributions of Eqs.~\eqref{eq:g_correction} and \eqref{eq:v_correction} are obtained as specific contractions of products of this tensor. 
As a demonstration of this function, some results are shown in Fig.~\ref{fig:contract}, corresponding to the following operations: $G_{ikil}$, $G_{iijj}$, $G_{ijij}$ and $G_{ijkl}G_{klmn}$.
% These resources were also needed in Chap. \ref{chap:shotnoise}.

\begin{figure}[h]
	\centering
	\includegraphics[width=\textwidth]{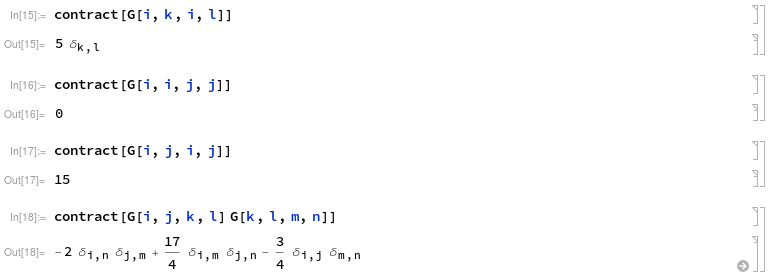}
	\caption
	[Examples of tensor contractions performed with the \texttt{contract} function]
	{A picture from the \texttt{Mathematica} software, with 
	examples of tensor contractions performed with the \texttt{contract} function on the tensor $G_{ijkl}$.}
	\label{fig:contract}
\end{figure}

As another example, the specific contraction which produces the result of Eq.~\eqref{eq:g_correction} is depicted in Fig.~\ref{fig:contract-noise}.

\begin{figure}[h]
	\centering
	\includegraphics[width=\textwidth]{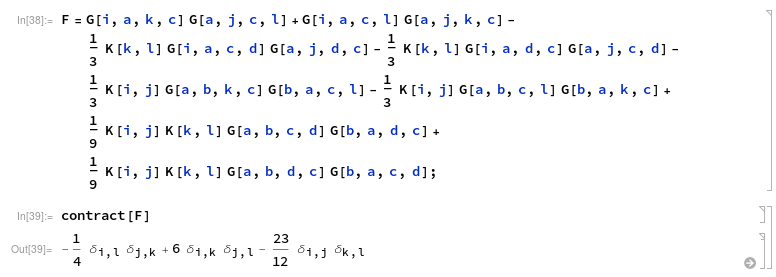}
	\caption
	[Tensor contraction for the noise renormalization in the RFD model.]
	{Tensor contraction for the noise renormalization in the RFD model.}
	\label{fig:contract-noise}
\end{figure}

As a last example, contractions of the tensor $G_{ijkl}$ with arbitrary vectors, denoted by $u$ and $v$, are shown, in Fig.~\ref{fig:contract-vector}.

\begin{figure}[ht]
	\centering
	\includegraphics[width=\textwidth]{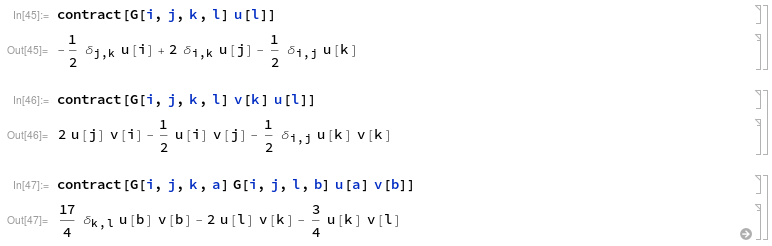}
	\caption
	[Tensor contractions including products of tensors and vectors.]
	{Tensor contractions including products of tensors and vectors.}
	\label{fig:contract-vector}
\end{figure}

A final remark is that this function can only perform contractions of repeated indices, while the value of individual elements of a tensor should be obtained manually, or with other tools.

The code for the \texttt{contract} function is available at https://git.io/fjQe8.
%\href{https://git.io/fjQe8}{https://git.io/fjQe8}.

\end{chapter}

\begin{chapter}{Cumulant Expansion}
\label{app:cumul}

\hspace{5 mm}

The cumulant expansion is a standard technique in statistics which is used to approximate a distribution or expectated value. They can be used as a simple way to describe the differences between a distribution and its Gaussian approximation. In quantum field theory, in particular, it is often necessary to compute $\langle e^x \rangle$, and the cumulant expansion offers a systematic way to obtain this result perturbatively in terms of the moments of the random variable $x$, which are $\langle x^p \rangle$.

%How to derive the formula for the cumulant expansion to arbitrary order. This was first seen in Chapter~\ref{chap:rfd}, in Eq.~\eqref{eq:cumul-gamma}.

This particular derivation is carried out to order $O(x^3)$, but it can be extended without further complications to arbitrary order. It is interesting to notice that there is no general way to obtain the coefficients of the cumulant expansion, other than following an algorithm, be it the one below or an equivalent.

To begin with, an expansion of the exponential produces
\begin{equation}
\begin{split}
	\langle e^x \rangle &= \left\langle 1 + x + \frac{x^2}{2!} + O(x^3) \right\rangle \\
	&= 1 + \langle x \rangle + \frac{ \langle x^2 \rangle}{2} + O(x^3) \ .
\end{split}
\end{equation}
Taking the logarithm of both sides and a further series expansion, the following result is obtained:
\begin{equation}
\begin{split}	
	\ln \langle e^x \rangle &= \ln \left( 1 + \langle x \rangle + \frac{\langle x^2 \rangle}{2} + O(x^3) \right) \\
	&= \langle x \rangle + \frac{\langle x^2 \rangle}{2} - \frac{ \langle x \rangle^2}{2} + O(x^3) \ .
\end{split}
\end{equation}
The desired result is then achieved as an exponential of the previous result,
\begin{equation}
	\langle e^x \rangle \simeq \exp\left\{ \langle x \rangle + \frac12 \big( \langle x^2 \rangle - \langle x \rangle^2 \big) + O(x^3) \right\} \ .
\end{equation}
By dropping the error term, this is a pragmatic way to perturbatively evaluate the given expected value.

Further references on cumulants can be found in \textcite{cumulants,novak2012}, and a list of coefficients for expansions of high order is provided as sequence A127671 of the Online Encyclopedia of Integer Sequences (OEIS), available in \textcite{oeis_cumul}.

\end{chapter}

%%%%%%%%%%%%%%%%%%%%%%%%%%%%%%%%%%%%%%%%%%%%%%%%%%%%%%%%%%%%%%%%%%%%%%%%%

\end{document}